\documentclass[aps,prc,superscriptaddress,11pt,showpacs,notitlepage]{revtex4}

	\usepackage{graphicx}
	\usepackage{amsmath,amssymb,mathrsfs}
\usepackage{epsf}
\usepackage{epsfig}

\usepackage{color}

\usepackage{amsmath}

\usepackage{amssymb}

\numberwithin{equation}{section}

\usepackage{subfigure}

\newcommand{\be}{\begin{equation}}
\newcommand{\ee}{\end{equation}}
\newcommand{\bea}{\begin{eqnarray}}
\newcommand{\eea}{\end{eqnarray}}

\newcommand{\bb}{\bibitem}
 
\newcommand{\eqn}{\begin{eqnarray}}
\newcommand{\eqnx}{\end{eqnarray}}

\begin{document}
\title{Skyrme models and nuclear matter equation of state}

\author{C. Adam}
\affiliation{Departamento de F\'isica de Part\'iculas, Universidad de Santiago de Compostela and Instituto Galego de F\'isica de Altas Enerxias (IGFAE) E-15782 Santiago de Compostela, Spain}
\author{M. Haberichter}
\affiliation{School of Mathematics, Statistics and Actuarial Science, University of Kent, Canterbury, CT2 7NF, UK}
\author{A. Wereszczynski}
\affiliation{Institute of Physics,  Jagiellonian University,
Lojasiewicza 11, Krak\'{o}w, Poland}

\begin{abstract}
We investigate the role of pressure in a class of generalised Skyrme models. We introduce pressure as the trace of the spatial part of the energy-momentum tensor and show that it obeys the usual thermodynamical relation. Then, we compute analytically the mean-field equation of state in the high and medium pressure regimes by applying topological bounds on compact domains. 
The equation of state is further investigated numerically for the charge one skyrmions. We identify which term in a generalised Skyrme model is responsible for which part in the equation of state. Further, we compare our findings with the corresponding results in the Walecka model. 
\end{abstract}
\maketitle 

\section{Introduction}
The Skyrme model \cite{skyrme} is a low energy effective model of QCD where baryons (atomic nuclei) are realised as emergent objects in a ``mesonic fluid''. In modern language, they are solitons in a field theory based entirely on some mesonic field $U$, in the simplest version, $U \in SU(2)$. The pertinent topological charge carried by a soliton (skyrmion) may be identified with the baryon charge. 
\\
The most general Lagrange density respecting Poincare invariance and leading to a standard Hamiltonian reads
\be
\mathcal{L}_{0246}=\mathcal{L}_0+\mathcal{L}_2+ \mathcal{L}_4+\mathcal{L}_6,
\ee
where
\be
\mathcal{L}_2=-\lambda_2 \mbox{Tr}\; (L_\mu L^\mu), \;\;\; \mathcal{L}_4=\lambda_4 \mbox{Tr} \; ([L_\mu , L_\nu]^2), \;\;\; \mathcal{L}_6 =-(24\pi^2)^2 \lambda_6 \mathcal{B}_\mu \mathcal{B}^\mu. 
\ee
Here, the left invariant current $L_\mu$ is given by
\be
 L_\mu = U^\dagger \partial_\mu U
\ee
and $\mathcal{L}_0$ is a non-derivative part, i.e., a potential. In its simplest version $\mathcal{L}_{024}$, where the sextic part is neglected, the model was shown to describe baryons and light nuclei with a very good accuracy \cite{sk}-\cite{Lau:2014baa}. However, this particular proposal leads to results which are difficult to reconcile with two important qualitative properties of atomic nuclei and nuclear matter. It gives rather large binding energies and describes crystal-like nuclear matter \cite{cryst}, whereas nuclear matter is only weakly bound in atomic nuclei and behaves more like a fluid. These unwanted properties of the usual Skyrme model can be resolved if one adds a larger amount of the sextic term. This is related to the observation that there exists a submodel within the family of Skyrme type models which has the BPS property (classical zero binding energies \cite{BPS}) and describes a fluid (has the energy-momentum tensor of a perfect fluid \cite{term} and is, in fact, a field theoretical realisation of the Eulerian fluid \cite{fluid grav}). This submodel is called {\it the BPS Skyrme model} \cite{BPS}  and reads
\be
\mathcal{L}_{BPS} \equiv \mathcal{L}_6+\tilde{\mathcal{L}}_0
\ee
where $\tilde{\mathcal{L}}_0$ is a further potential.
Then, an improved proposal which might also be valid for the high baryon charge regime, is given by {\it the near-BPS Skyrme model} \cite{nearBPS}
\be
\mathcal{L}=\epsilon\left(\mathcal{L}_0+\mathcal{L}_2+\mathcal{L}_4\right) + \mathcal{L}_{BPS}, \label{full}
\ee  
where the usual Skyrme model only provides rather small corrections to masses (binding energies), as $\epsilon$ is assumed to be a small parameter. Let us remark that the sextic term $\mathcal{L}_6$ can be effectively introduced by a coupling with the $\omega$ meson \cite{jackson}-\cite{ding}. Here we consider the general Skyrme theory with the $SU(2)$ fields only, where some effects induced by further mesons emerge due to a particular form of the action rather than by the inclusion of some new explicit degrees of freedom. 
\\
Although the BPS limit is physically very well motivated and provides rather accurate results for binding energies for larger nuclei (after a careful treatment of semiclassical quantisation of (iso)-rotational modes, inclusion of the Coulomb interaction and iso-spin breaking \cite{nearBPS} (see also Ref. \cite{Marl})), it is still necessary to investigate the full near-BPS version. The inclusion of the non-BPS part leads to the appearance of physical pions and determines the proper geometry (shape) of skyrmions \cite{sp1}, which is of high importance as it results in particular patterns for the iso-rotational excitations. Hopefully, one can get a model which unifies the convincing results for baryons and light nuclei with the crucial properties of the BPS model. This is, in principle, a difficult task since the solvability property of the BPS model is lost once the perturbative part (usual Skyrme action) is added. Moreover, such a perturbation is non-analytical as the non-BPS part is the dominating part close to the vacuum. Hence, no small $\epsilon$ expansion exists at the level of the field equations. Some recent findings for solitons with first few baryon charges, however in a not so ``near-BPS'' regime, can be found in Ref. \cite{sp2} (see also \cite{nitta}). 

The aim of the present work is to investigate the issue of the equation of state (EoS) in the general Skyrme model with a special focus on the near-BPS model. The BPS submodel allows for a rather complete analysis of its thermodynamics at zero temperature, as a consequence of the BPS property and of its perfect fluid form which, however, no longer hold for the general (non-BPS) model. The first difficulty, therefore, is to properly define the pressure, since in the absence of a perfect fluid description no obvious definition exists. This is analysed in section 2. In section 3, using some topological energy bounds on a compact domain, an asymptotic equation of state is derived. Then, in section 4, we make a conjecture about subleading terms in the EoS by exploiting some properties of generalised BPS Skyrme models. In section 5, the medium and large pressure regimes are checked numerically for the charge one, hedgehog configuration. Section 6 is devoted to a comparison with the equation of state (EoS) of the Walecka model. Finally, we summarise our findings. We shall frequently use the standard Skyrme field parametrization
\be \label{stand-par} 
U= \cos \xi +i\sin \xi \; \vec n \cdot \vec \sigma \; , \quad \vec n = (\sin \chi \cos \Phi , \sin \chi \sin \Phi ,\cos \chi)
\ee
where $\xi ,\chi ,\Phi$ are real field variables and $\vec \sigma$ are the Pauli matrices. Further, we shall always assume that $U= {\bf 1}$, i.e., $\xi =0$, is the vacuum value of the Skyrme field which it must approach at spatial infinity.
\section{Average pressure and average chemical potential}
The energy-momentum tensor in the BPS Skyrme model takes the form of a perfect fluid energy-momentum tensor and, therefore, it directly defines the field theoretical (microscopic) pressure and baryon chemical potential (i.e., the pressure and chemical potential densities). For a generic Skyrme Lagrangian, there is no such fluid description and one usually has a non-trivial energy-momentum tensor with different stresses $ T^{ij}$ in different directions. However, a pressure density $p= \frac{1}{3}\sum_i T^{ii}$ and the corresponding {\it average pressure} $P$, i.e., the volume average of $\frac{1}{3} T^{ii}$ (summation assumed), as well as an average chemical potential $\bar{\boldsymbol{\mu}}$ may still be defined as proper thermodynamic (macroscopic) variables, i.e., obeying the required thermodynamical relations
\be
\left( \frac{\partial E}{\partial V} \right)_B=-P, \label{p1}
\ee
\be
\left( \frac{\partial E}{\partial B} \right)_V= \bar{\boldsymbol{\mu}}\label{mu1}.
\ee
Here $E=E(V, P, B, \bar{\boldsymbol{\mu}})$ is the static energy of the Skyrme model, depending on the volume $V$, pressure $P$, baryon charge $B$ and chemical potential $\bar{\boldsymbol{\mu}}$.
For this purpose, we first define the following generalized step function
\be 
\tilde \Theta (U) =
\left\{
\begin{array}{c}
1 \quad \mbox{for} \quad U\not=1 \\
0 \quad \mbox{for} \quad U=1
\end{array}
\right.\,,
\ee
(for the standard parametrization (\ref{stand-par}), $\tilde \Theta (U)$ may be replaced by the standard step function $\Theta (\xi)$). Further, we define the locus function of a static skyrmion configuration $U=U_0 (\vec x)$ as the pullback $U^*_0 (\tilde \Theta (U))$ of $\tilde \Theta (U)$ under $U_0$, and the locus set of the skyrmion $U_0 (\vec x)$,
\be
\Omega = \{ \vec x \in \mathbb{R}^3 \; | \;  U^*_0 (\tilde \Theta (U)) \equiv \tilde \Theta (U_0 (\vec x))=1 \},
\ee
i.e., the set $\Omega \subset \mathbb{R}^3$ where the skyrmion is located (deviates from the vacuum).
Now, let us consider the general Skyrme static energy functional 
\be \label{en-func}
E (V,P,B,\bar{\boldsymbol{\mu}})=\int d^3x \; \varepsilon [U] + P\left( \int d^3 x  \tilde \Theta (U(\vec x))-V \right) - 
\bar{\boldsymbol{\mu}} \left( \int d^3 x \mathcal{B}_0 - B\right) , 
\ee
where $\varepsilon [U]$ is the energy density and $P$ and $\bar{\boldsymbol{\mu}} $ are Lagrange multipliers. In particular, $P$ imposes the condition that all possible solutions of the variational problem (\ref{en-func}) must have  volume $V$, i.e., $\int d^3 x  \,\tilde \Theta (U(\vec x)) = \int_\Omega d^3 x =V$. 
Obviously, $P$ and $\bar{\boldsymbol{\mu}}$ obey the thermodynamical relations (\ref{p1}) and (\ref{mu1}) by construction. 
We still have to show that the Lagrange multiplier $P$ is indeed the average pressure as defined above. For that we consider scaling transformations $x^i \rightarrow e^\lambda x^i=(1+\lambda)x^i$. Then $\delta_\lambda U=\lambda x^i \partial_i U$.  For the energy functional (\ref{en-func}) we get off-shell, in first order in $\lambda$, 
\be
\delta_\lambda E = \lambda  \int T_{ii}d^3 x -3\lambda P\int d^3x \tilde \Theta (U(\vec x)),
\ee
where we use the fact that the variation of the energy  $\int d^3 x \; \varepsilon [U]$ under the scaling transformation gives the integral of the trace of the spatial part of the energy-momentum tensor \cite{sp1}. But any solution of the Euler-Lagrange equations is a stationary point and, therefore, 
$\delta_\lambda E =0$ on-shell and
\be
P=\frac{\frac{1}{3} \int_\Omega T_{ii}d^3 x}{\int_\Omega d^3x} \label{p2},
\ee 
which is exactly the average pressure definition. We remark that the lagrange multiplier imposing a fixed value for the volume does not uniquely determine the compact domain $\Omega$. In fact, using the volume preserving diffeomorphisms we get infinitely many different sets $\Omega$ which result in different values of the trace integral (i.e., the pressure $P$). In other words, one has to fix the compact domain $\Omega$ and, therefore, the skyrmion solution up to target space symmetries, to get a unique average pressure. In general, this is a very difficult task. However, for a spherically symmetric unit charge skyrmion, the obvious choice is such that the symmetry of the equilibrium solution is preserved while squeezed. Then, $\Omega $ is a three dimensional ball. Moreover, in the BPS limit the static energy functional is invariant under the volume preserving diffeomorphisms of the base space and therefore all {\em SDiff} related $\Omega$ give exactly the same pressure - as is the case for a perfect fluid. 

Let us also remark that the proper (thermodynamical) volume appearing in relation (\ref{p1}) is the geometrical volume of a topological soliton. It means that in a typical case such a volume tends to infinity in the equilibrium (zero pressure) limit since typical skyrmions are infinitely extended solutions. This leads to some difficulties if one wants to consider the mean-field energy density $\bar{\varepsilon}=E/V$ at the equilibrium. Obviously, one gets $\bar{\varepsilon}=0$ which is the same as for the vacuum configuration.  Note, however, that compact skyrmions with finite volumes are known (for example, they are quite common in the BPS Skyrme model).

\section{Asymptotic equation of state}
Now we will use these definitions to compute thermodynamical functions in the asymptotic regime, i.e., in the limit of high pressure (density) or equivalently in the limit of small volume. In order to do that we need an expression that shows how the energy of skyrmions changes with volume and baryon density. This is a quite difficult problem as, in principle, it requires the solution of the full three dimensional problem. However, in the asymptotic regime we can use some previously found topological bounds. 

The relevant bounds for the quartic (Skyrme) and sextic terms are \cite{bound1,bound2}
\be
E_4 = \frac{1}{16} \int d^3 x \; \mbox{Tr} \; [L_i, L_j]^2 \geq 3(2\pi^2)^{4/3} \frac{B^{4/3}}{V^{1/3}}, \label{E4bound}
\ee
\be
E_6 = \int d^3 x \; (\epsilon^{ijk} \; \mbox{Tr} L_iL_jL_k)^2 \geq \frac{B^2}{V}. 
\ee
There is no bound for the quadratic part
\be
E_2 = \int d^3 x \; \mbox{Tr} \; L_i L_i,
\ee
or for a general potential. Therefore, the asymptotic static energy for any Skyrme model can be approximated as ($\lambda_6=\lambda^2/(24)^2$ and $\lambda_0=\mu^2$)
\be
E = \lambda_2E_2+\lambda_4 E_4 +\lambda^2 \pi^4 E_6 + \lambda_0 E_0 \geq \lambda_4 3(2\pi^2)^{4/3} \frac{B^{4/3}}{V^{1/3}} + \pi^4\lambda^2  \frac{B^2}{V} \geq \pi^4\lambda^2  \frac{B^2}{V}.
\ee
In the small volume limit, the quartic term provides a subleading contribution. The same holds for the potential term and the sigma model part, which affect the total energy even less. However, since in the small volume limit the sextic term governs the masses of skyrmions, one recovers the BPS limit. Hence, the bound is, in fact, saturated for sufficiently small volumes (high pressures). This is not the case for the quartic term. As we will see, the corresponding bound is not saturated. All this allows us to predict the asymptotic formula for the energy as
\be
E=\pi^4\lambda^2 \frac{ B^2}{V} + \alpha \frac{B^{4/3}}{V^{1/3}} + o(V^{-1/3}), \;\;\; V \rightarrow 0, \label{pred}
\ee
with 
\be
\alpha \geq 3 (2\pi^2)^{4/3}\lambda_4. 
\ee
Then, the average pressure is
\be
P = \pi^4\lambda^2 \bar{\rho}_B^2  + \frac{\alpha}{3} \bar{\rho}_B^{4/3} + o( \bar{\rho}_B^{4/3}),
\ee
and the average baryon chemical potential
\be
\bar{\boldsymbol{\mu}} = 2\pi^4\lambda^2 \bar{\rho}_B + \frac{4\alpha}{3} \bar{\rho}_B^{1/3} + o( \bar{\rho}_B^{1/3}),
\ee
where $\bar{\rho}_B=\frac{B}{V}$ is the average baryon (particle) density. 
The average energy density is
\be
\bar{\varepsilon} = \pi^4\lambda^2 \bar{\rho}_B^2  + \alpha \bar{\rho}_B^{4/3} + o( \bar{\rho}_B^{4/3}).
\ee
In the leading approximation, we re-obtain the BPS Skyrme model asymptotic behaviour 
\be
P = \pi^4\lambda^2 \bar{\rho}_B^2, 
\ee
\be
\bar{\boldsymbol{\mu}} = 2\pi^4\lambda^2 \bar{\rho}_B ,
\ee
\be
\bar{\varepsilon} =   \pi^4\lambda^2 \bar{\rho}_B^2.
\ee
The resulting equation of state reads
\be
\bar{\varepsilon} = P .
\ee

Let us summarise the main findings:
\begin{enumerate}
\item The sextic term gives the main contribution in the high pressure limit. This means that this term should not be omitted if dense nuclear matter is considered. The asymptotic equation of state always has a universal (potential independent) form $\bar{\varepsilon}=P$. 
\item The quartic, usual Skyrme term, gives a subleading contribution which modifies the equation of state at moderate pressures. The functional dependence is known, which is not the case for the multiplicative constant $\alpha$, for which we have derived a lower bound.  
\item The potential and the sigma model part give contributions which are even subleading in comparison to the $E_4$ contribution. On the other hand, they may be significant close to nuclear saturation density. 
\end{enumerate}

For a deeper insight into the role played by the sigma model part as well as by the potential term, we need to perform numerical computations. Nonetheless, one can get some help from the BPS Skyrme model which, as we will see below, can give some understanding on the functional mass-volume relation originating from the $E_2$ and $E_0$ terms.
\section{The BPS fluid toy models}
Each part of the generalised Skyrme model has a specific number of spatial derivatives. This is a main - but clearly not the only - difference between them. As we have previously commented, the sextic term is special as it leads to a perfect fluid model. Here, we want to learn how a specific number of derivatives can change the equation of state, assuming that all  terms are based on the same thermodynamical quantity - here the baryon density. 
Within this approach, we replace the sigma model term and the quartic Skyrme term by energy functionals based only on the baryon density, which is a natural quantity for the sextic term (Note, that similar non-linear models have been recently considered in \cite{krusch} in the context of the baby Skyrme model in $2+1$ dimensions). 

We start with the Skyrme (quartic) term
\be
\mathcal{L}_4=\lambda_4 \mbox{Tr} \; ([L_\mu , L_\nu]^2) \longrightarrow \mathcal{L}_4^f=\lambda_4 ( \mathcal{B}^\mu \mathcal{B}_\mu)^{2/3},
\ee
which now is represented by a "four derivative" term (in the sense that under Derrick scaling $x^\mu \to \Lambda x^\mu$ it scales like $\Lambda^{-4}$) constructed from the baryon density (current). 
Such a new term leads to the following energy density (for static solutions)
\be
\varepsilon=\lambda_4 \rho_B^{4/3},
\ee
and pressure
\be
P=\frac{1}{3}\lambda_4 \rho_B^{4/3},
\ee
where we use that, for a general energy density based on baryon (particle) density, the pressure reads
\be
P=\rho_B \frac{\partial \varepsilon}{\partial \rho_B} -\varepsilon.
\ee
The corresponding equation of state reads 
\be
P= \frac{1}{3} \bar{\varepsilon}, 
\ee
which exactly coincides with the mean-field equation of state for the usual Skyrme term. Hence, the substitution completely reproduces the thermodynamical properties of the $\mbox{E}_4$ model. This leads to the conjecture that, at least to some extent, thermodynamical properties related to each term of the generalised Skyrme model can be similar to properties generated by the BPS fluid counterpart, i..e, by terms built out of the baryon density elevated to a certain power. Now, let us apply the same strategy to extract information about possible equations of state that may emerge from $\mbox{E}_2$ and $\mbox{E}_0$. 
In the case of the sigma model term we get 
\be
\mathcal{L}_2=-\lambda_2 \mbox{Tr}\; (L_\mu L^\mu)  \;\; \longrightarrow \;\;  \mathcal{L}_{2}^f=- \lambda_2 (\mathcal{B}_\mu \mathcal{B}^\mu)^{1/3} . 
\ee
Thus, we find
\be
\varepsilon=\lambda_2 \rho_B^{2/3} , 
\ee
\be
P=-\frac{\lambda_2}{3} \rho_B^{2/3} , 
\ee
and the equation of state
\be
P=-\frac{1}{3}   \varepsilon.  
\ee
As expected, this part is responsible for a negative pressure, i.e., an attractive force (which is balanced by the inclusion of other stabilising terms). 

The issue of the potential is more subtle as it does not depend on the baryon density but, in general, on the Skyrme field. One finds that 
\be
\varepsilon = \mathcal{U}, \;\;\; P=-\mathcal{U},
\ee
and
\be
P=-\varepsilon.
\ee
However, for the step-function potential $\mathcal{U}=\tilde \Theta (U)$, it only contributes as a numerical constant to thermodynamical quantities. Then, for example, the energy density is shifted by such a constant. Since in a first approximation, we can always model a potential by the step-function potential, we may expect that a constant density term shows up.   

Hence, finally we conjecture a theoretically motivated energy density-baryon density relation which can be valid not only for asymptotically high pressure but also in a medium pressure regime 
\be
\bar{\varepsilon}=\pi^4\lambda^2 \bar{\rho}_B^2 + \alpha \bar{\rho}_B^{4/3} +\beta + \tilde{\beta} \; \bar{\rho}_B^{2/3} +O(1).
\ee
This leads, for the charge one sector, to the following energy-volume formula
\be
E=\pi^4\lambda^2 \frac{1}{V}+ \alpha \frac{1}{V^{1/3}} +\beta V +\tilde{\beta} \; V^{1/3} +O(V) \label{EV-theor}.
\ee
In the subsequent sections we will compare formula (\ref{EV-theor}) with numerical computations. One should keep in mind that, in contrast to the first two leading terms, which have been proven to emerge in the high pressure regime, the last two subleading terms are conjectured (or at best heuristically motivated) by analysing similar BPS models. 
\section{Numerical results in the charge one sector}
The equation of state as well as the equivalence of the two pressure definitions can be tested by direct numerical computations in the charge one sector where the hedgehog ansatz can be assumed
\be
U= \cos \xi +i \sin \xi \vec{n} \cdot \vec{\sigma},
\ee
where $\xi = \xi (r)$, $\vec{n}=(\sin \theta \cos \phi, \sin \theta \sin \phi, \cos \theta)$ is a unit vector covering the $\mathbb{S}^2$ target subspace once, and $\vec{\sigma}$ are the Pauli matrices. Then, the energy contributions are given by
\be
E_0=4\pi \int r^2 dr 2(1-\cos \xi),
\ee
\be
E_2=4\pi \int r^2 dr 2 \left(\xi_r^2+\frac{2\sin^2\xi}{r^2} \right),
\ee
\be
E_4=4\pi \int r^2dr \left( 2 \frac{\sin^2 \xi}{r^2} \xi^2_r + \frac{\sin^4 \xi}{r^4}\right),
\ee
\be
E_6=4\pi \int r^2 dr \frac{1}{4r^4} \sin^4 \xi \xi^2_r.
\ee
The aim is not only to verify the asymptotic formula for the EoS but also to get some insight into the role that is played by the quadratic and potential parts, for which we do not have any analytical (asymptotic) expression. For $B=1$ skyrmions we solve a one-dimensional ODE with boundary conditions which guarantee the nontrivial topology and pressure
\be
\xi (r=0)=\pi, \;\;\; \xi (r=R)=0.
\ee
In other words, we enclose the skyrmion in a finite volume $V=(4\pi/3) R^3$.  

We use the collocation method \cite{Ascher:1981} to determine the profile function $\xi$ which minimises the 
Skyrme energy functional $E$ with boundary conditions $\xi(0)=\pi$ and $\xi(R)=0$. The associated 
Euler-Lagrange equations are solved on the interval $0<r<R$. For the high pressure regime, we consider the interval $0.1<R<1$. 
We start with $R=0.1$ as  right boundary point and then increase in each collocation run the right boundary value by $\text{d}r=0.0001$. 
The increment $\text{d}r$ is chosen to be sufficiently small which allows us to check numerically the equivalence of field-theoretical and 
thermodynamical pressure. For the low pressure regime, we run simulations up to a maximal boundary point value of 10 with $\text{d}r=0.001$.

We use the nonlinear curve-fitting function {\texttt optimize.curve\_fit} from Python's {\texttt SciPy} package \cite{SciPy} to fit the asymptotic energy and pressure formulas to our numerical data in the high pressure regime. 

The energy and length units have to be fixed by comparison with experimental nuclear physics data. In this article, we calibrate the usual Skyrme model without the sextic part using the approach of Refs.~\cite{sk}. 
The energy and length units are tuned to match the experimental nucleon ($M_N=939\,\text{MeV}$) and 
delta ($M_\Delta=1231\,\text{MeV}$) masses. This fixes the energy and length scale as follows
\begin{align}
12\pi^2\times\left[\frac{F_\pi}{4e_{\text{Sky}}}\right]=12\pi^2\times 5.58\,\text{MeV}\,,\quad \left[\frac{2}{e_{\text{Sky}}F_\pi}\right]=0.755\,\text{fm}\,,
\label{Cali_ANW}
\end{align}
and gives the well-known parameter values
\begin{align}
e_{\text{Sky}}=4.84,\quad F_\pi=108 \,\text{MeV},\quad \hbar=46.8,\quad \text{ and } \quad m_\pi= 138 \,\text{MeV}\,.
\label{Cali_ANW_para}
\end{align}
In this article, if not explicitly stated otherwise, the rescaled pion mass $\mu$ is set to 1. Note that the limitations of the parameter set (\ref{Cali_ANW_para}) have been discussed previously \cite{Battye:2005nx,Houghton:2005iu,Battye:2014qva} and different parameter choices have been explored in the literature, e.g. Refs.~ \cite{light,Battye:2005nx,Lau:2014baa,Foster:2015cpa}.

Our numerical results in section~\ref{Sec_NumRes_nearBPS} on BPS and near BPS Skyrme models are expressed in terms of the same energy and length units (\ref{Cali_ANW}). However, note that for the full near-BPS Skyrme model $E_{0246}$, discussed in section~\ref{Sec_NumRes_nearBPS_E0246}, the coupling parameters of the BPS part are fixed by matching the BPS Skyrmion mass to one-fourth of the helium nucleus mass and the Skyrmion radius to the nucleon radius (see  section~\ref{Sec_NumRes_nearBPS_E0246} and Ref.~\cite{Adam:2014dqa} for a detailed discussion).
\subsection{The perturbative Skyrme model}
We start with the usual, old Skyrme model without the sextic part. The reason for this is to find a clear signal from the quartic term which now gives the leading high pressure behavior and then understand the influence of the potential and sigma model terms as first subleading effects. The energy functional we minimise reads
\be
\mbox{E}_{024}=\lambda_2 E_2 + \lambda_4 E_4 + \lambda_0 E_0, 
\ee
where the constants are 
\be
\lambda_2= \frac{1}{24\pi^2}, \;\;\; \lambda_4= \frac{1}{12 \pi^2}, \;\;\; \lambda_0 = \frac{1}{12 \pi^2},
\ee
or zero in the case where the pertinent term is omitted. Here we consider the usual Skyrme potential which is relevant for the conventional (perturbative) Skyrme model
\be
E_0 = \; \mbox{Tr} \;  (1-U).
\ee
For all possible submodels we first checked the equivalence between the average field-theoretical pressure and the thermodynamical pressure. We always found a perfect agreement between the pressures computed by Eqs. (\ref{p1}) and (\ref{p2}). It is also a precision test for our numerical computations. 
\subsubsection{$\mbox{E}_4$ model}
\begin{figure}
\subfigure[]{\includegraphics[totalheight=6.0cm]{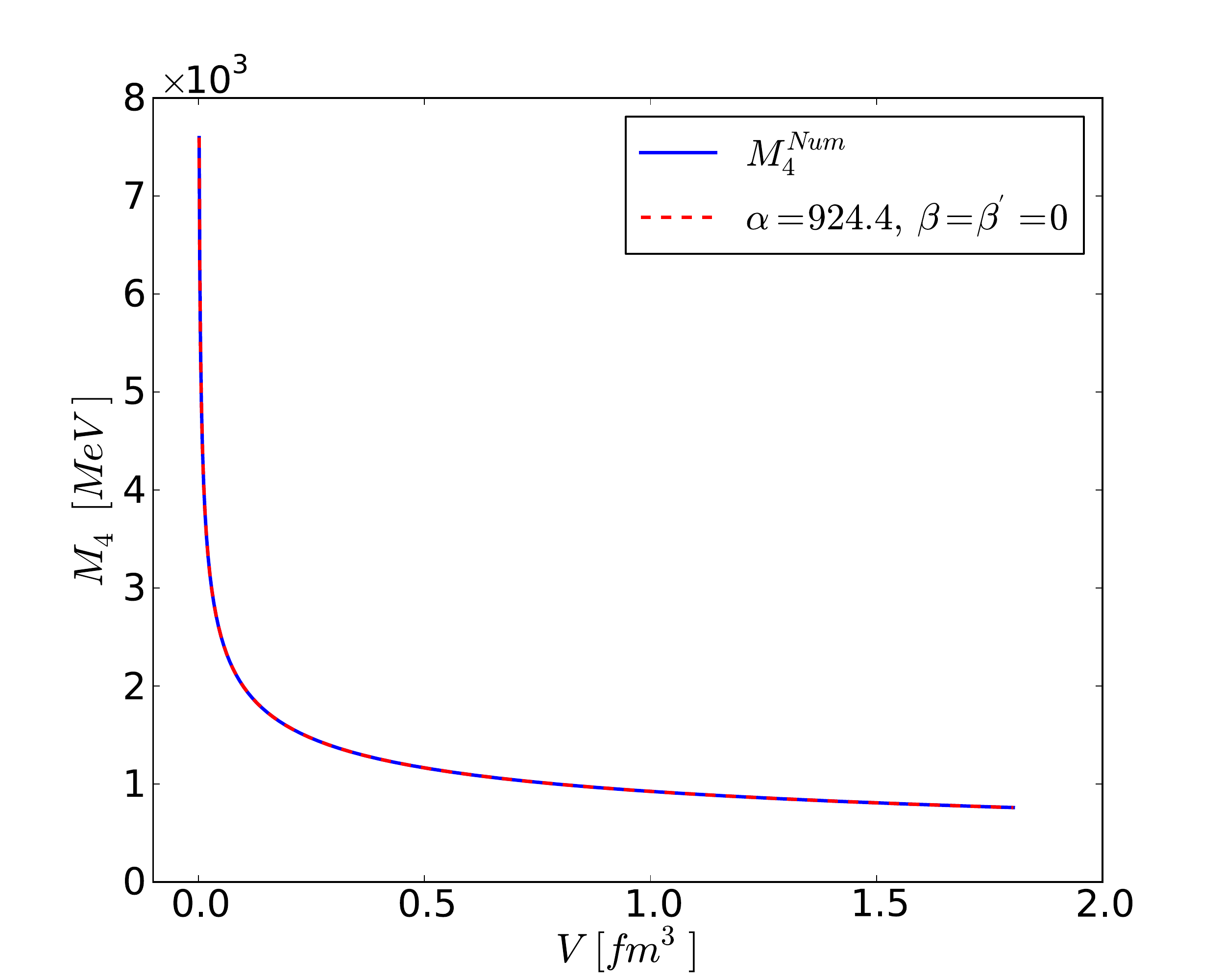}}
\subfigure[]{\includegraphics[totalheight=6.0cm]{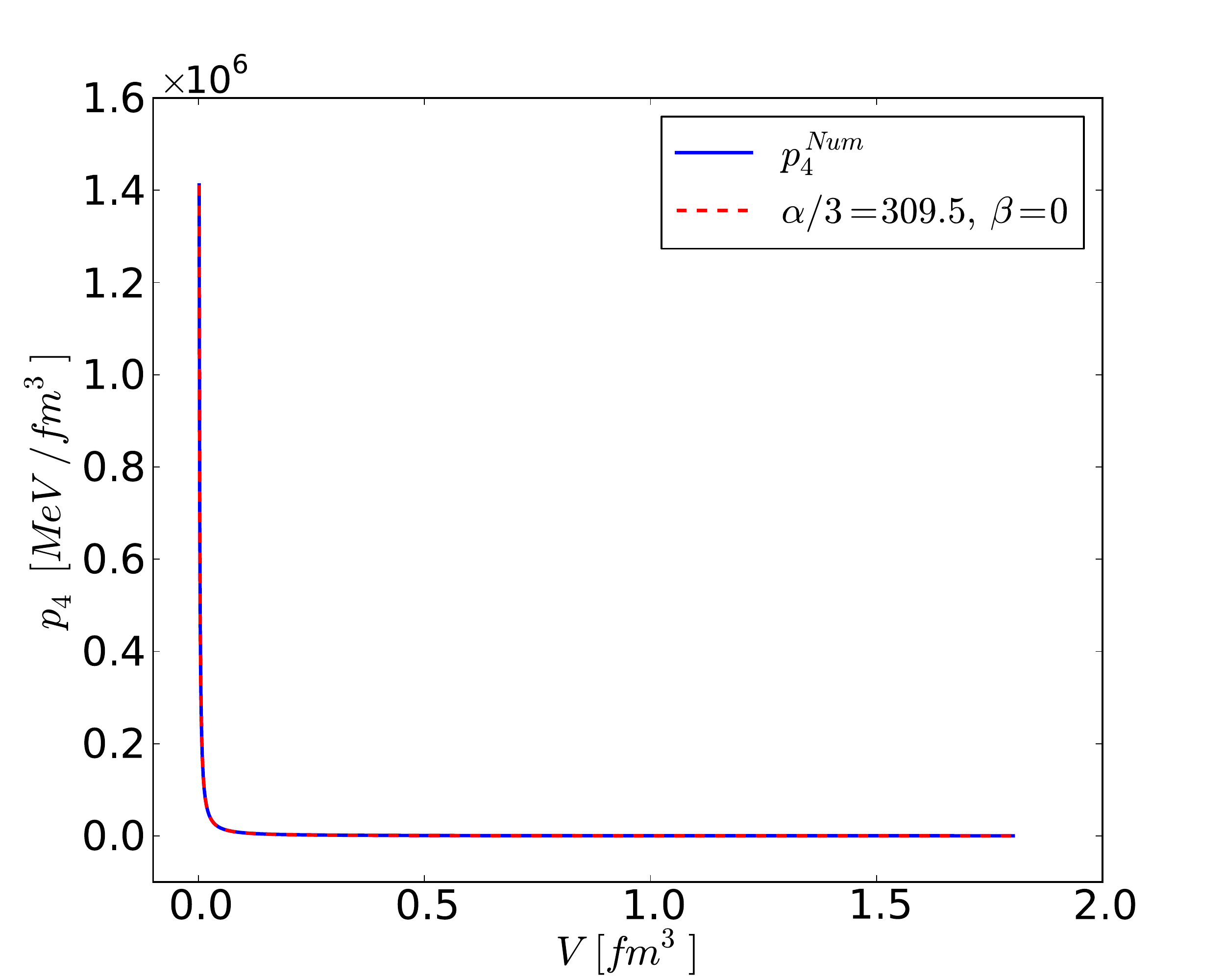}}
\caption{(Color online) (a) Energy and (b) pressure  of the B=1 skyrmion as a function of the volume for $E_4$ model.}
\label{E4}
\end{figure}

We begin with the simplest submodel which contains only the quartic part. Such a limit of the Skyrme model has been considered before, however, from the instanton point of view \cite{sp3}. Obviously, due to the Derrick theorem there are no stable solitonic solutions at equilibrium (zero pressure), i.e., in the full infinite $\mathbb{R}^3$. However, if we close the system in a finite volume $V$, then skyrmions do exist. For a charge one soliton we have the following energy integral
\be
\mbox{E}_4=\lambda_4 E_4= \frac{1}{12 \pi^2} 4\pi \int_0^R dr r^2 \left( 2 \frac{\sin^2 \xi}{r^2} \xi^2_r + \frac{\sin^4 \xi}{r^4}  \right).
\ee
This leads to a relatively simple equation of motion
\be
\frac{d^2}{dr^2} \cos \xi + \frac{1}{r^2} \sin^2 \xi \cos \xi=0
\ee
accompanied by the pertinent boundary conditions. Unfortunately, we were not able to solve it analytically. 
It is straightforward to notice that a solution on the segment $[0,R]$ can be related to a solution on a different segment $[0,\Lambda R]$ by the scale transformation $r \rightarrow \Lambda r$. Then, the energy scales with the factor $\Lambda^{-1}$, which leads to
\be
\mbox{E}_4[V] = \frac{1}{V^{1/3}} \mbox{E}_4[V=1] \equiv \alpha \frac{1}{V^{1/3}},
\ee
where the coefficient $\alpha$ can be understood as the energy in the unit volume.
Hence, the pressure is a simple function of the volume 
\be
P= \frac{\alpha}{3} \frac{1}{V^{4/3}}.  
\ee
Such an equation of state is confirmed by numerical computations. Indeed, we find that 
\be
\alpha = 1.853 \;\;\;\; \mbox{or} \;\;\;\; \alpha = 924.4 \; \mbox{MeV fm},
\ee
where the first value is in Skyrme units while the second is in physical units. It is worth to compare this with the previously derived bound. Namely, $3(2\pi^2)^{4/3} / (12\pi^2)= 1.351$. This shows that the bound is {\it not} saturated. The true solution is significantly above the bound. 

Finally, we get an exact density-pressure equation of state which is valid for any value of the pressure
\be
\bar{\varepsilon}=3P. \label{eos-e4}
\ee
This result can be easily generalised to any topological sector and to any given compact $\Omega$. Indeed, let us consider the energy functional
\be
\mbox{E}_4[\Omega]= \frac{1}{12 \pi^2} \frac{1}{16} \int_\Omega d^3 x\;  \mbox{Tr} \; [L_i,L_j]^2.
\ee
Obviously, it scales homogeneously under scaling transformations. Then again
\be
\mbox{E}_4 [V; \Omega]= \frac{\alpha_{B, \Omega}}{V^{1/3}},
\ee
where the constant $\alpha$ depends on the topological charge and a particular form of the compact manifold (assumed now to have unit volume). From bound (\ref{E4bound}) we know that
\be
\alpha_{B,\Omega} \geq 3(2\pi^2)^{4/3} B^{4/3}.
\ee
However, as the energy functional is not invariant under the volume preserving transformations this constant should depend on the choice of $\Omega$. In any case, it proves that the equation of state (\ref{eos-e4}) is an off-shell universal equation of state for the pure quartic model. 

Finally let us remark that the bound {\it is} saturated if the geometries of the base and target spaces coincide. Indeed, if the base space is a three-sphere with a radius $R$ then we get
\be
\mbox{E}_4^{\mathbb{S}^3}= \frac{1}{12 \pi^2} 4\pi \frac{1}{R} \int_0^\pi d\psi  \left( 2 \sin^2 \xi \xi^2_\psi + \frac{\sin^4 \xi}{\psi^2}  \right),
\ee
where $\psi$ is the third angular coordinate on the sphere. This variational problem has a solution $\xi =\psi$. Hence, the charge one skyrmionic solution is just the identity map between the base and target spheres. One can easily verify that in this case the bound is equal to the value of the last integral. This agrees with a classical result of Manton \cite{manton-s3} where he shows that the identity map is a stable solution for the $\mbox{E}_{24}$ Skyrme model for a sufficiently small radius $R$ of the base space sphere. The geometrical reason why the $\mathbb{R}^3$ and $\mathbb{S}^3$ cases are so distinct is the following. The bound can be saturated if and only if the three eigenvalues of the corresponding strain tensor are 1) all equal and 2) constant which means that the corresponding map must be an isometry \cite{bound1}. This is a very restrictive condition which, for the target space $\mathbb{S}^3$, may be fulfilled only if the base space is also a three-sphere. 
\begin{figure}
\subfigure[]{\includegraphics[totalheight=6.cm]{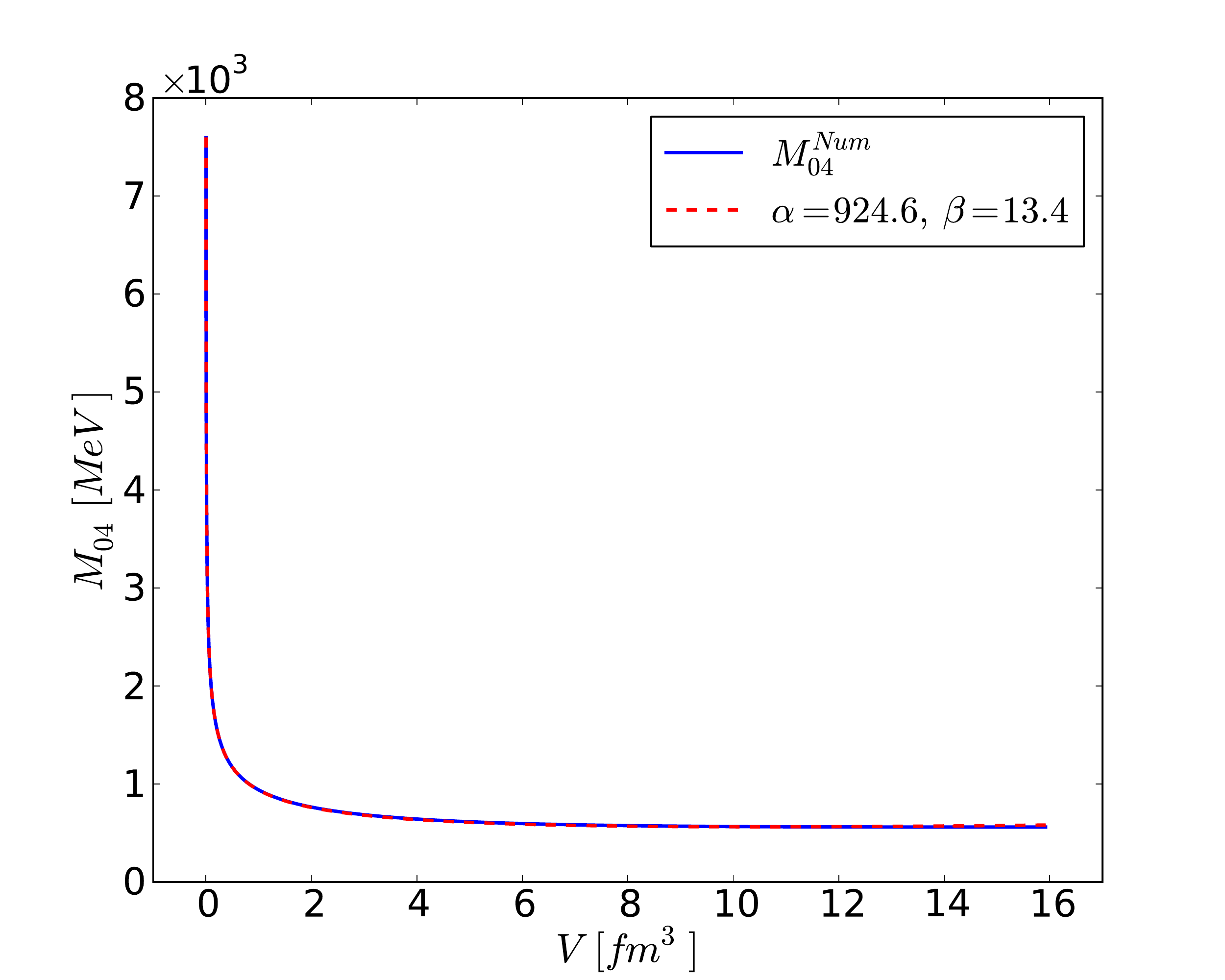}}
\subfigure[]{\includegraphics[totalheight=6.cm]{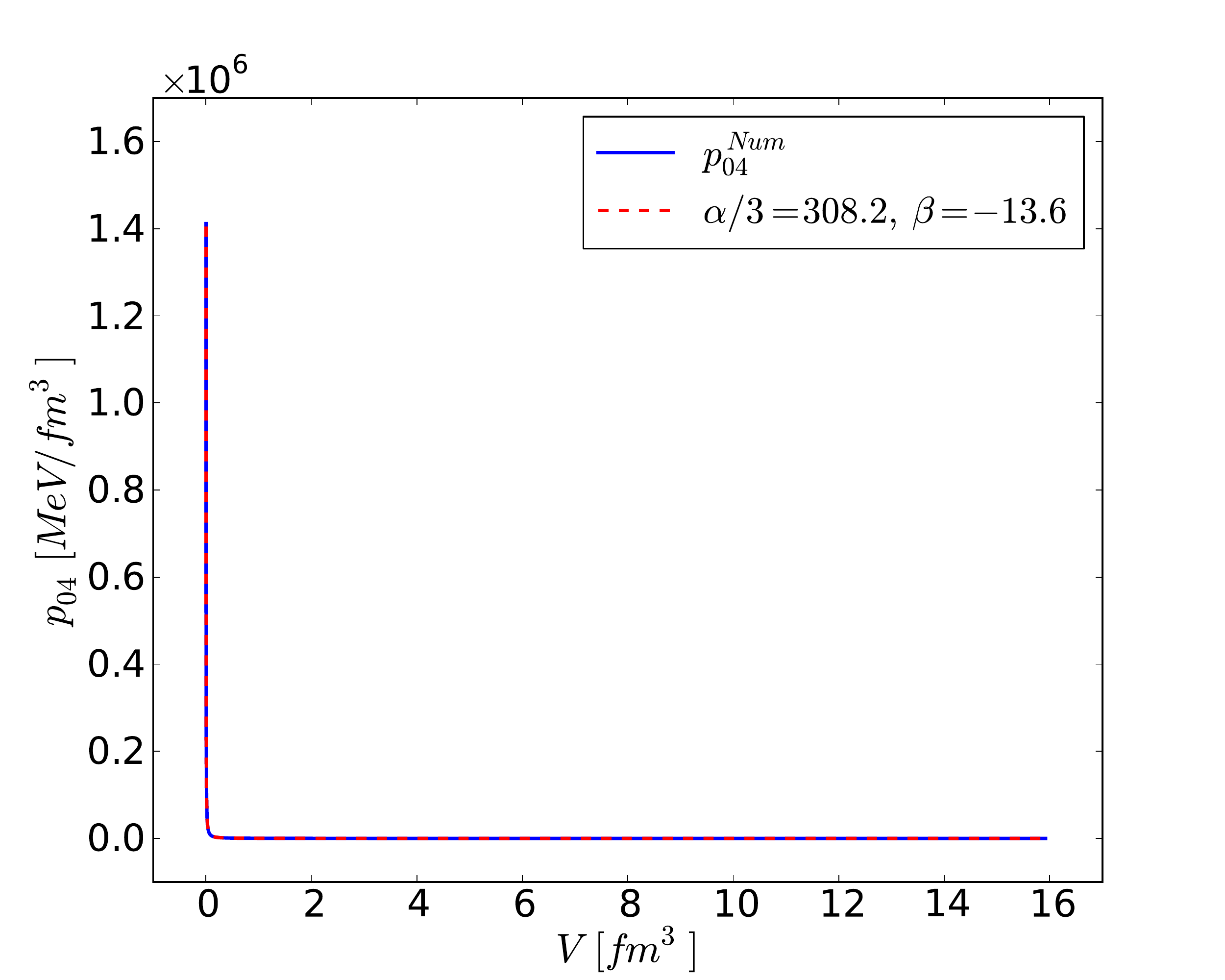}}\\
\subfigure[]{\includegraphics[totalheight=6.cm]{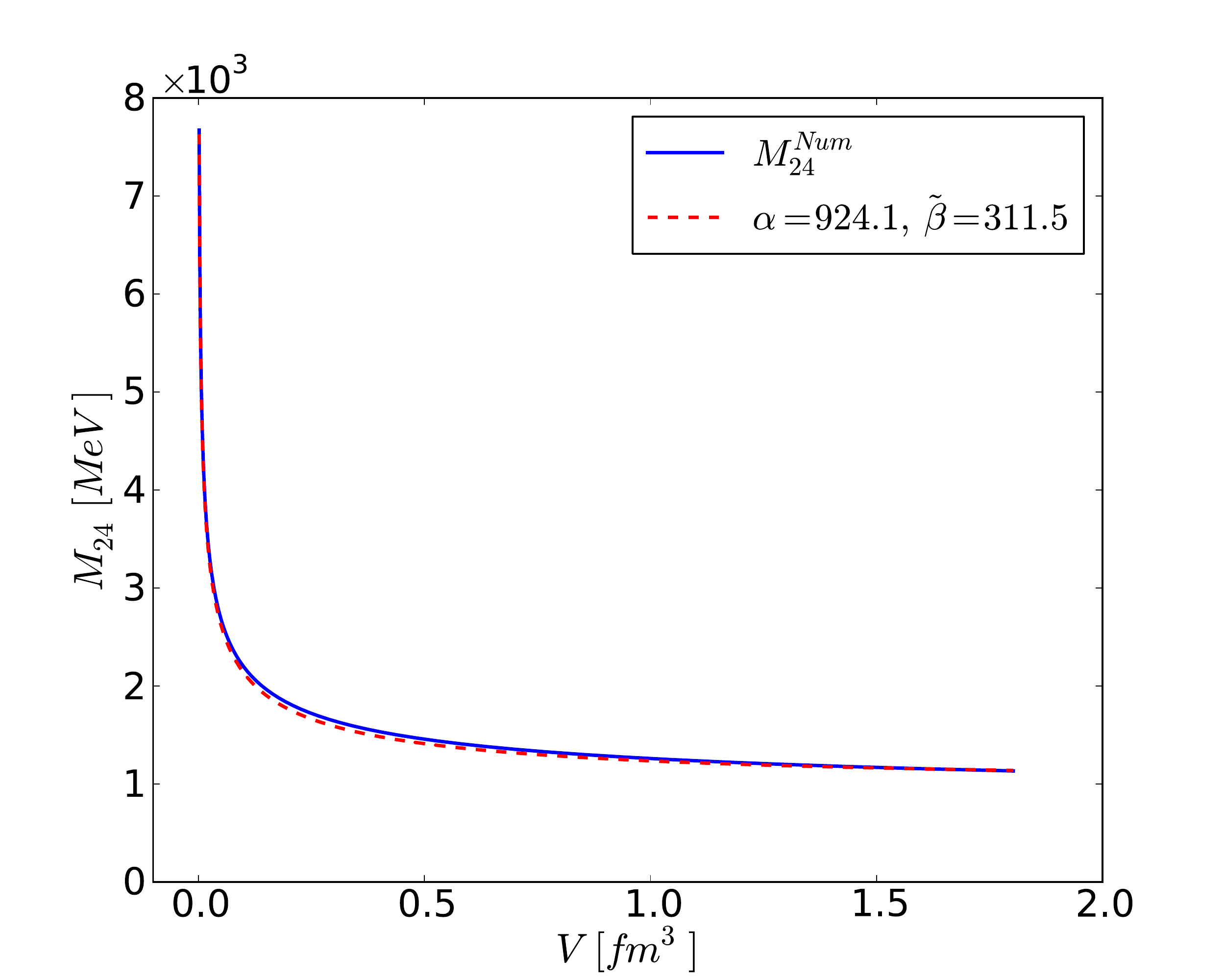}}
\subfigure[]{\includegraphics[totalheight=6.cm]{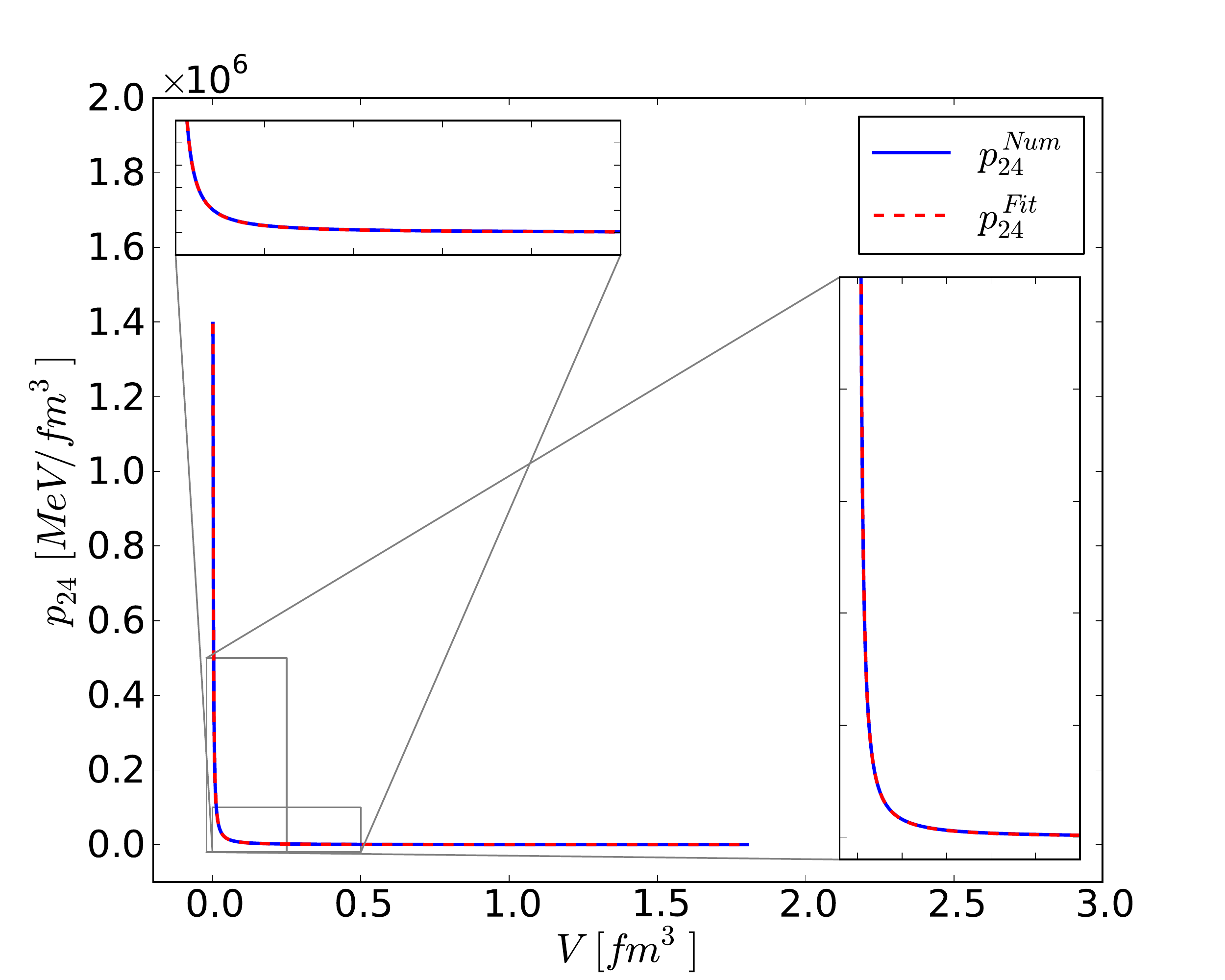}}
\caption{(Color online) Energy and pressure  of the B=1 skyrmion as a function of the volume for $E_{04}$ and $E_{24}$ model.}
\label{E04}
\end{figure}
\subsubsection{$\mbox{E}_{04}$ model}
The pure quartic Skyrme model can be stabilised by the addition of a potential term. Then, the Derrick argument is evaded and an equilibrium solution exists. One peculiarity of this submodel (with the usual Skyrme potential) is that one gets a compact skyrmion, i.e., a topological soliton which achieves its vacuum value at a finite distance.  The size of the compact charge one skyrmion at equilibrium is $$R=2.07 \;\;\; \mbox{or} \;\;\; R=1.563 \; \mbox{fm},$$ where the first number is in Skyrme units while the second in physical units.
This sets the maximum volume (i.e., the volume at zero pressure) to $V_{max}=15.9 \; \mbox{fm}^3$. 

The general asymptotic energy formula
\be
\mbox{E}= \alpha \frac{1}{V^{1/3}} + \tilde{\beta} V^{1/3} +\beta V +o(V) \label{E04-theor},
\ee
gives the following values for the parameters
\be
\alpha_{04}=924.6 \; \mbox{MeV fm}, \;\; \beta_{04}=13.4\; \mbox{MeV fm}^{-3} , \;\; \tilde{\beta} _{04}=0.
\ee
In Fig.~\ref{E04} we plot the numerically computed energy (mass) and pressure as a function of the volume together with the theoretical formulas. We find a perfect agreement. We can also use the theoretical formula (\ref{E04-theor}) to get the mass of the equilibrium skyrmion. Then, we get $M_{04}=580 \; \mbox{MeV}$ which is only $3.5 \%$ above the true mass.  Undoubtedly, the theoretical formula works very well for this submodel. 
\subsubsection{$\mbox{E}_{24}$ model}
\begin{figure}
\subfigure[]{\includegraphics[totalheight=6.cm]{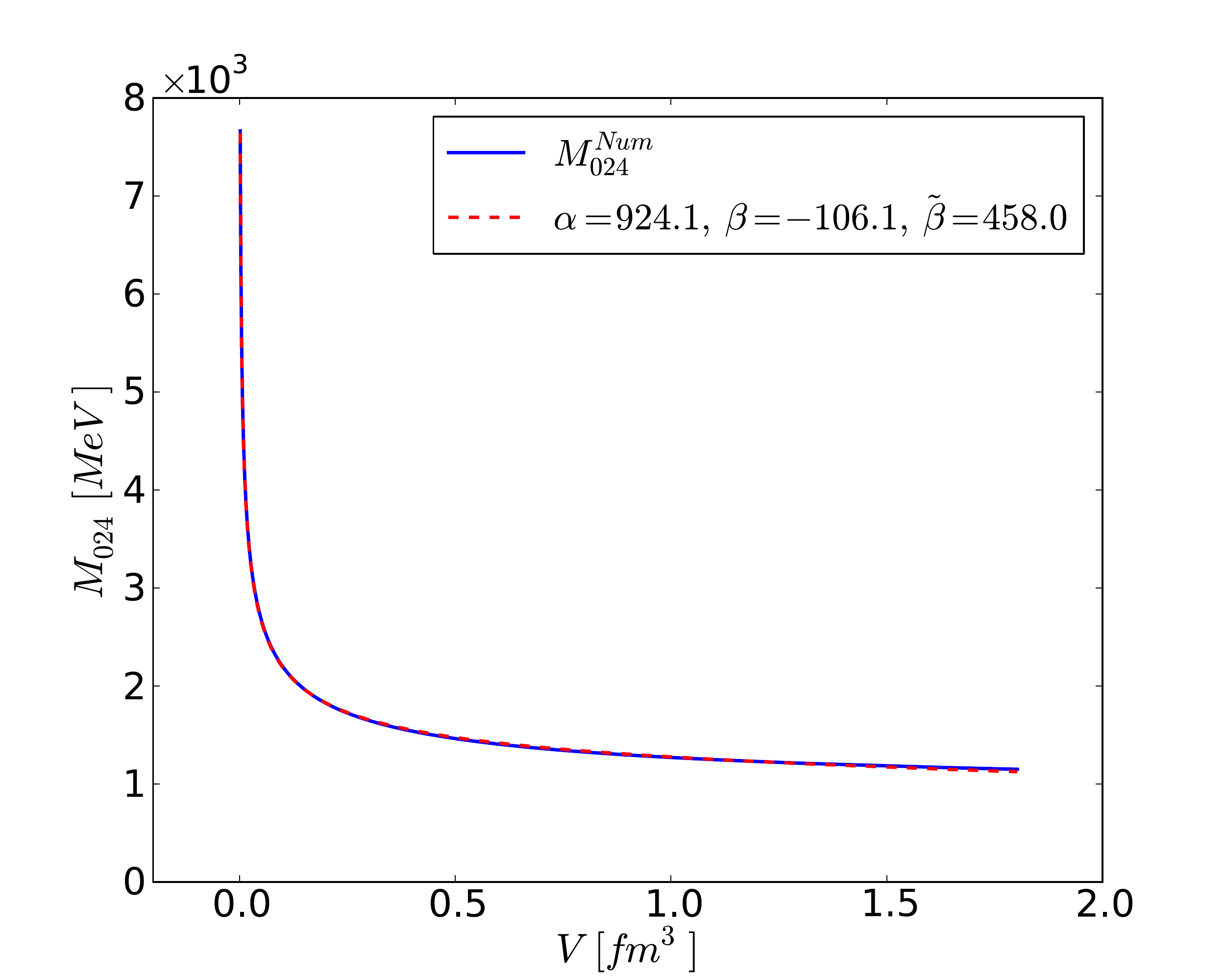}}
\subfigure[]{\includegraphics[totalheight=6.cm]{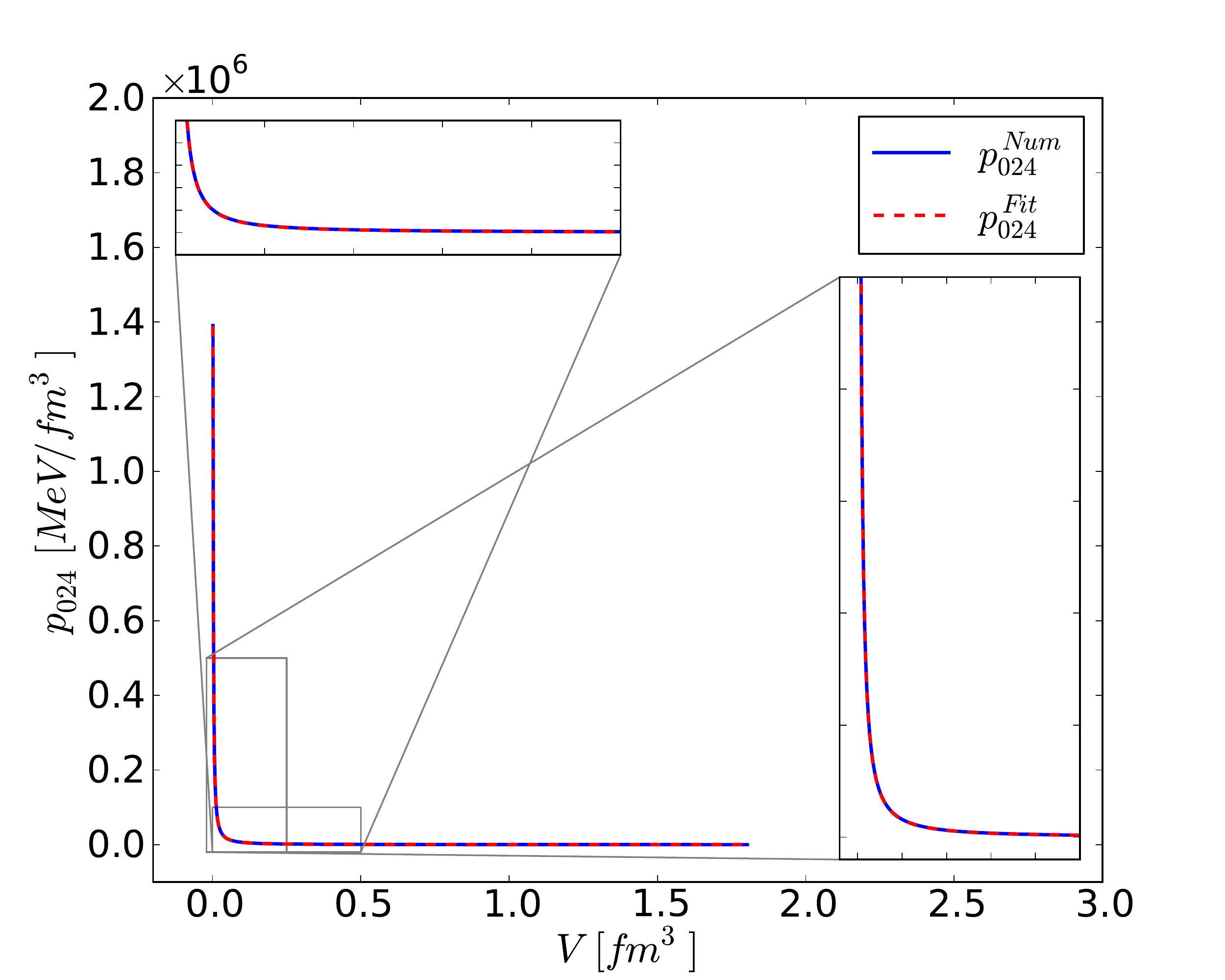}}
\subfigure[]{\includegraphics[totalheight=6.cm]{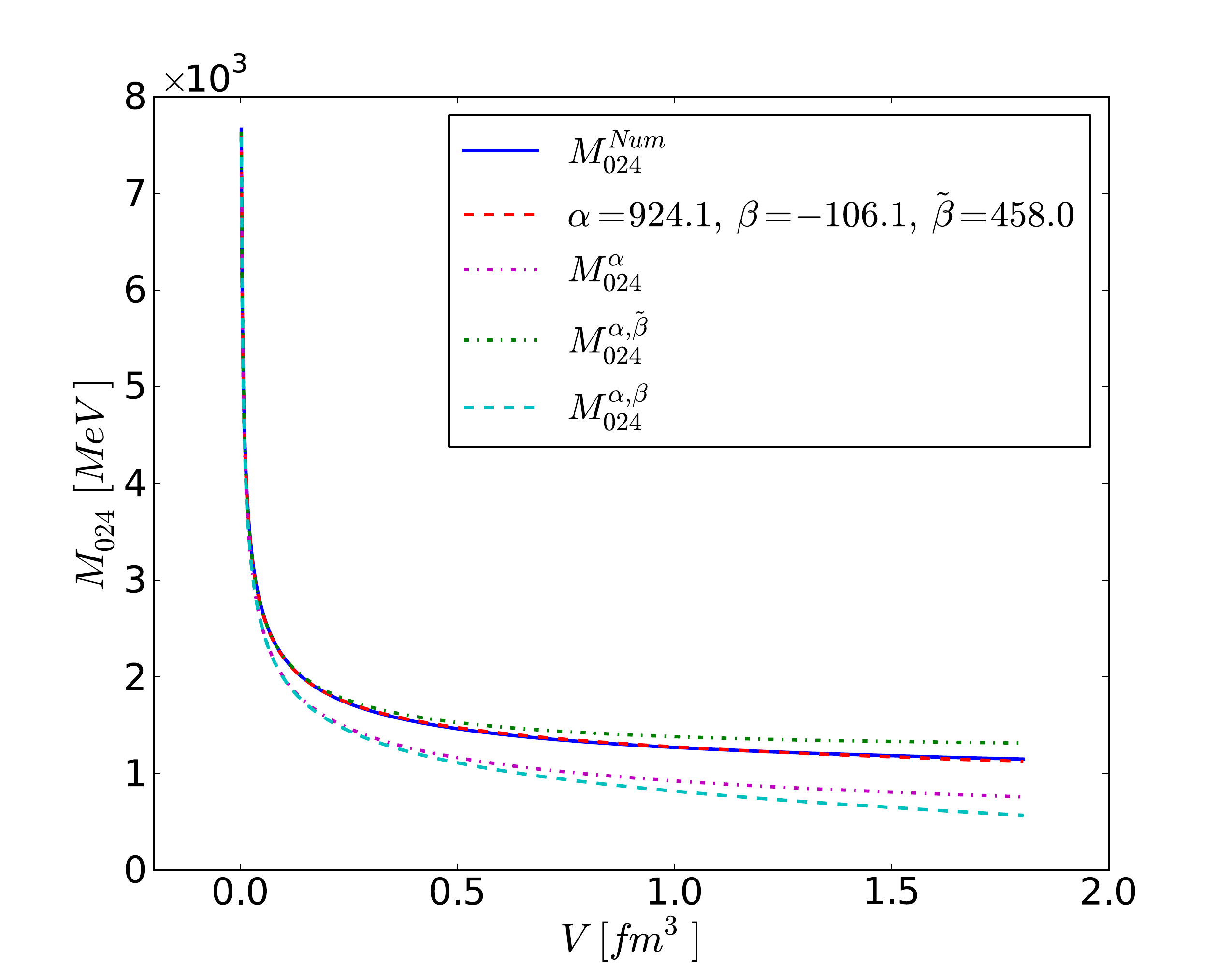}}
\subfigure[]{\includegraphics[totalheight=6.cm]{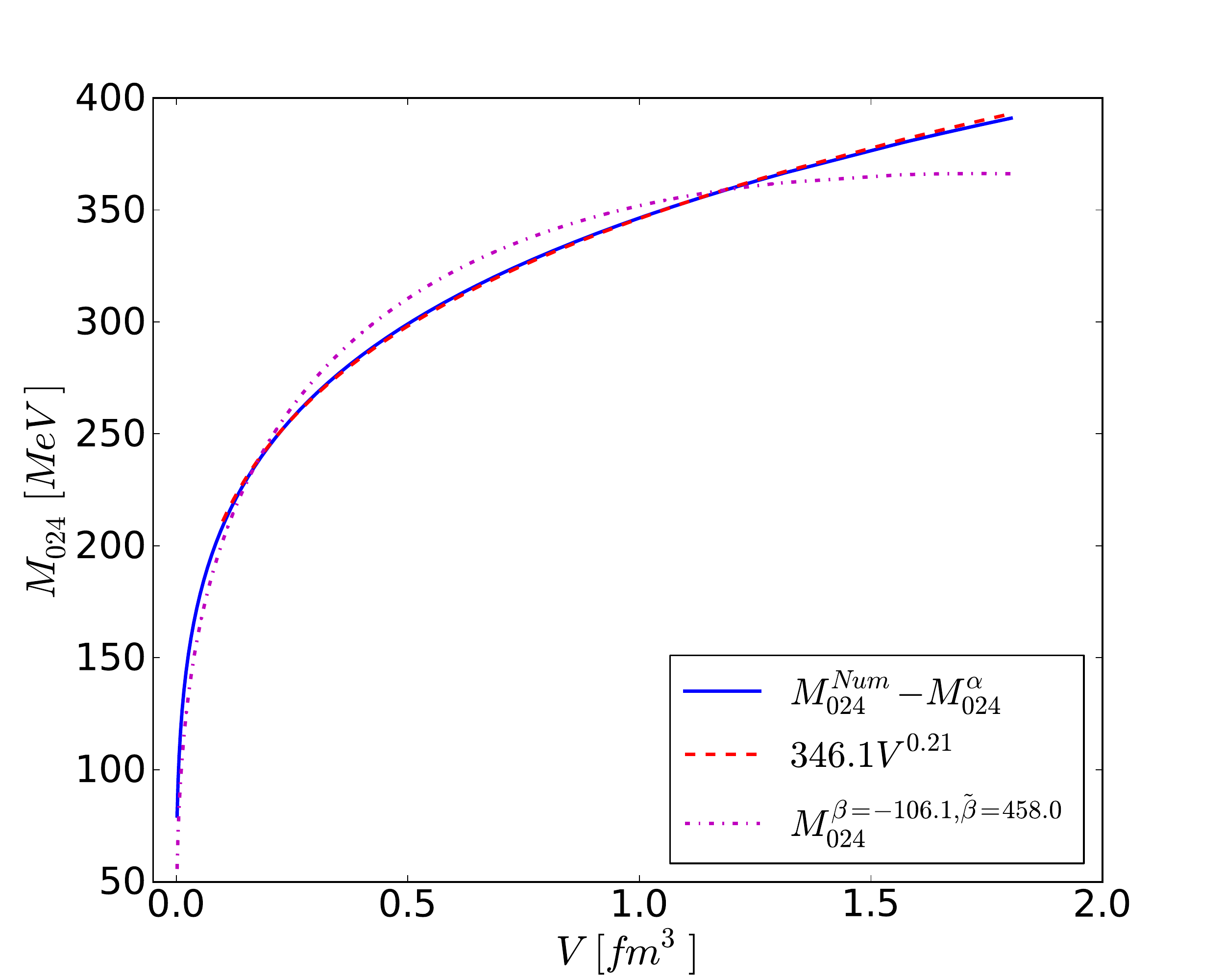}}
\caption{(Color online) Energy and pressure  of the B=1 skyrmion as a function of the volume for $E_{024}$ model. Fig. (a)-(c) with the theoretical fit. Fig (d) with the best fitted curve.}
\label{E024}
\end{figure}
\begin{figure}
\subfigure[]{\includegraphics[totalheight=5.cm]{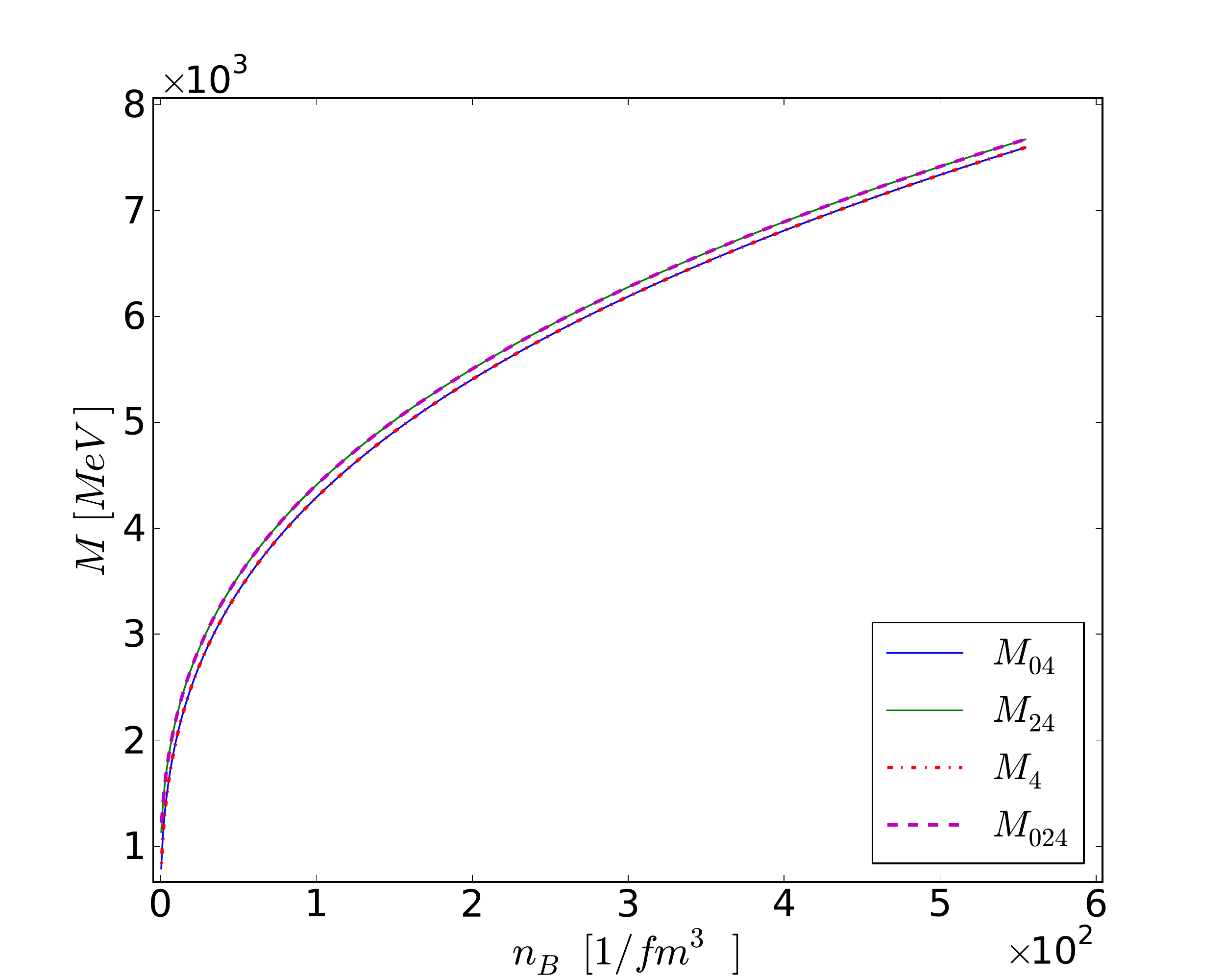}}
\subfigure[]{\includegraphics[totalheight=5.cm]{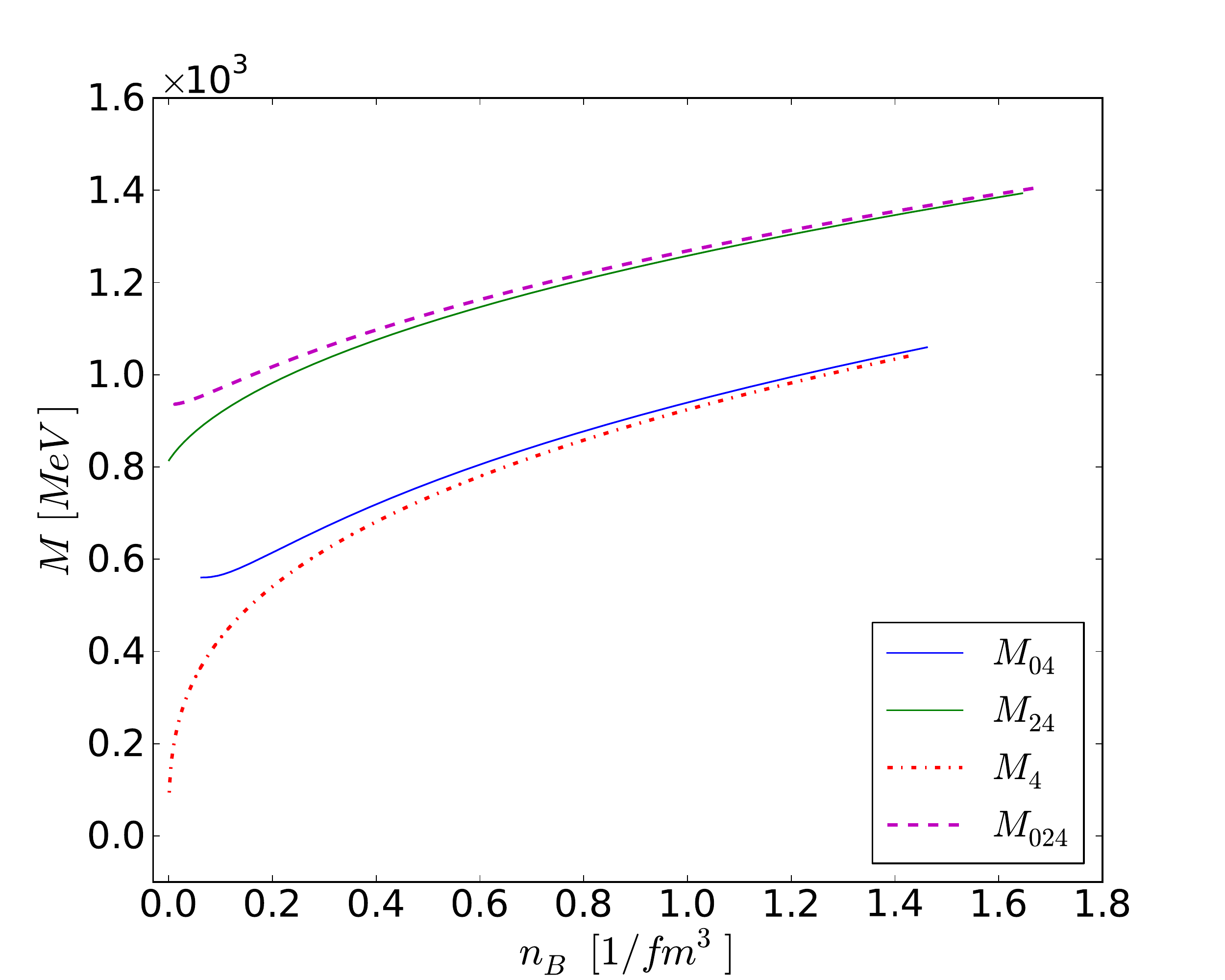}}
\subfigure[]{\includegraphics[totalheight=5.cm]{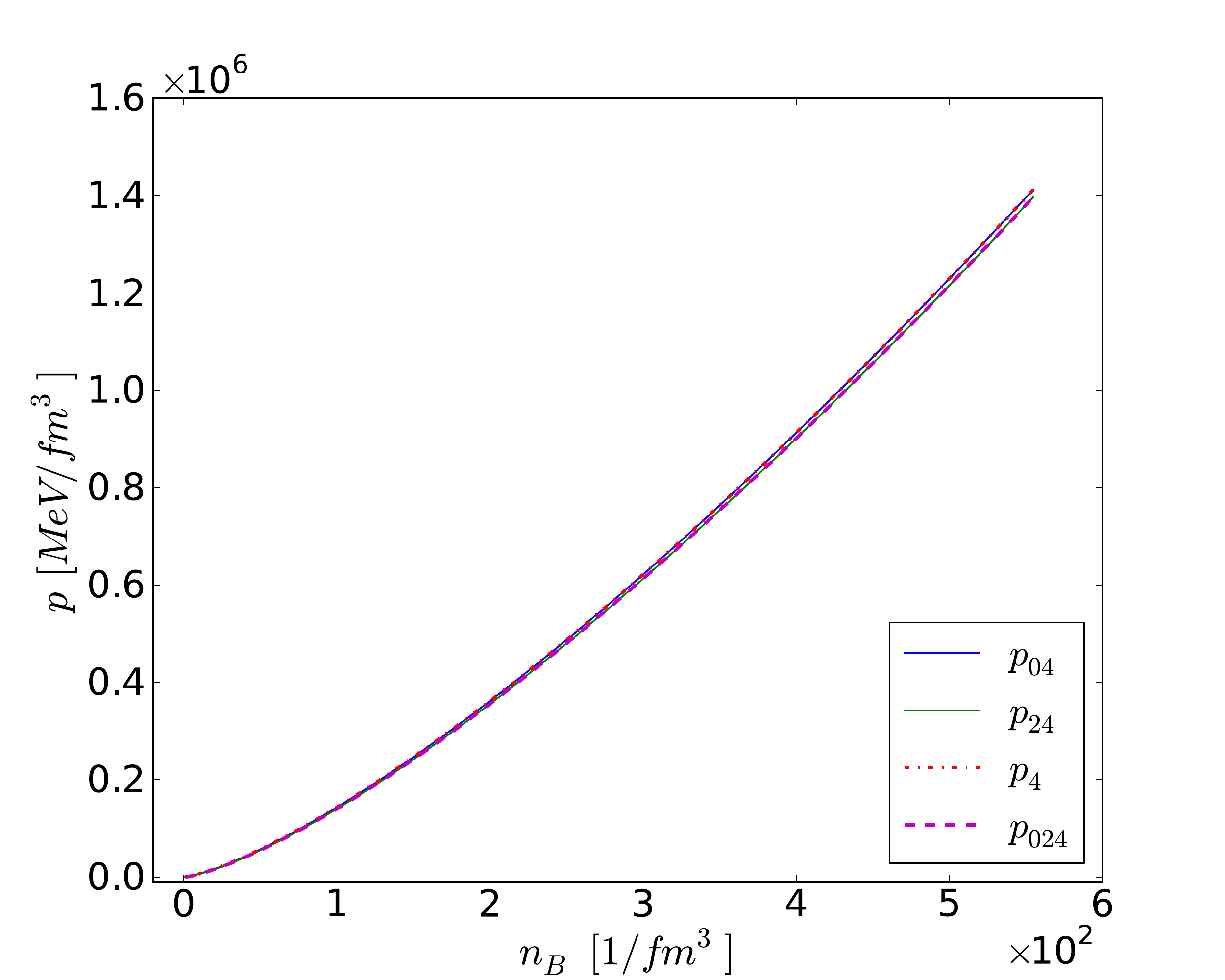}}
\subfigure[]{\includegraphics[totalheight=5.cm]{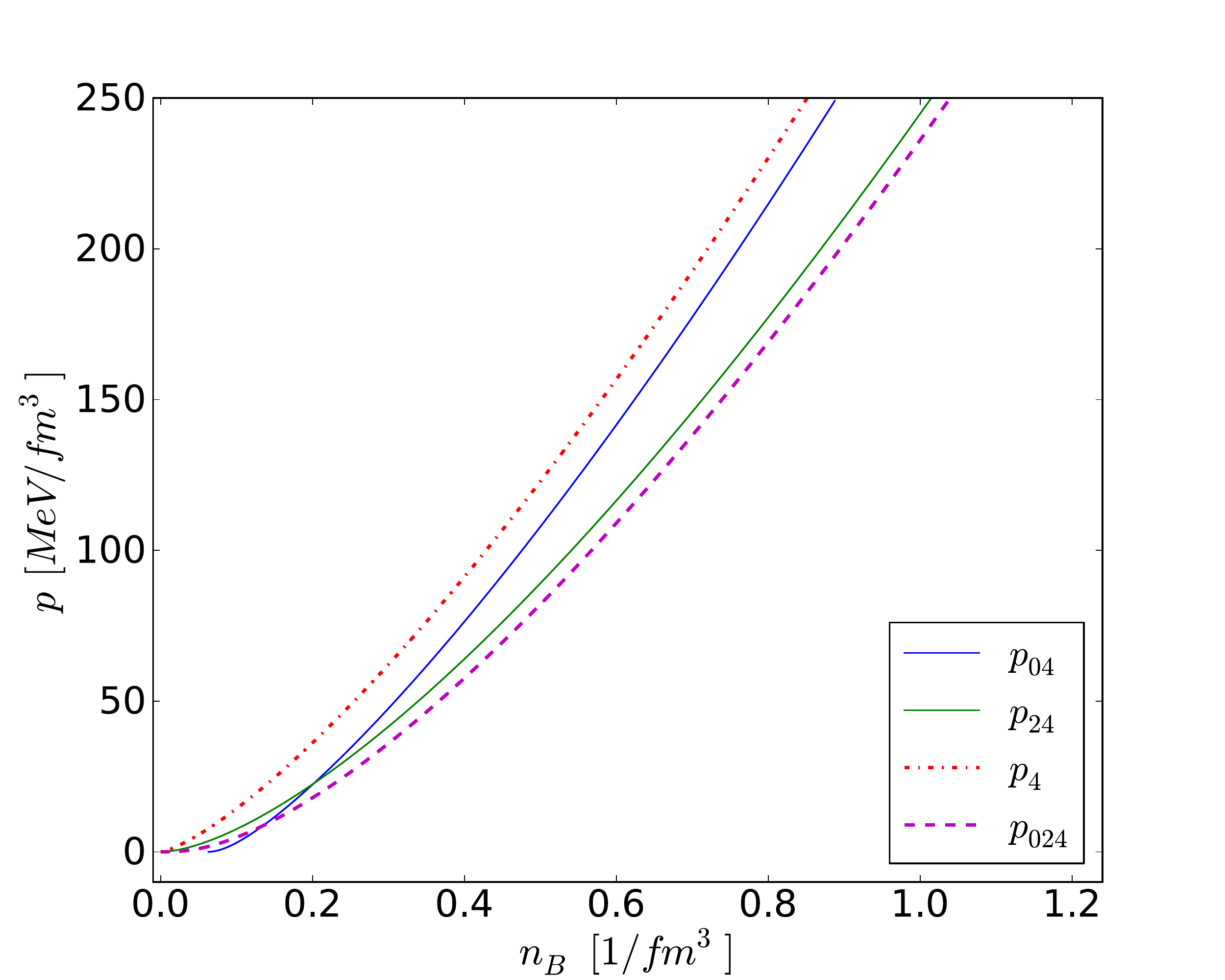}}
\subfigure[]{\includegraphics[totalheight=5.cm]{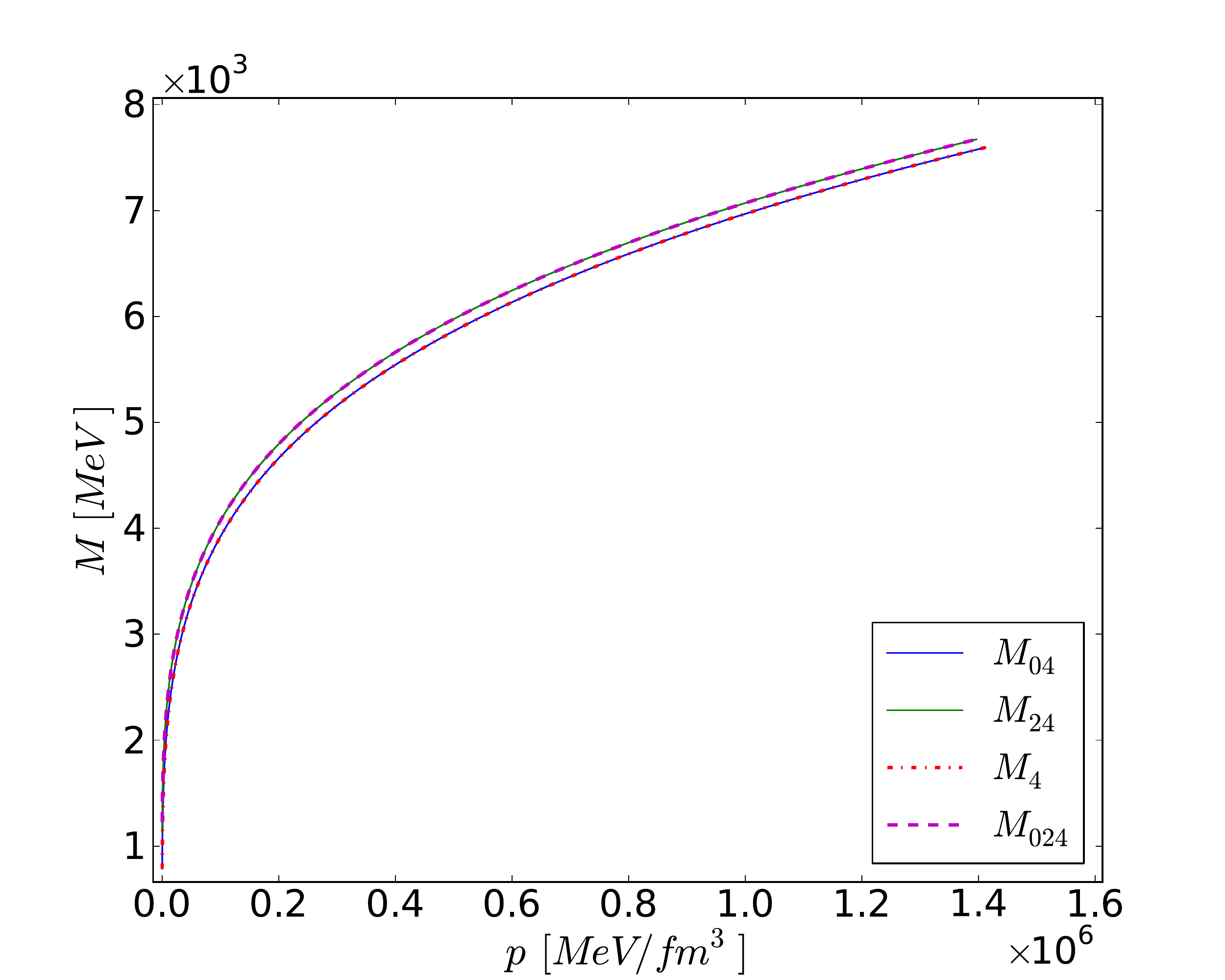}}
\subfigure[]{\includegraphics[totalheight=5.cm]{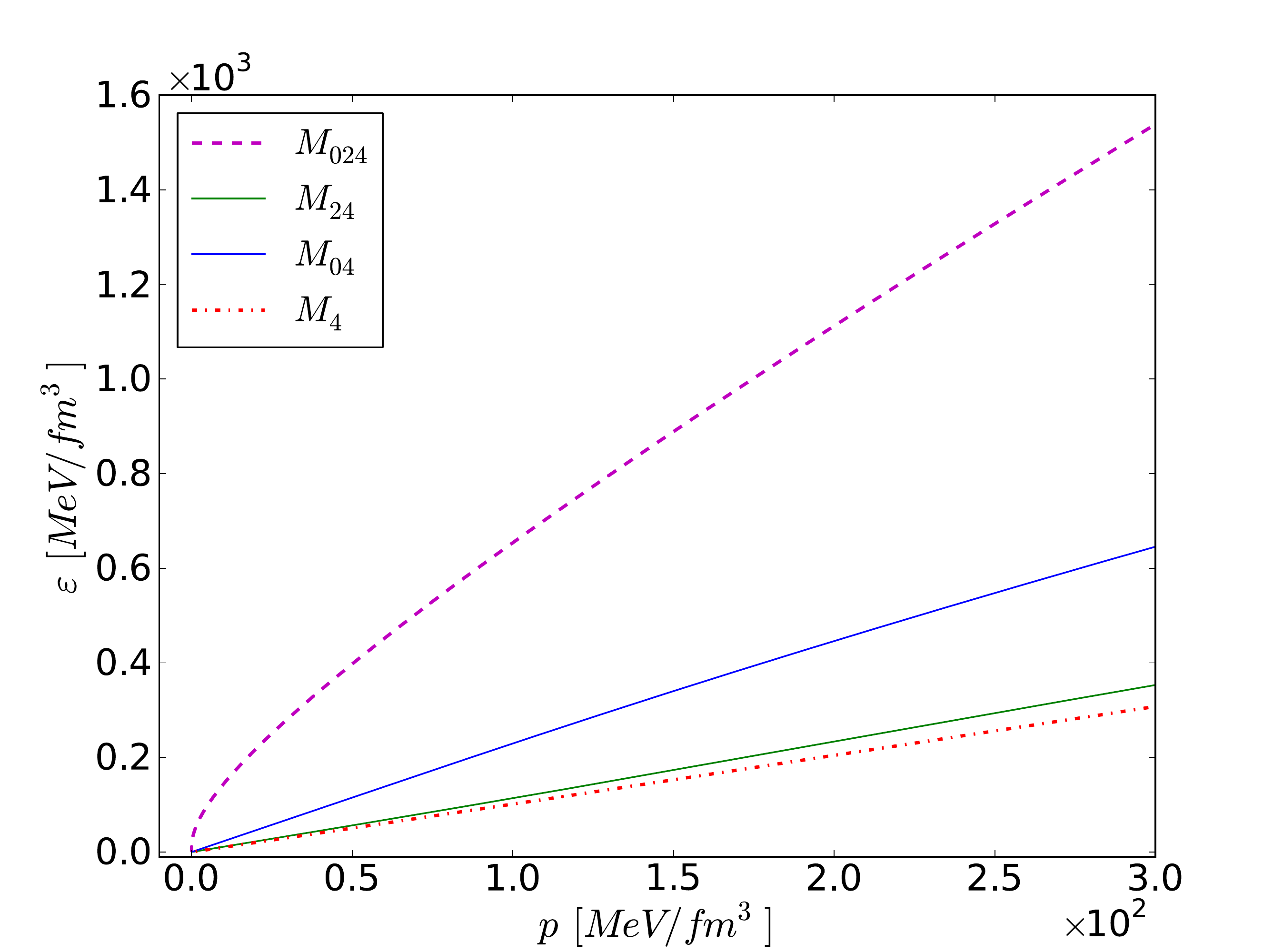}}
\subfigure[]{\includegraphics[totalheight=5.cm]{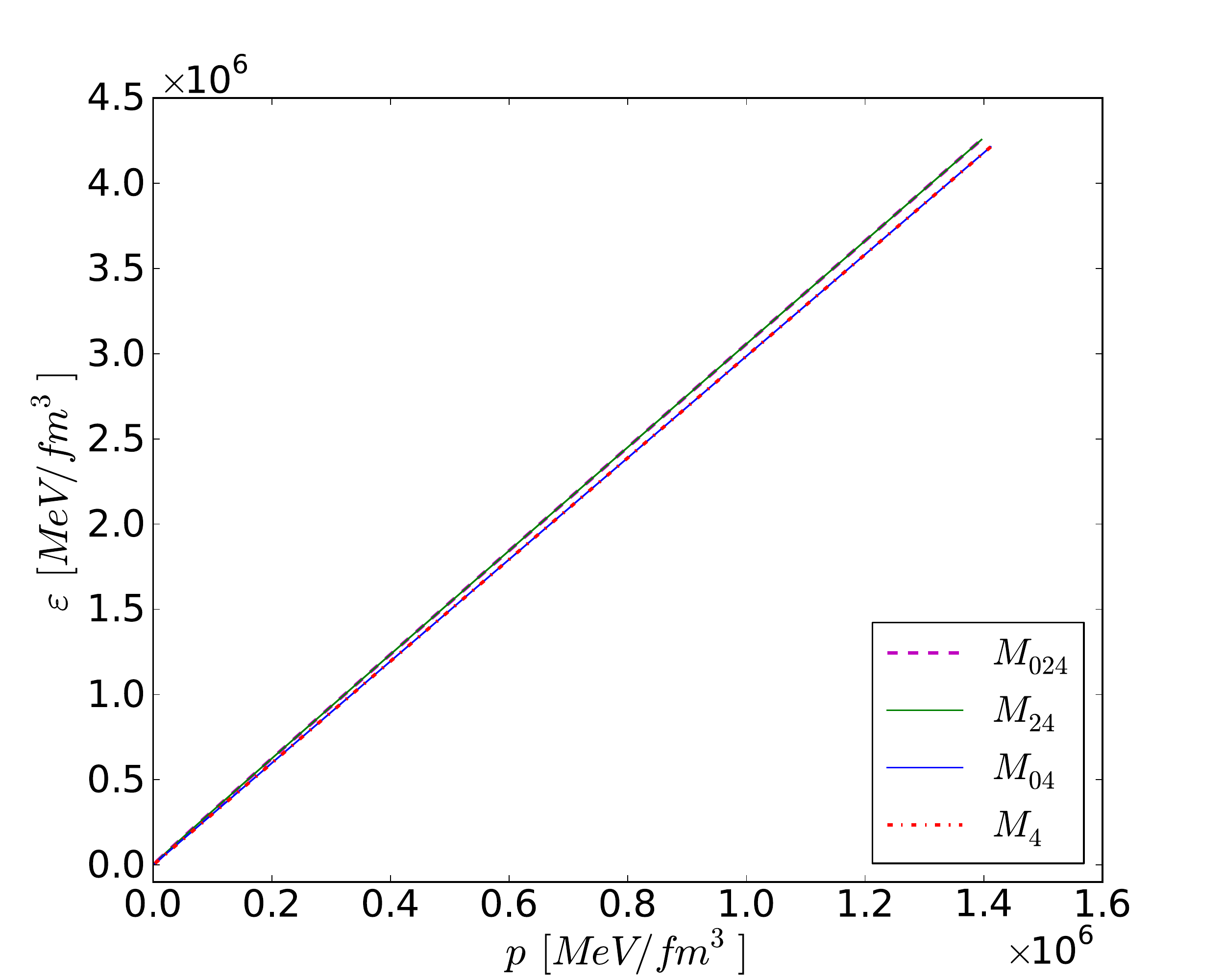}}
\caption{(Color online) Comparison plots for the B=1 skyrmion: (a)-(d)  mass and pressure as a function of the average baryon density; (e) mass as a function of pressure; (f)-(g) equation of state.}
\label{MP_n}
\end{figure}

The conventional way for stabilising the $\mbox{E}_4$ model is to add the sigma model part. This is the usual massless Skyrme model. Now the equilibrium solution in the charge one sector is an infinitely extended skyrmion. Again, we want to  fit the theoretical energy formula. The first issue is that we get a slightly smaller value of the leading term. Namely, $\alpha_{24}=915.9 \; \mbox{MeV fm}$. This means that we are still a bit away from the exact asymptotic regime. Apparently, one has to squeeze the skyrmion to higher pressures to reach the proper value for $\alpha$. It shows that the sigma model part provides a significantly stronger attractive force than the potential term, which influences the squeezed configurations and has a stronger impact on the energy-volume relation. 
\\
In the fitting procedure we take this into account by imposing that $\alpha_{24}= 924.1 \; \mbox{MeV fm}$.  Then,
\be
\alpha_{24}=924.1 \; \mbox{MeV fm}, \;\; \beta_{24}=0, \;\; \tilde{\beta}_{24}= 311.5 \; \mbox{MeV fm}^{-1} .
\ee
In Fig.~\ref{E04} we  compare the numerical results and the theoretical curve. Again we find nice agreement in the high and medium pressure regime. For sufficiently small pressure (large volume) the theoretical formula is not valid any longer as the $V^{1/3}$ term gives a diverging contribution. 
\subsubsection{$\mbox{E}_{024}$ model}
Finally, one can include all three terms. Then, after fitting the theoretical curve we get (with the same remark on the $\alpha$ constant which again does not take precisely its asymptotic value but $\alpha=915.9 \; \mbox{MeV}$)
\be
\alpha_{024}=924.1 \; \mbox{MeV fm}, \;\; \beta_{024}=-106.1\; \mbox{MeV fm}^{-3} , \;\; \tilde{\beta} _{024}=458.0 \; \mbox{MeV fm}^{-1}.
\ee
Note that the constant $\beta$, which is related to the potential part of the action, takes a negative value. In fact, the potential and sigma model term contribute together to the subleading part of the energy-volume relation, which leads to this effect. It also suggests that the theoretical formula (the last two terms motivated by the BPS model analysis) is not quite adequate for this model. In fact, if we subtract the (corrected) leading term from the numerical energy then the best fit is found for $V^a$ where $a\approx 1/5$ - see Fig. \ref{E024}. This leads to the following effective mass-volume formula
\be
E_{024}=\alpha \frac{1}{V^{1/3}} + \gamma V^{1/5} +o(V^{1/5}),
\ee
where $\gamma= 346.1 \mbox{MeV fm}^{-1/5}$. This shows that in the full perturbative Skyrme model there is a rather strong mixing between the sigma model and potential part (and probable also with the quartic term) which effectively provides new subleading terms in the energy-volume relation. Therefore the theoretical curve proposed previously does not seem to work quite well in the full perturbative Skyrme model. 
\\
In Fig. \ref{MP_n} we plot numerical results describing how the energy (mass) and pressure depend on the average baryon density $n_B=1/V$. Furthermore, we plot the resulting equation of state relating the energy and pressure.  

\begin{figure}
\subfigure[]{\includegraphics[totalheight=6.cm]{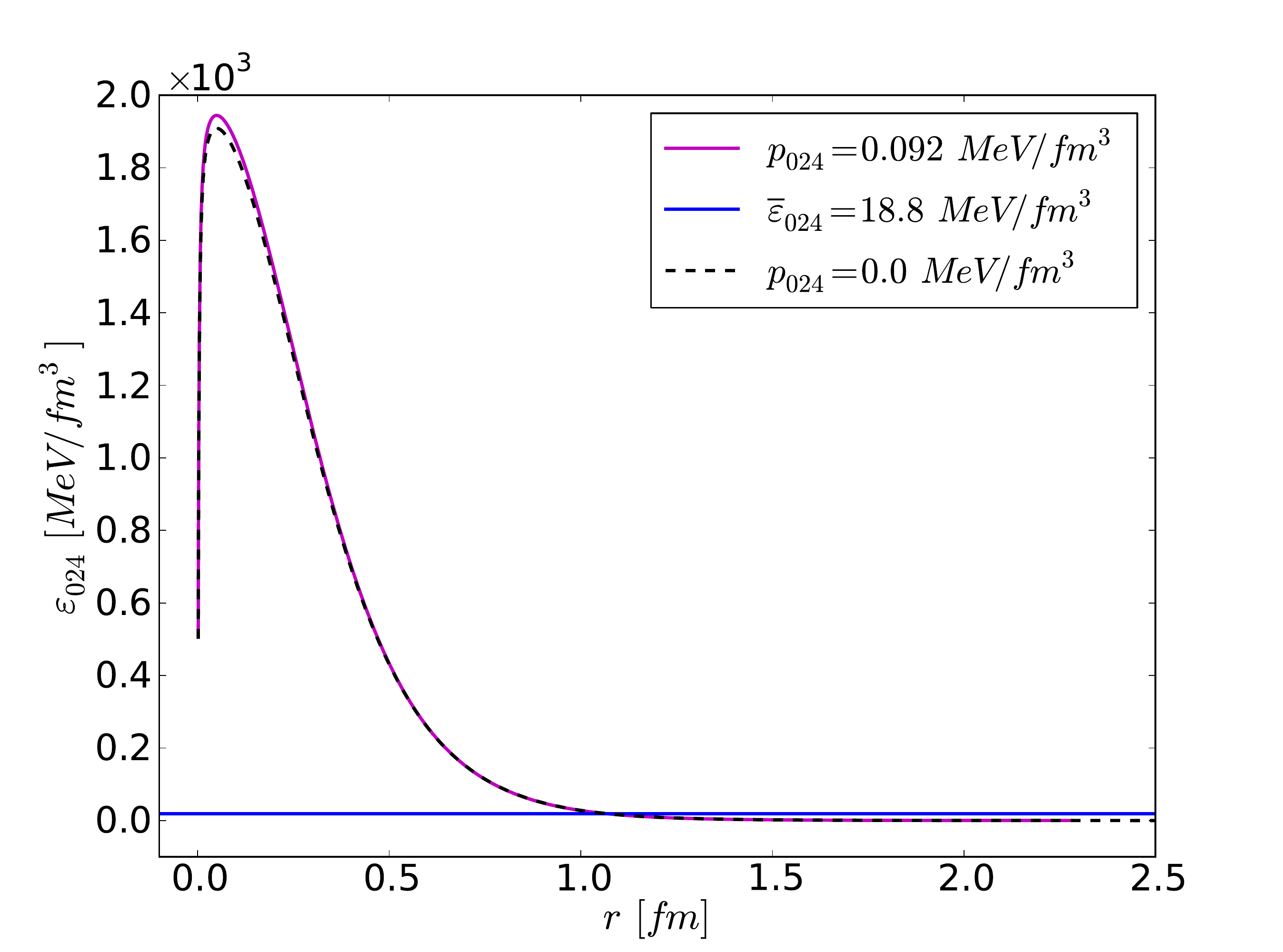}}
\subfigure[]{\includegraphics[totalheight=6.cm]{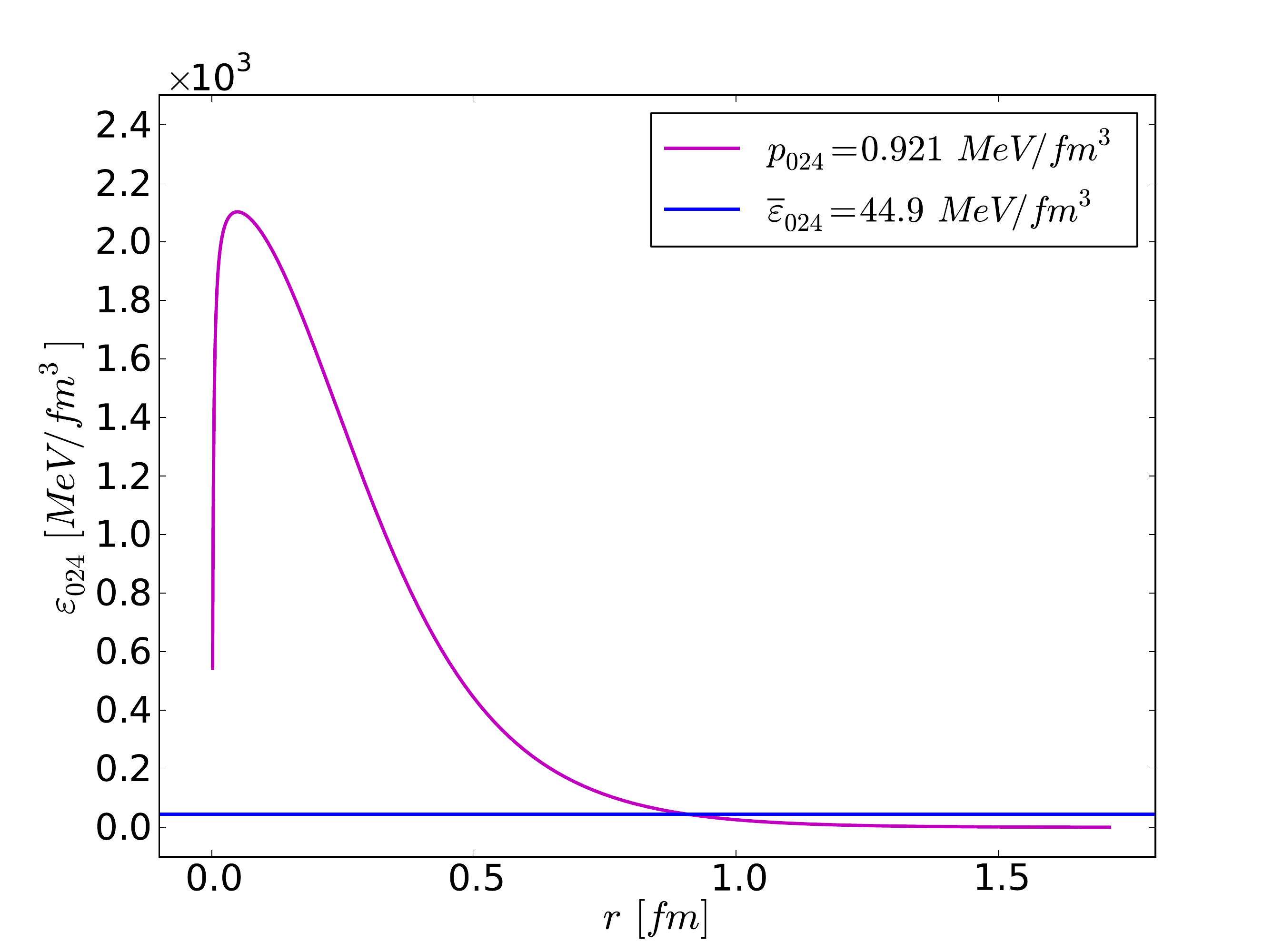}}
\subfigure[]{\includegraphics[totalheight=6.cm]{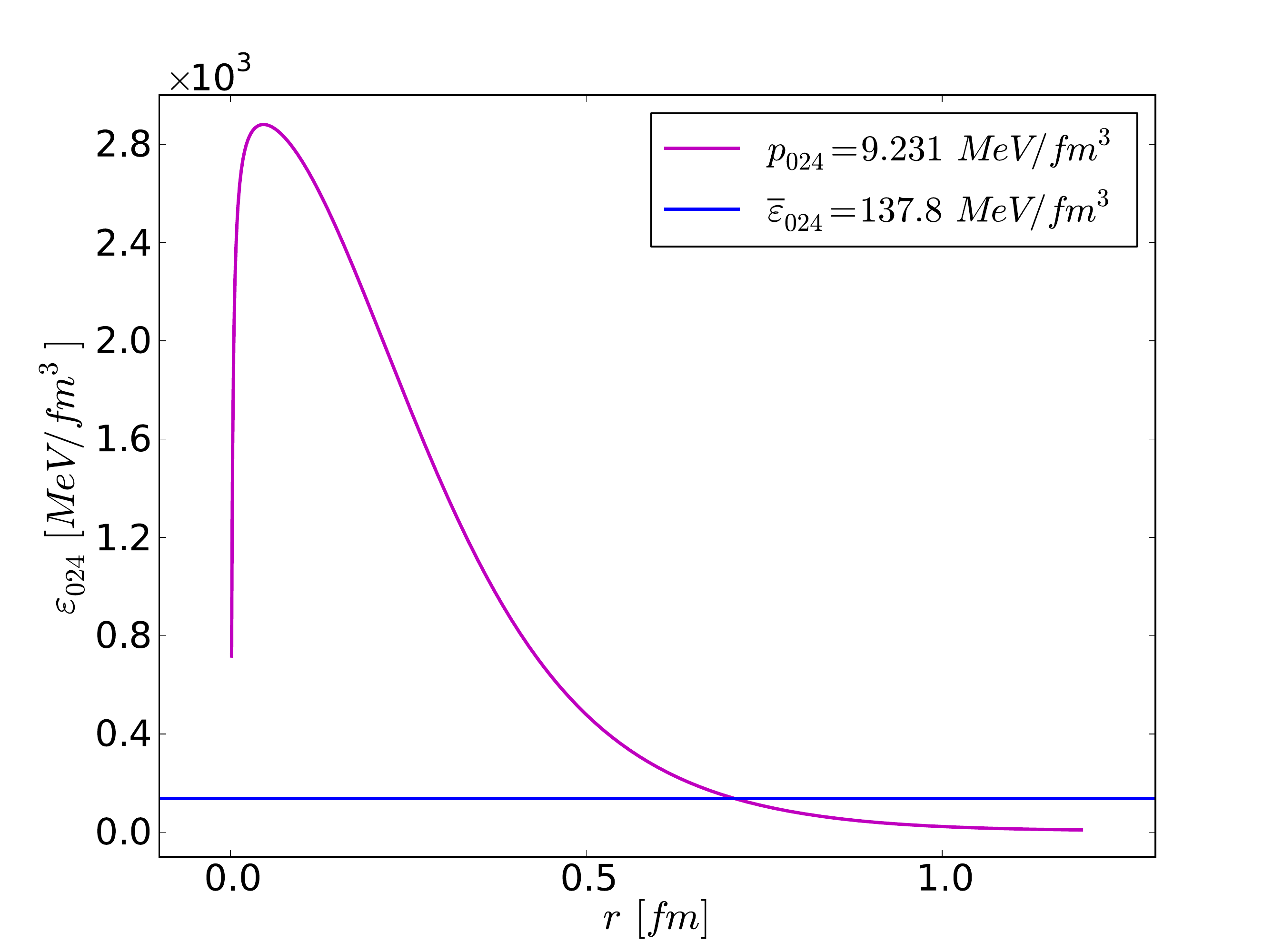}}
\subfigure[]{\includegraphics[totalheight=6.cm]{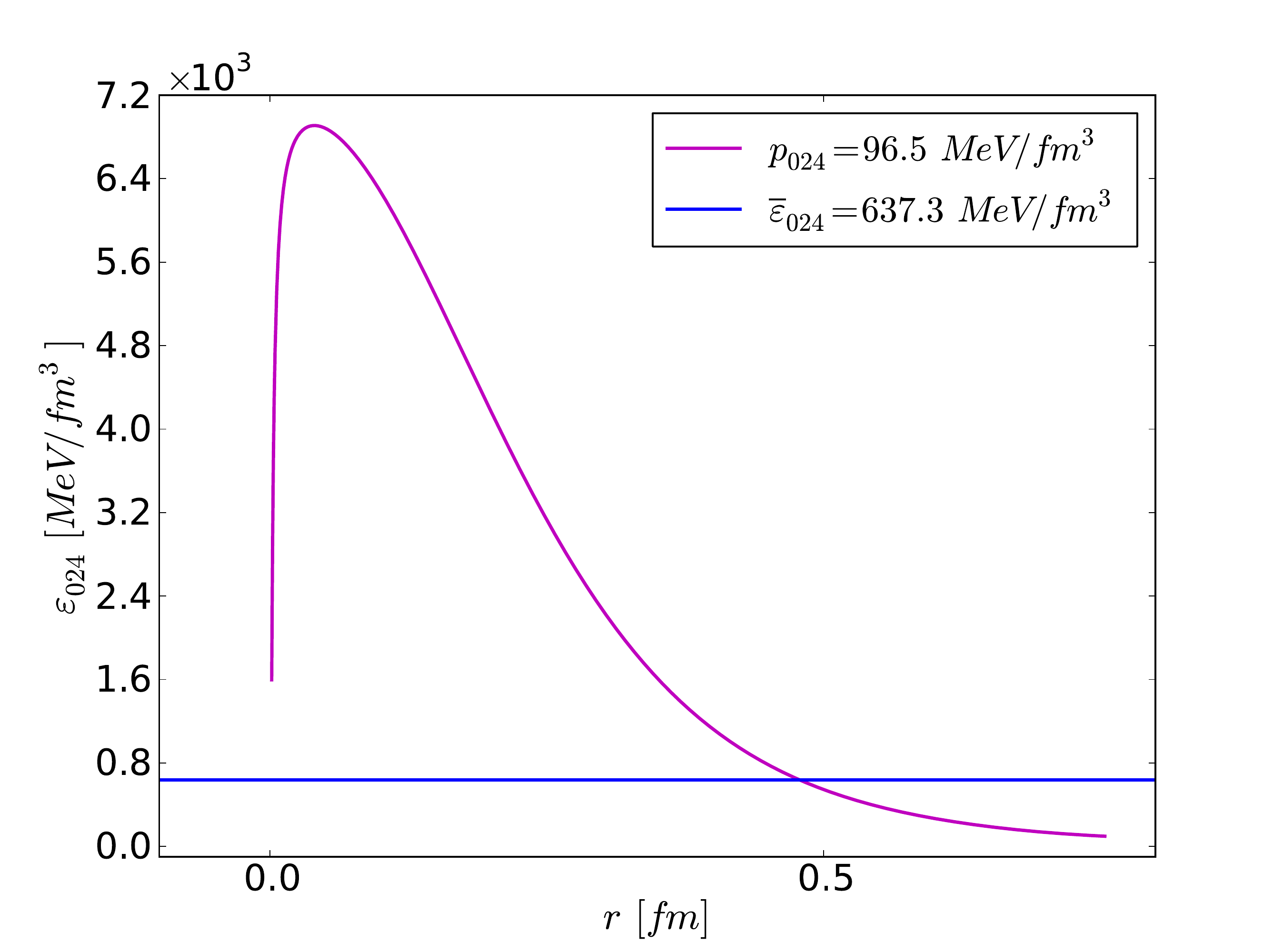}}
\subfigure[]{\includegraphics[totalheight=6.cm]{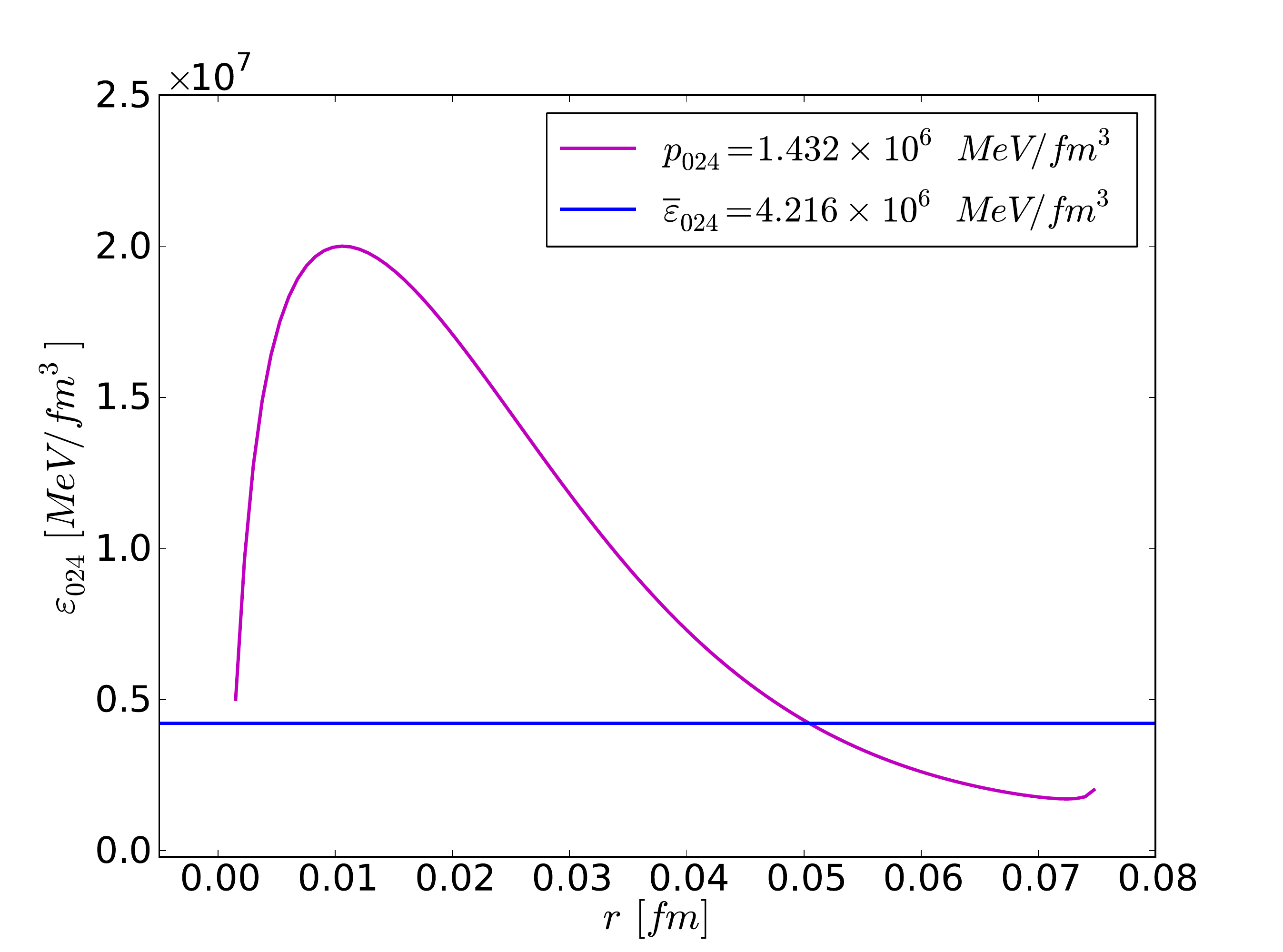}}
\caption{(Color online) Local energy density distribution for the B=1 skyrmion in the $\mbox{E}_{024}$ model for different values of the pressure together with the corresponding mean-field energy density.}
\label{en_den}
\end{figure}

It is also instructive to compare the mean-field energy density $\bar{\varepsilon}$ with the true energy density distribution. This comparison makes no sense at the equilibrium ($P=0$) where the soliton is infinitely extended. This means that the geometric volume is infinity $V=\infty$ and the corresponding mean-field energy density is simply zero. However, for squeezed configuration the volume is always finite. Then, at least for the medium and high pressure regime the mean-field values obtained here for the charge one sector should coincide with their counterparts derived for a skyrmion on a torus (crystal). In Fig. \ref{en_den} we plot the local energy density for different values of the pressure together with the corresponding mean-field energy density. 
It is visible that the energy density has a global maximum whose value is significantly bigger than the average mean-field energy density. For small pressure, this is not surprising, as the corresponding solution has a rather big volume. However, even for very large pressure ($P=1.432 \times 10^6\; \mbox{MeV fm}^{-3}$) the maximum of the energy density is approximately five times bigger that the average value. For medium pressure this difference increases. For example for $P=9.231 \; \mbox{MeV fm}^{-3}$ the maximum energy density is 20 times bigger than the average value. 

Let us finally  remark that the leading term coefficient found in our numerical computations, $\alpha=1.853$ in Skyrme units (or $\alpha = 924.4 \; \mbox{MeV}$ in physical units), provides a bigger number (a better bound) not only than the analytical bound  but also than the value found by Kutschera et. al. for a multi-skyrmion configuration being a cubic lattice of skyrmions, which in the Skyrme units is $\alpha_{kut} =  1.837$ \cite{kut}. As in the asymptotic regime the multi-skyrmion should behave identical to the charge one skyrmion, we think that this constant gives the proper bound for high pressure. 

\subsection{The near BPS Skyrme model}\label{Sec_NumRes_nearBPS}
In this subsection, we switch on the sextic term which, as we already know, governs the high density regime. The total energy has the following form
\be
\mbox{E}_{0246}= \epsilon\left( \lambda_2 E_2 + \lambda_4 E_4 + \lambda_0 E_0 \right) + \lambda_6 E_6 +\tilde{\lambda}_0 \tilde{E}_0.
\ee
The constants are
\be
\lambda_2= \frac{1}{24\pi^2}, \;\;\; \lambda_4= \frac{1}{12 \pi^2}, \;\;\; \lambda_0 = \frac{1}{12 \pi^2}, \;\;\; \lambda_6=\lambda^2 \pi^4 \frac{1}{12 \pi^2}, \;\;\; \tilde{\lambda}_0=\frac{\mu^2}{12\pi^2},
\ee
where $\lambda$ and $\mu^2$ will be fixed later (or set to zero if the corresponding term is omitted). For the usual (perturbative) Skyrme model part, this corresponds to the same choice as before, but which the overall multiplication by the parameter $\epsilon$, which is assumed to have three values $\epsilon=1, 0.1, 0.01$. 
Again, the potential entering the perturbative Skyrme part is the standard potential (providing the physical mass for pions) 
\be
E_0 = \; \mbox{Tr} \;  (1-U),
\ee
while the BPS part of the potential is $\tilde{E}_0= E_0^2$. 
\subsubsection{The BPS model - $\mbox{E}_{06}$ model}
\begin{figure}
\subfigure[]{\includegraphics[totalheight=5.cm]{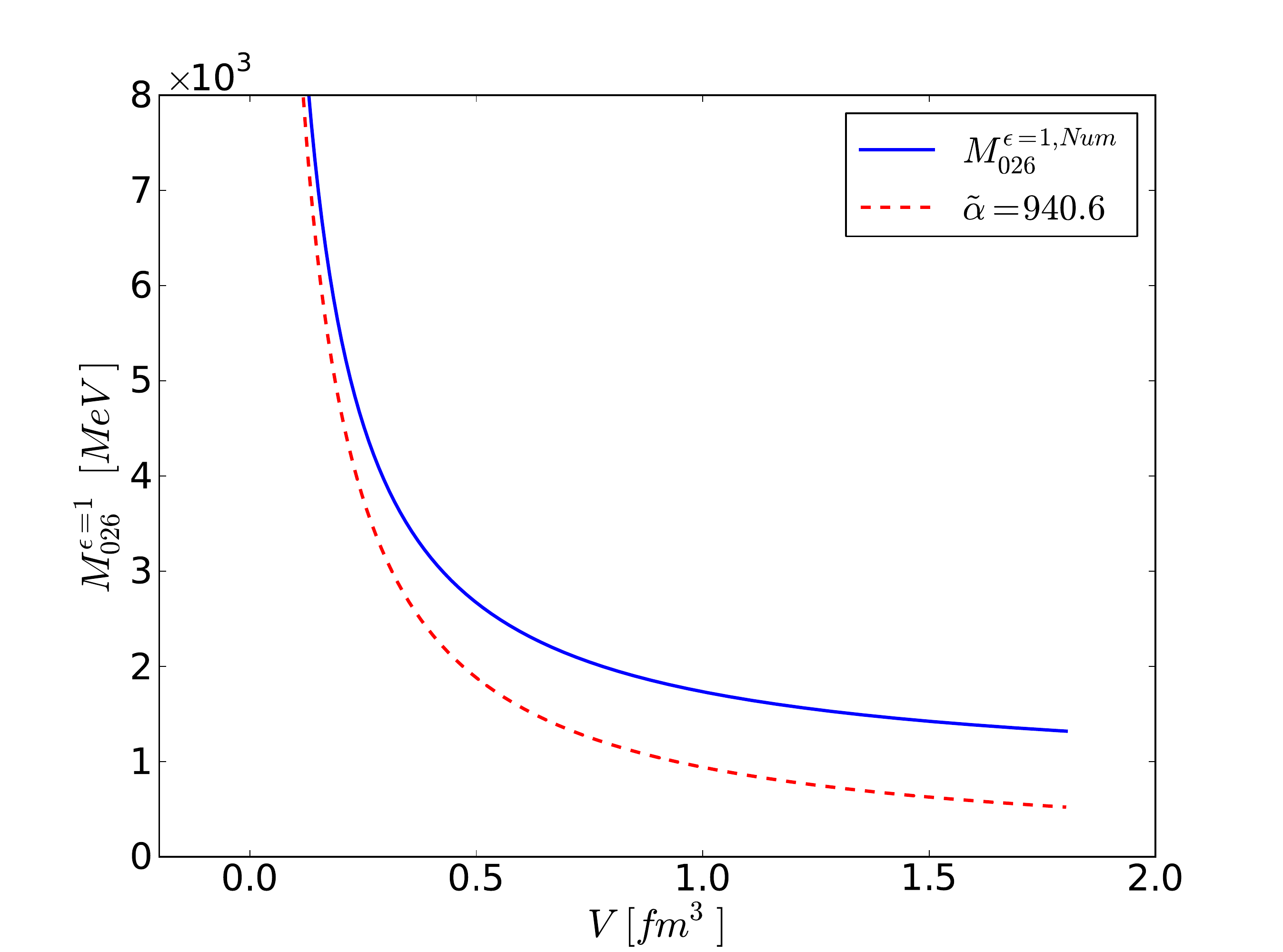}}
\subfigure[]{\includegraphics[totalheight=5.cm]{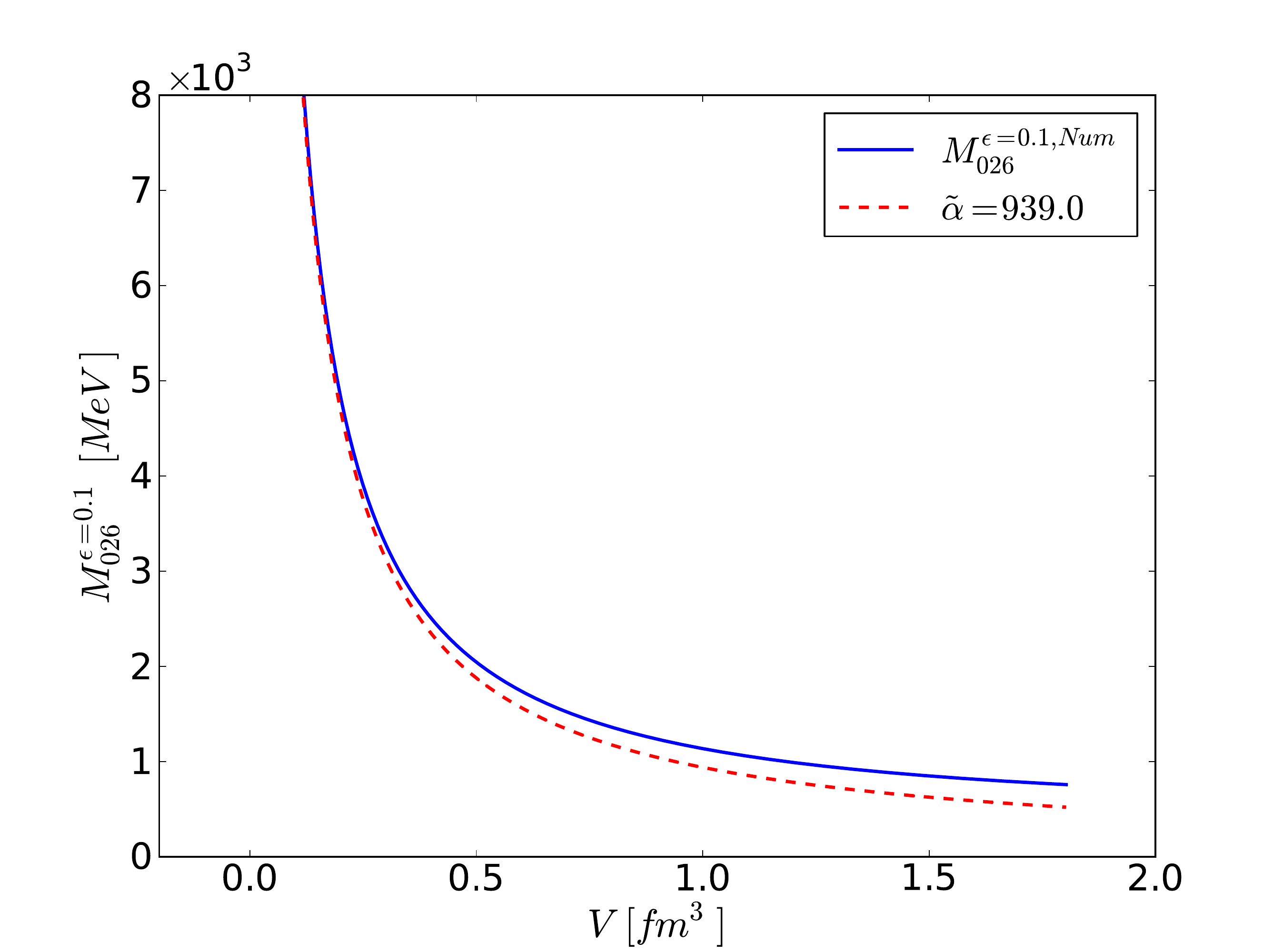}}
\subfigure[]{\includegraphics[totalheight=5.cm]{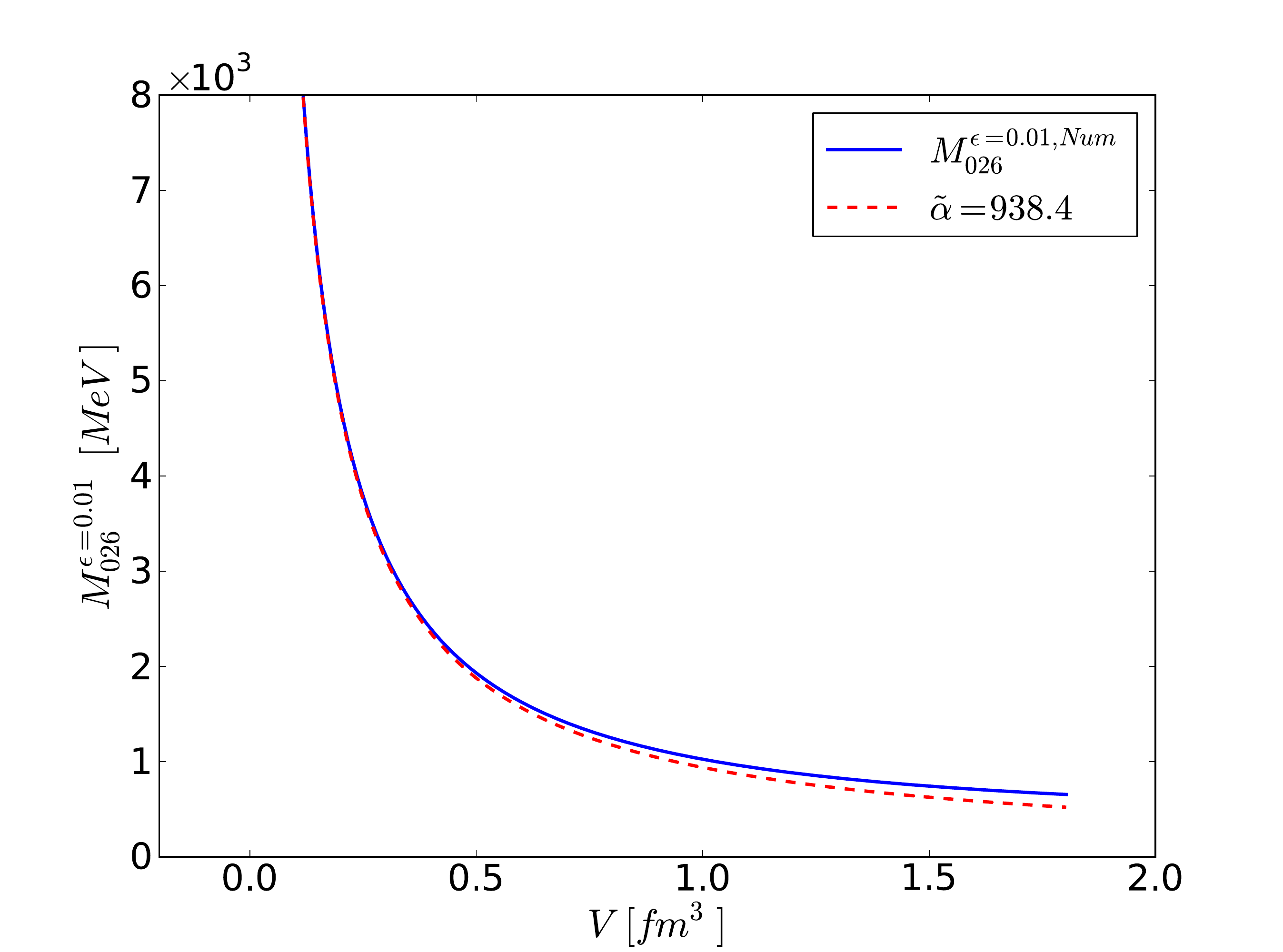}}
\subfigure[]{\includegraphics[totalheight=5.cm]{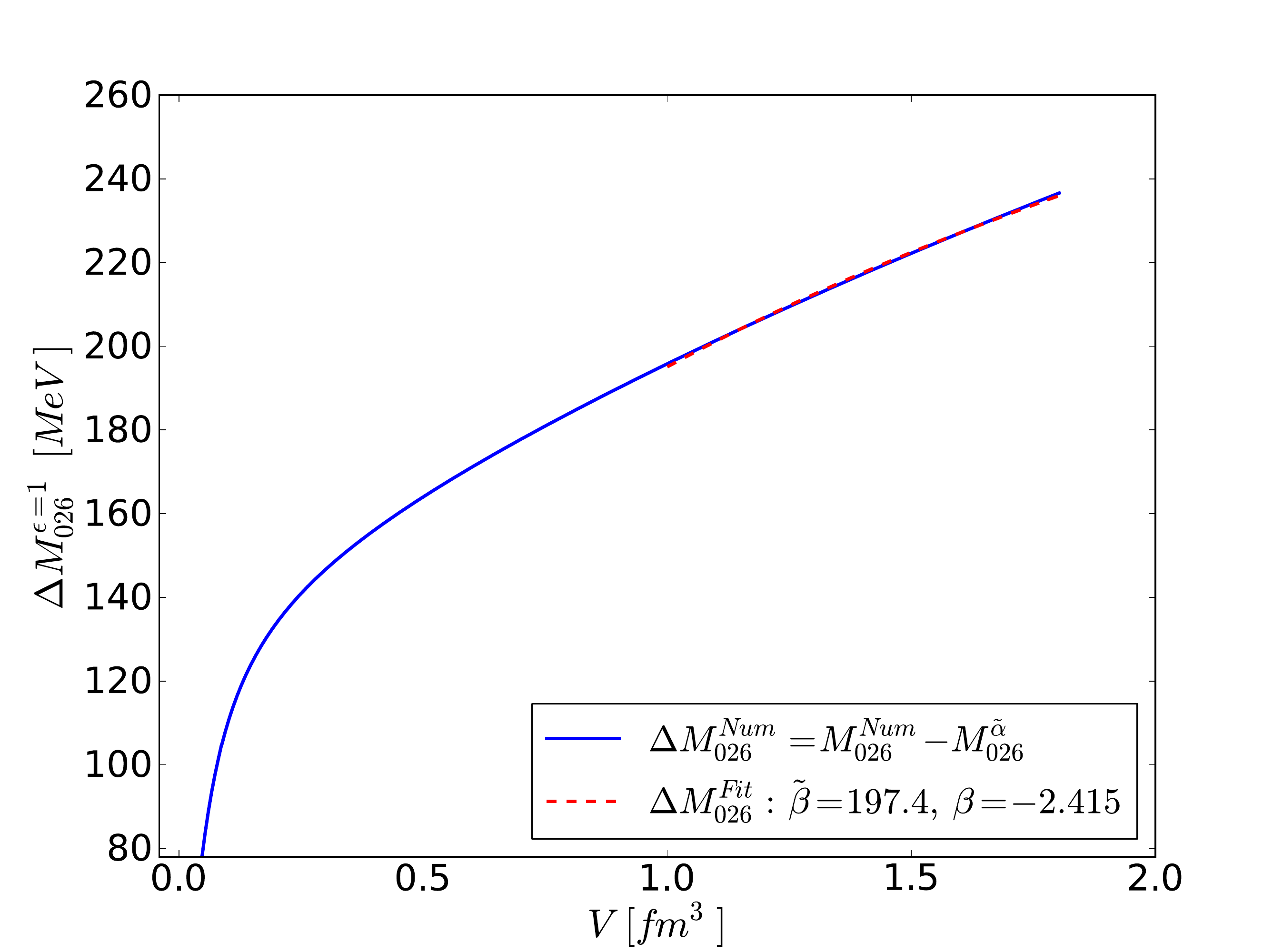}}
\subfigure[]{\includegraphics[totalheight=5.cm]{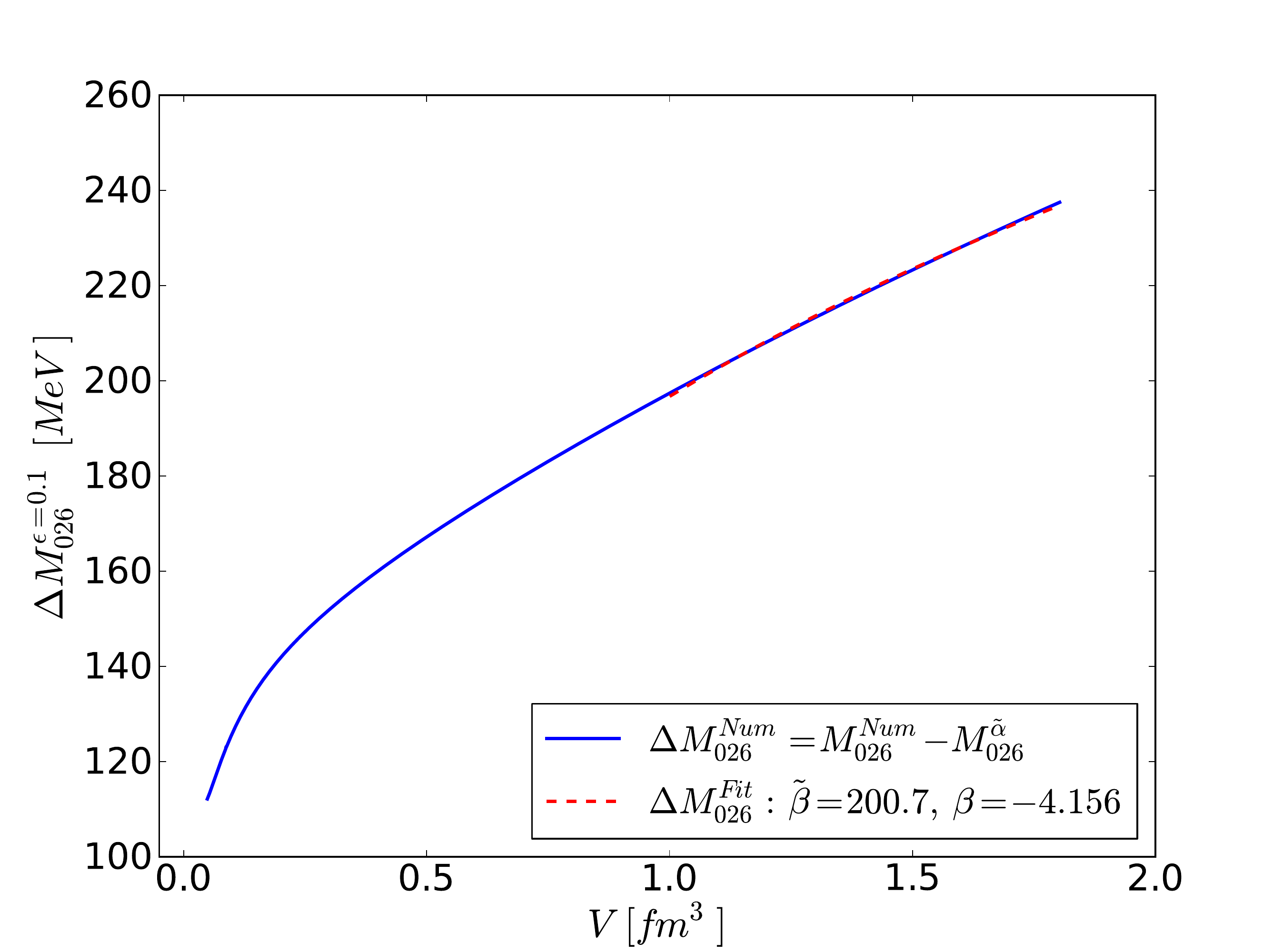}}
\subfigure[]{\includegraphics[totalheight=5.cm]{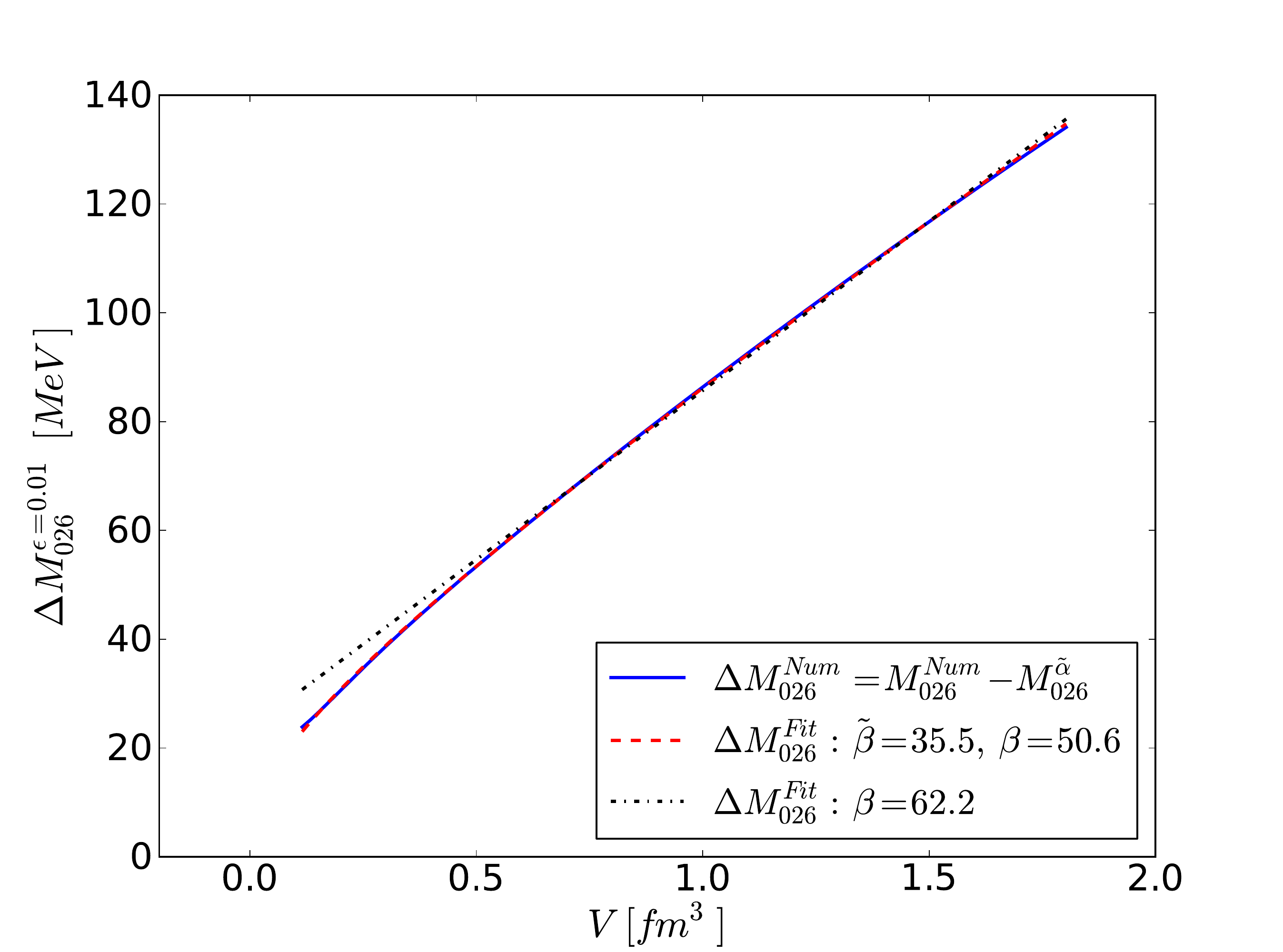}}
\subfigure[]{\includegraphics[totalheight=5.cm]{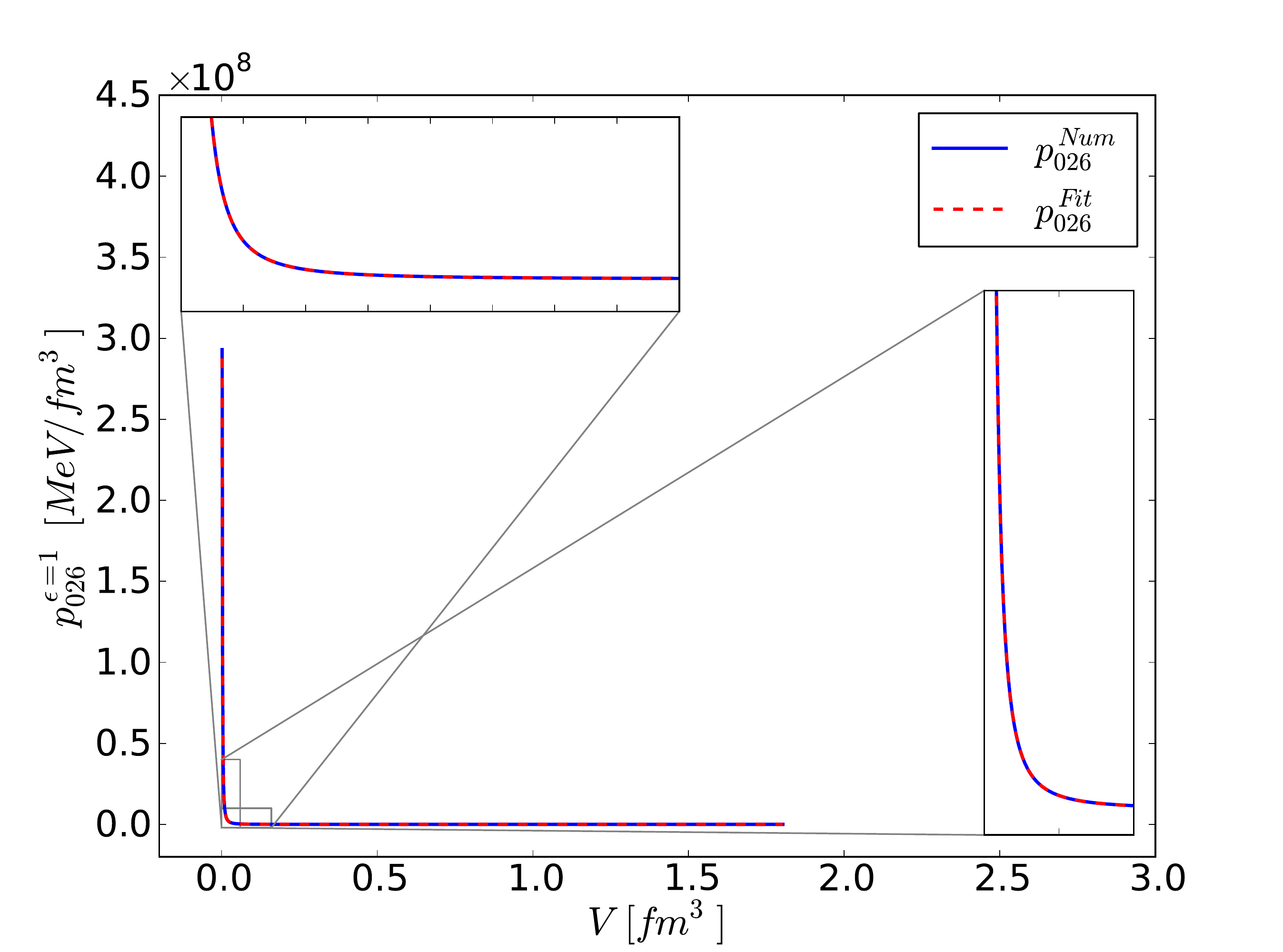}}
\subfigure[]{\includegraphics[totalheight=5.cm]{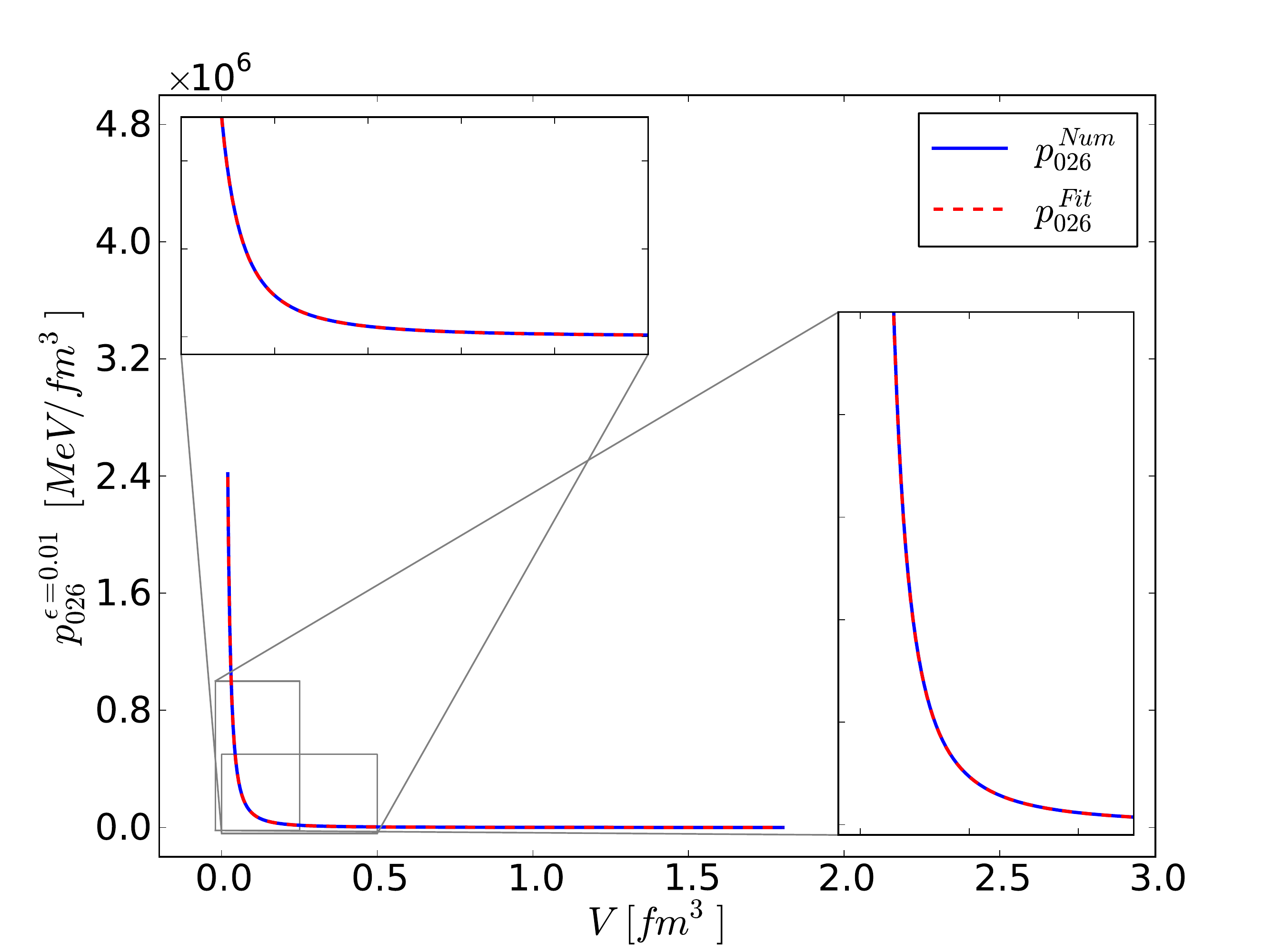}}
\caption{(Color online) Energy and pressure  of the B=1 skyrmions as a function of the volume for $E_{026}$ model.}
\label{E026}
\end{figure}
In order to test the numerics we begin with the BPS Skyrme model, which can be solved analytically. The field equations of motion can be integrated to the following first order equation \cite{term}
\be
\frac{\lambda}{2r^2} \sin^2 \xi \xi_r = - \mu \sqrt{ \tilde{E}_0+ P \frac{12\pi^2}{\mu^2} },
\ee
where we assumed the hedgehog ansatz for the charge one soliton. Of course, the BPS equation can be found as the zero pressure condition \cite{baz}. The  potential reads $\tilde{E}_0=4(1-\cos \xi)^2$. Using this equation, it is possible to compute the total energy integral and find the volume of the solution with a given pressure. Indeed, we find \cite{fluid grav}
\be
E(P)=\frac{2\pi}{12 \pi^2} \lambda \mu \int_0^\pi d\xi \sin^2 \xi \frac{8(1-\cos \xi)^2 +  P \frac{12\pi^2}{\mu^2}}{\sqrt{4(1-\cos \xi)^2 +  P \frac{12\pi^2}{\mu^2}}},
\ee
\be
V(P)=2\pi \frac{\lambda}{\mu} \int_0^\pi d\xi \sin^2 \xi \frac{1}{\sqrt{4(1-\cos \xi)^2 +  P \frac{12\pi^2}{\mu^2}}}.
\ee
This gives the following mean-field equation of state 
\begin{equation}
\bar{\varepsilon}=P + \tilde{\mu}^2 \left( \frac{5}{2} \frac{{}_3F_2 [\{ \frac{1}{2}, \frac{7}{4}, \frac{9}{4} \}, \{ \frac{5}{2}, 3 \}, -\frac{4\tilde{\mu}^2}{P} ]  }{{}_3F_2 [\{ \frac{1}{2}, \frac{3}{4}, \frac{5}{4} \}, \{ \frac{3}{2}, 2 \}, -\frac{4\tilde{\mu}^2}{P} ] }  \right),
\end{equation}
where ${}_pF_q[\{a_1,\ldots,a_p\}, \{b_1,\ldots,b_q\},z] $ is a generalised hypergeometric function, and $\tilde{\mu}^2=1/3\pi^2$. Note, however, that the true energy density is not spatially constant and, in fact, differs quite significantly from its mean-field approximation. This is of some importance, e.g.,  for the application to neutron stars. For example, the mass-radius relation changes significantly if we switch from the true non-mean field energy density to the mean-field average energy density \cite{fluid grav}. 
\\
The mean-field equation of state results, in the asymptotic region, in a simple energy-volume formula 
\be
E = \tilde{\alpha} \frac{1}{V}+ \beta_\infty V + o(V),
\ee
where
\be
\tilde{\alpha}=\frac{1}{12\pi^2} \lambda^2 \pi^4=  \frac{\pi^2}{3}=935.6 \; \mbox{Mev fm}^3, \;\;\;\;\;  \beta_\infty = \frac{5}{6\pi^2} = 64.8 \; \mbox{MeV fm}^{-3}. 
 \ee
in Skyrme units and physical units (with the ANW calibration assumed). Here we chose $\lambda=2$ and $\mu=1$. This asymptotical off-set can be compared with the equilibrium energy density
 \be
\bar{\varepsilon}_0 = \frac{1}{3\pi^3} = 26.0 \; \mbox{MeV fm}^{-3}.
 \ee
 The non-zero value is due to the compact nature of the equilibrium solution. In fact, it has a finite size and a volume
 \be
 R=\left( \frac{3\pi}{2}\right)^{1/3} \;\; \Rightarrow \;\; V = 2\pi^2 = 8.49 \; \mbox{fm}^3.
 \ee
Let us comment that for the most interesting full near-BPS Skyrme model we apply a different choice for parameters and different calibration scheme, which was previously used for neutron stars. 
\\
In numerical computation we reproduce the mean-field equation of state for the BPS Skyrme model. For instance, we find
\be
\tilde{\alpha}=938.2 \; \mbox{MeV fm}^3 \;\;\; \mbox{and} \;\;\; \beta = 61.5 \; \mbox{Mev fm}^{-3},
\ee
which is in agreement with the analytical results within $0.3 \%$ and $5\%$ respectively. 
\subsubsection{$\mbox{E}_{026}$ model}
Now we consider a submodel for which the attractive force is provided entirely by the sextic term i.e., the quartic term is absent. Additionally we assume $\lambda=2$ and $\mu=1$. The leading term in the energy-volume relation is found in the numerical computations. Specifically, we reach a regime where
\be
\tilde{\alpha}_{\epsilon=0.01} = 938.4, \;\;\; \tilde{\alpha}_{\epsilon=0.1} = 939.0, \;\;\; \tilde{\alpha}_{\epsilon=1} = 940.6. 
\ee
Obviously, the leading part more correctly describes also a medium volume regime if we are closer to the BPS limit i.e., if we reduce the $\epsilon$ parameter (see Fig. \ref{E026}). We also get the subleading (linear in the volume) term. For the nearest BPS Skyrme considered here ($\epsilon=0.01$) we find
 \be
 \beta_{\epsilon=0.01} =  62. 2 \; \mbox{MeV fm}^{-3},
 \ee
 with, however, a non-zero value for the off-set. It is also possible to fit $\tilde{\beta}V^{1/3}+\beta V$ curve. For $\epsilon=0.1$ and $\epsilon=1$ it leads to a negative value for $\beta$, which seems to indicate that such a curve is rather not the right one. 
 \begin{figure}
\subfigure[]{\includegraphics[totalheight=5.cm]{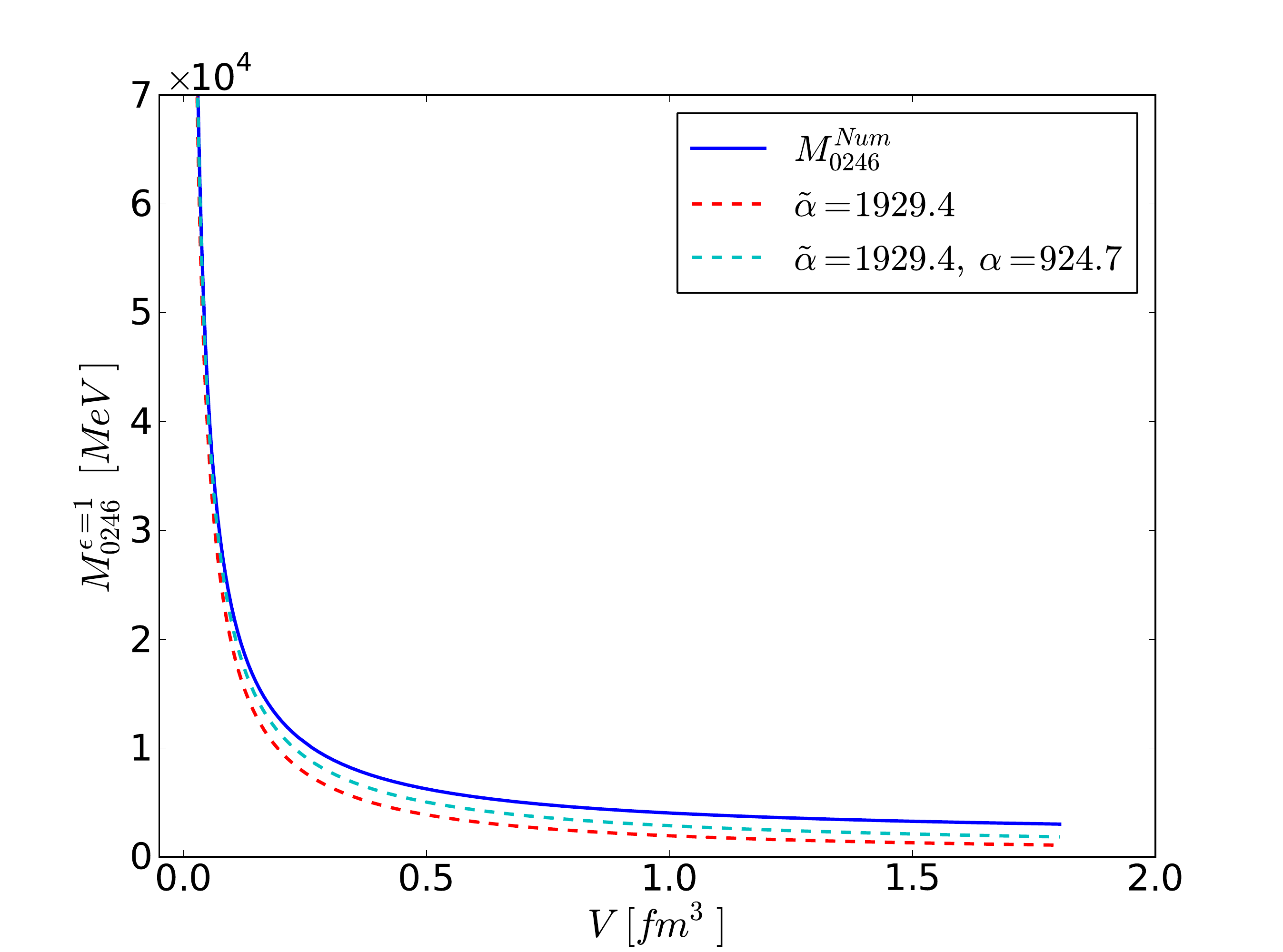}}
\subfigure[]{\includegraphics[totalheight=5.cm]{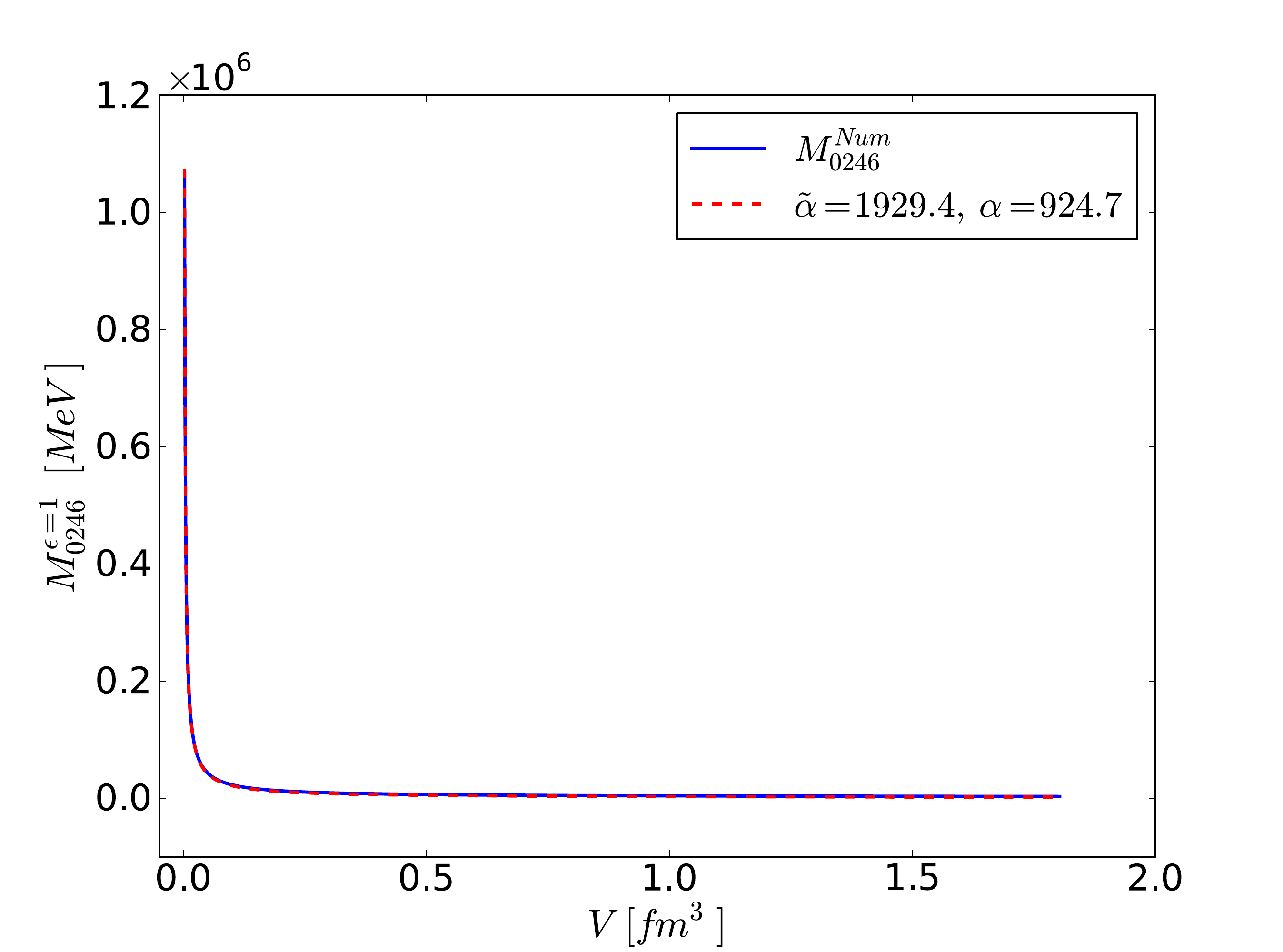}}
\subfigure[]{\includegraphics[totalheight=5.cm]{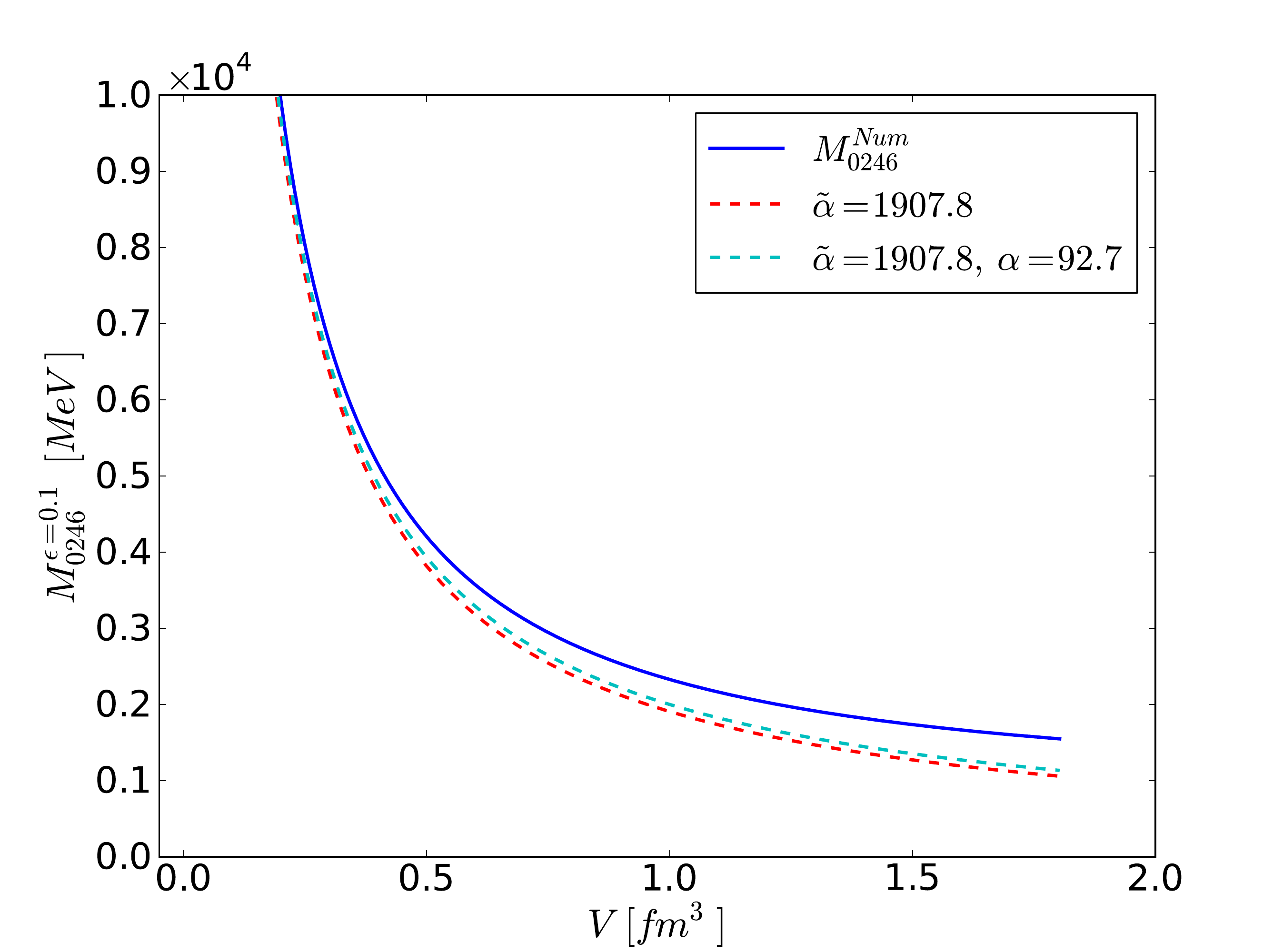}}
\subfigure[]{\includegraphics[totalheight=5.cm]{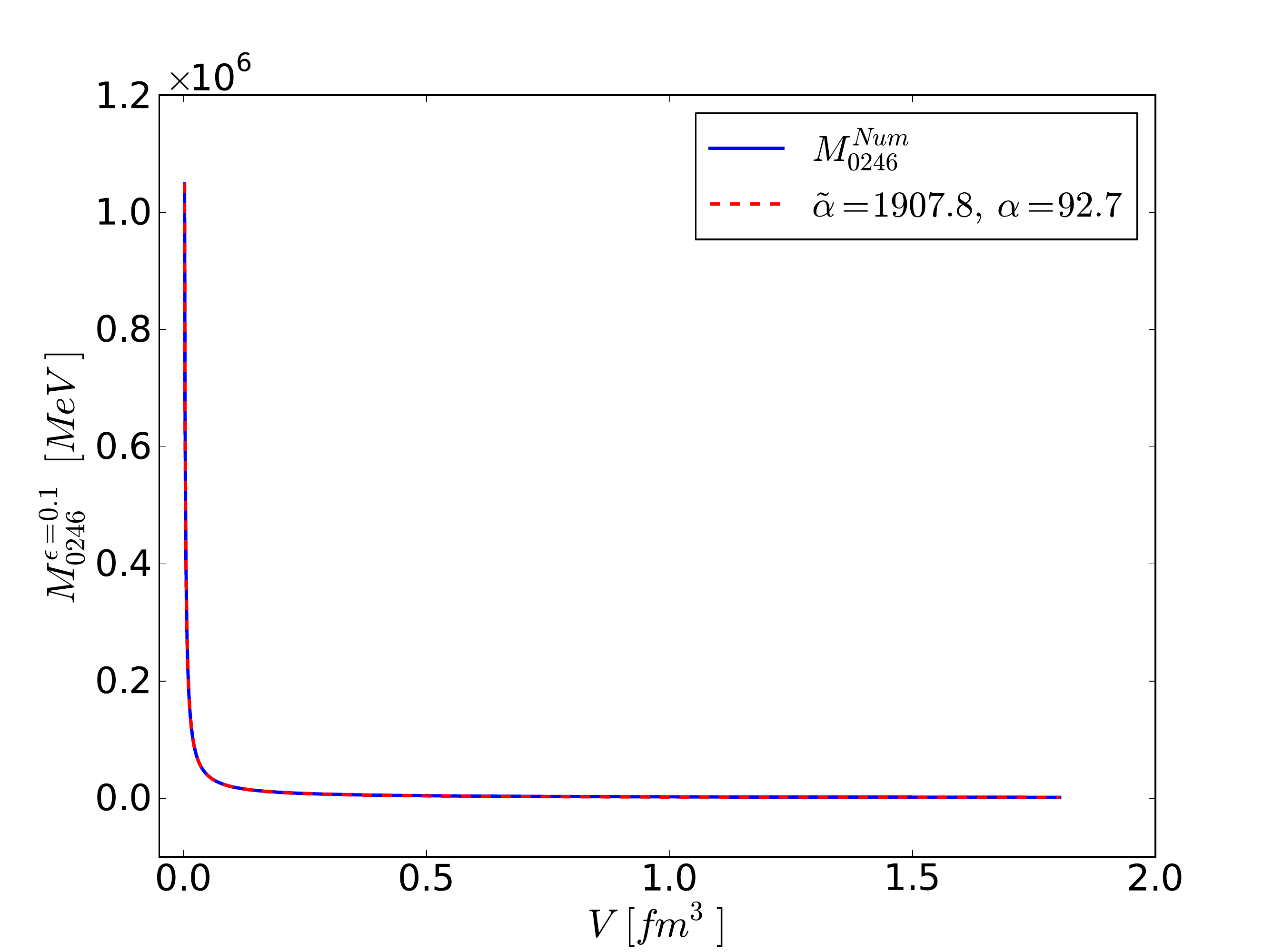}}
\subfigure[]{\includegraphics[totalheight=5.cm]{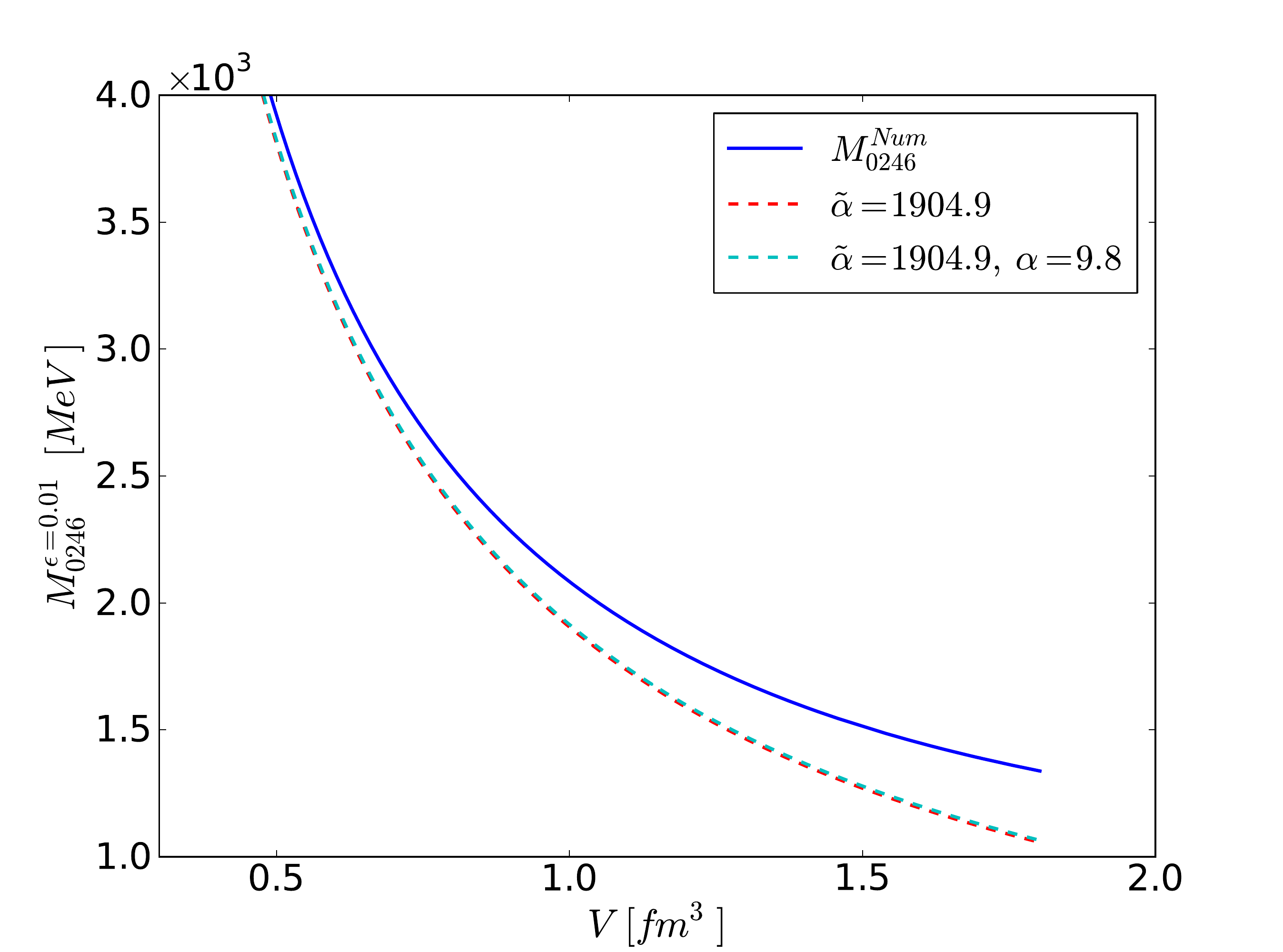}}
\subfigure[]{\includegraphics[totalheight=5.cm]{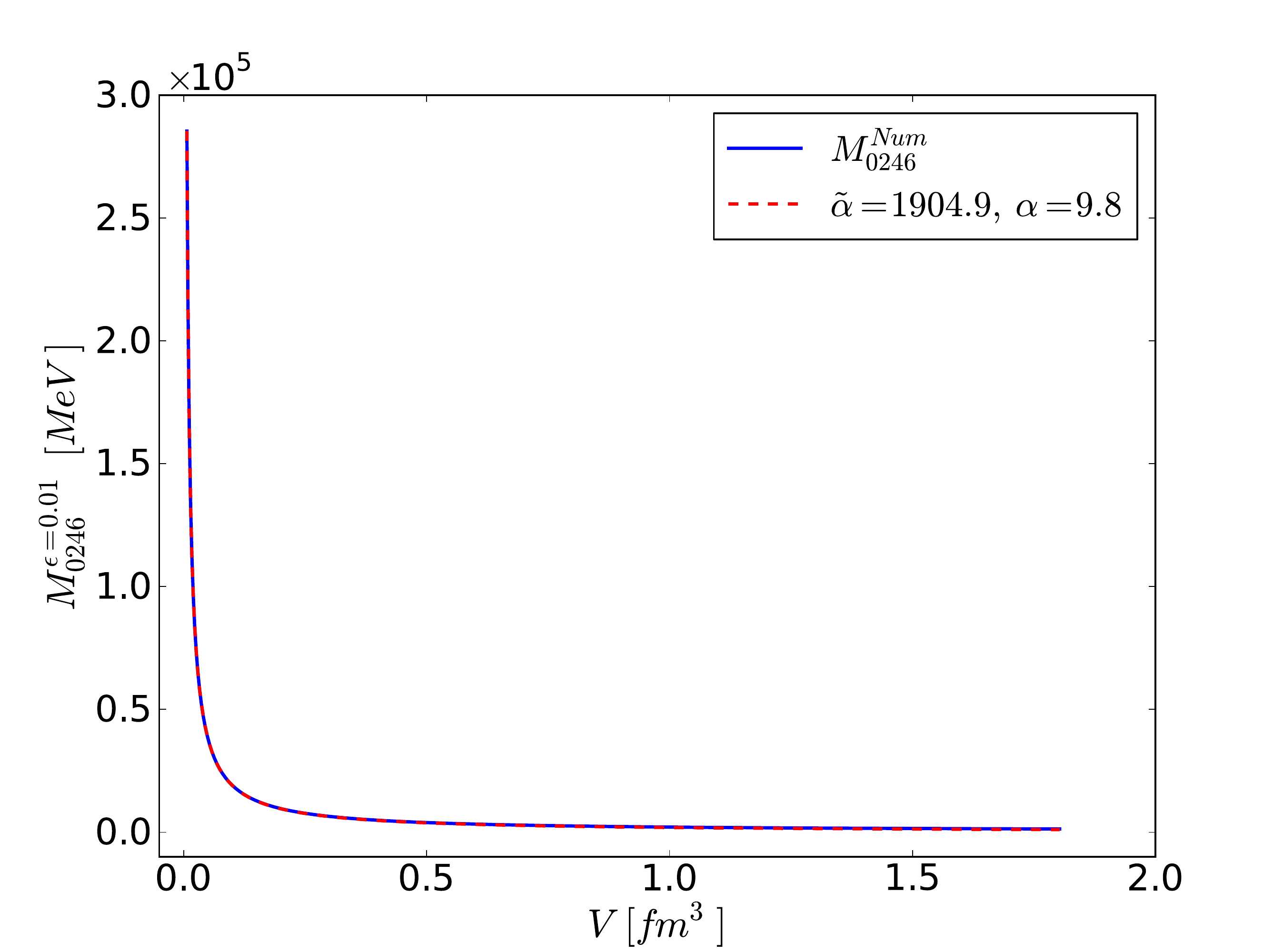}}
\subfigure[]{\includegraphics[totalheight=5.cm]{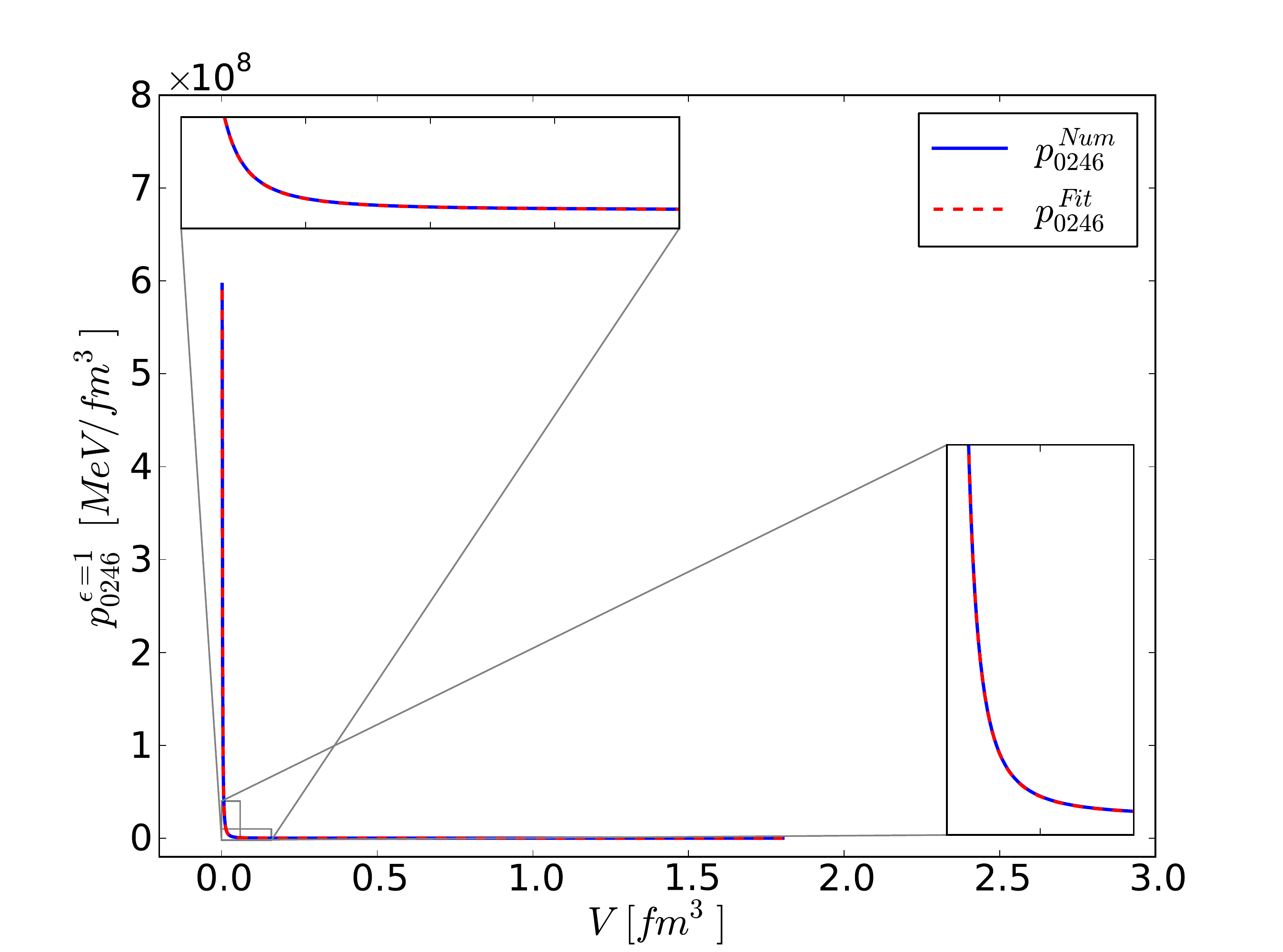}}
\subfigure[]{\includegraphics[totalheight=5.cm]{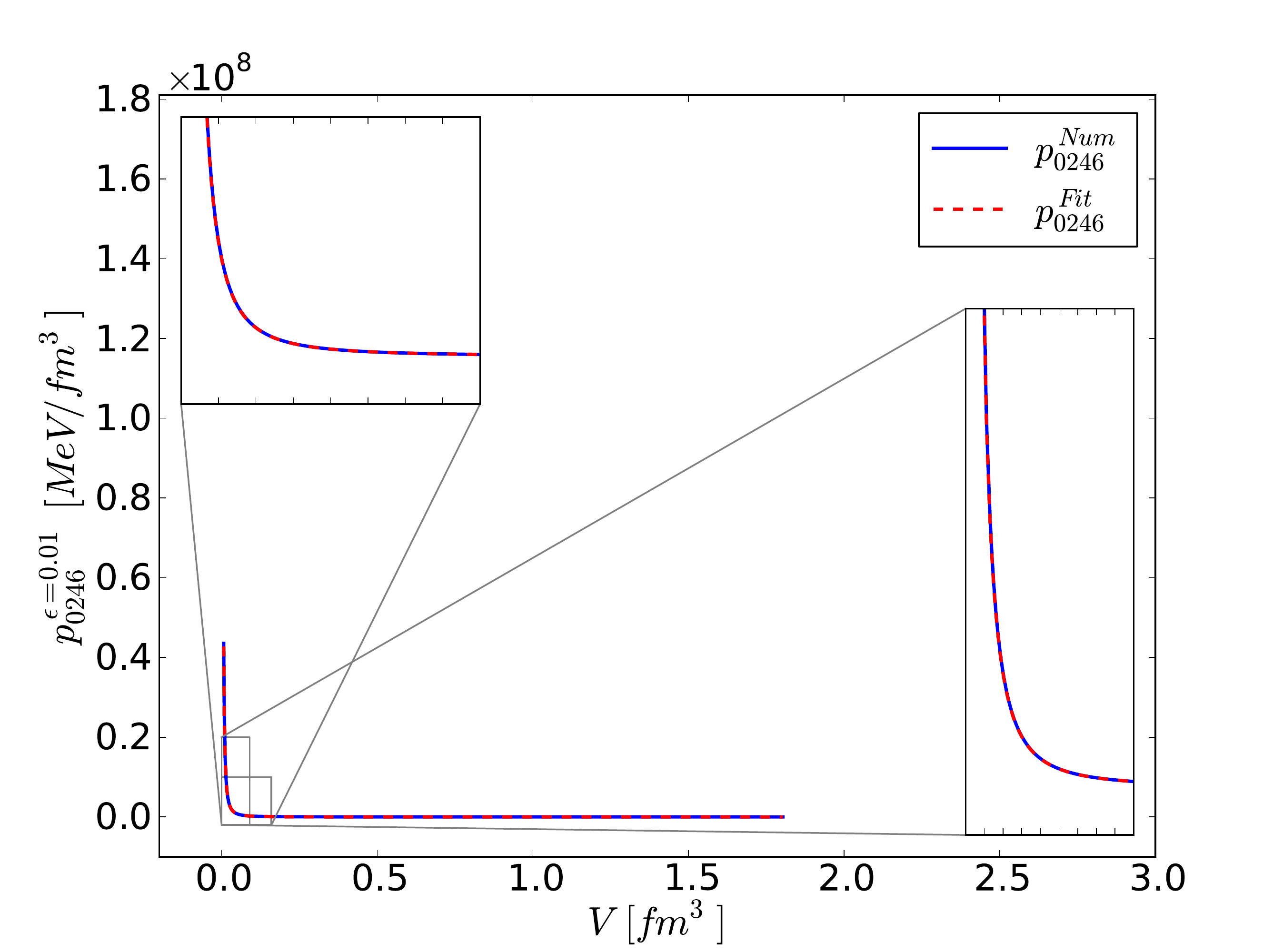}}
\caption{(Color online) Energy and pressure of B=1 skyrmion as a function of the volume for $E_{0246}$ model.}
\label{E0246}
\end{figure}
\begin{figure}
\subfigure[]{\includegraphics[totalheight=5.cm]{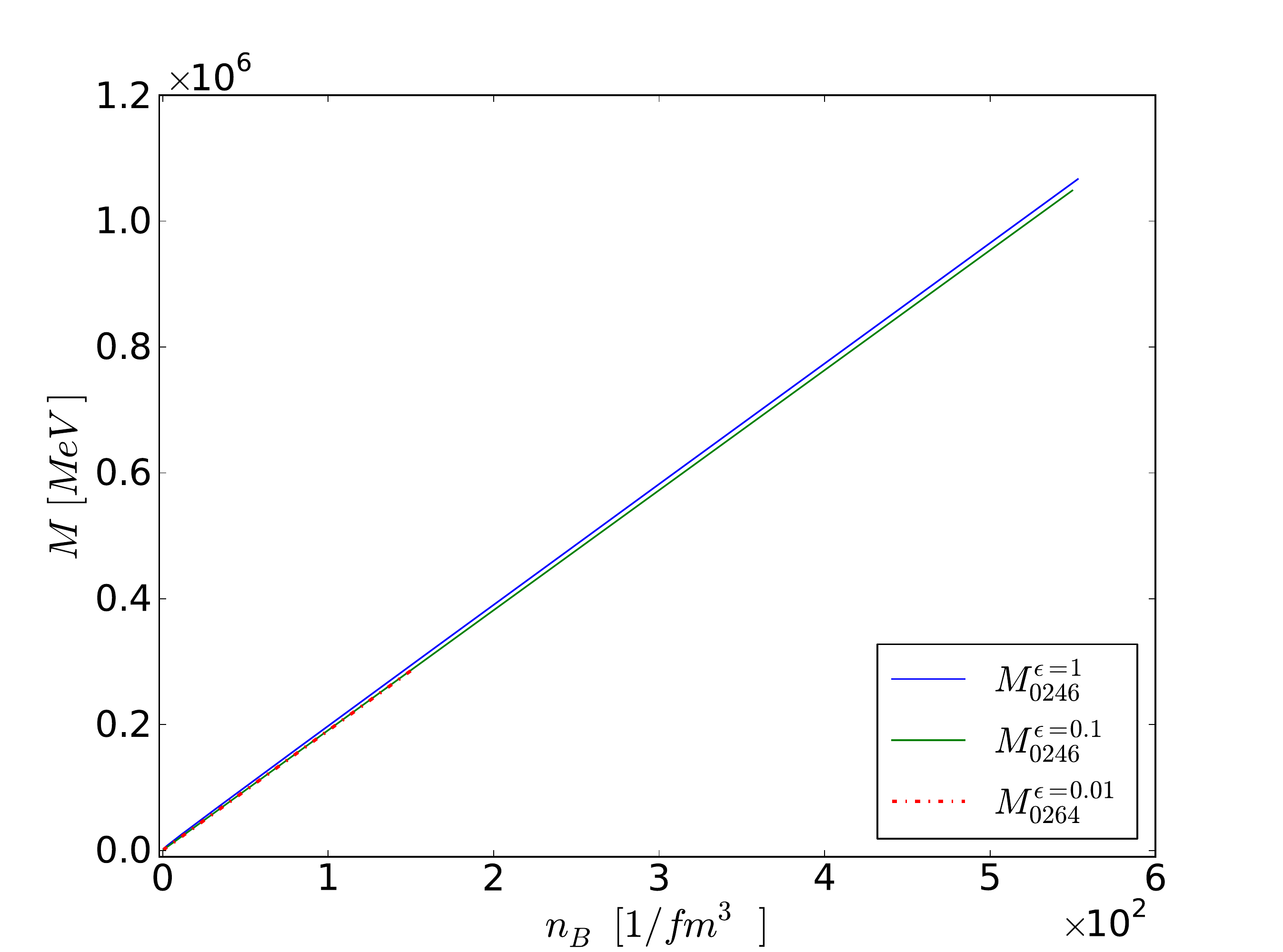}}
\subfigure[]{\includegraphics[totalheight=5.cm]{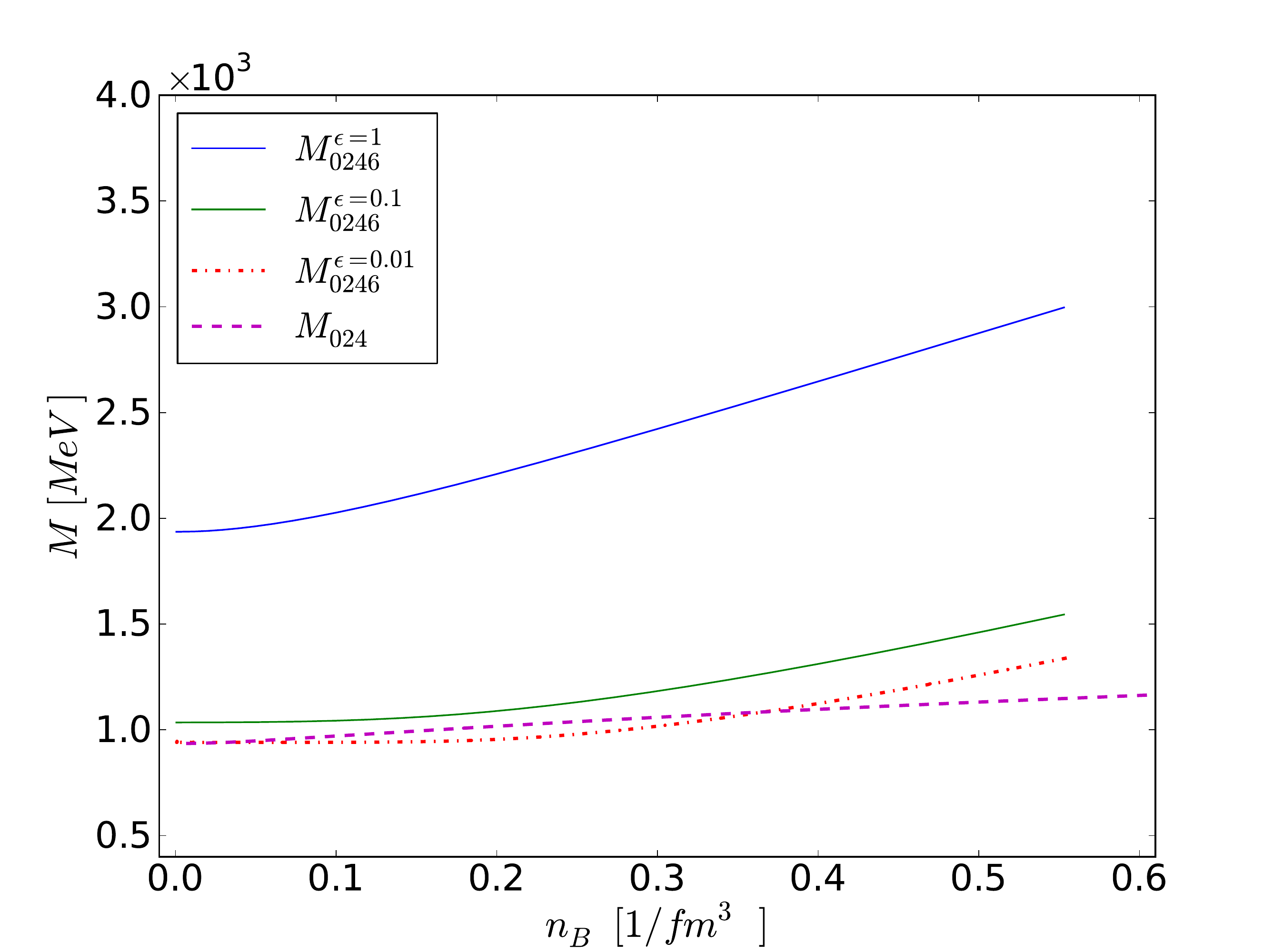}}
\subfigure[]{\includegraphics[totalheight=5.cm]{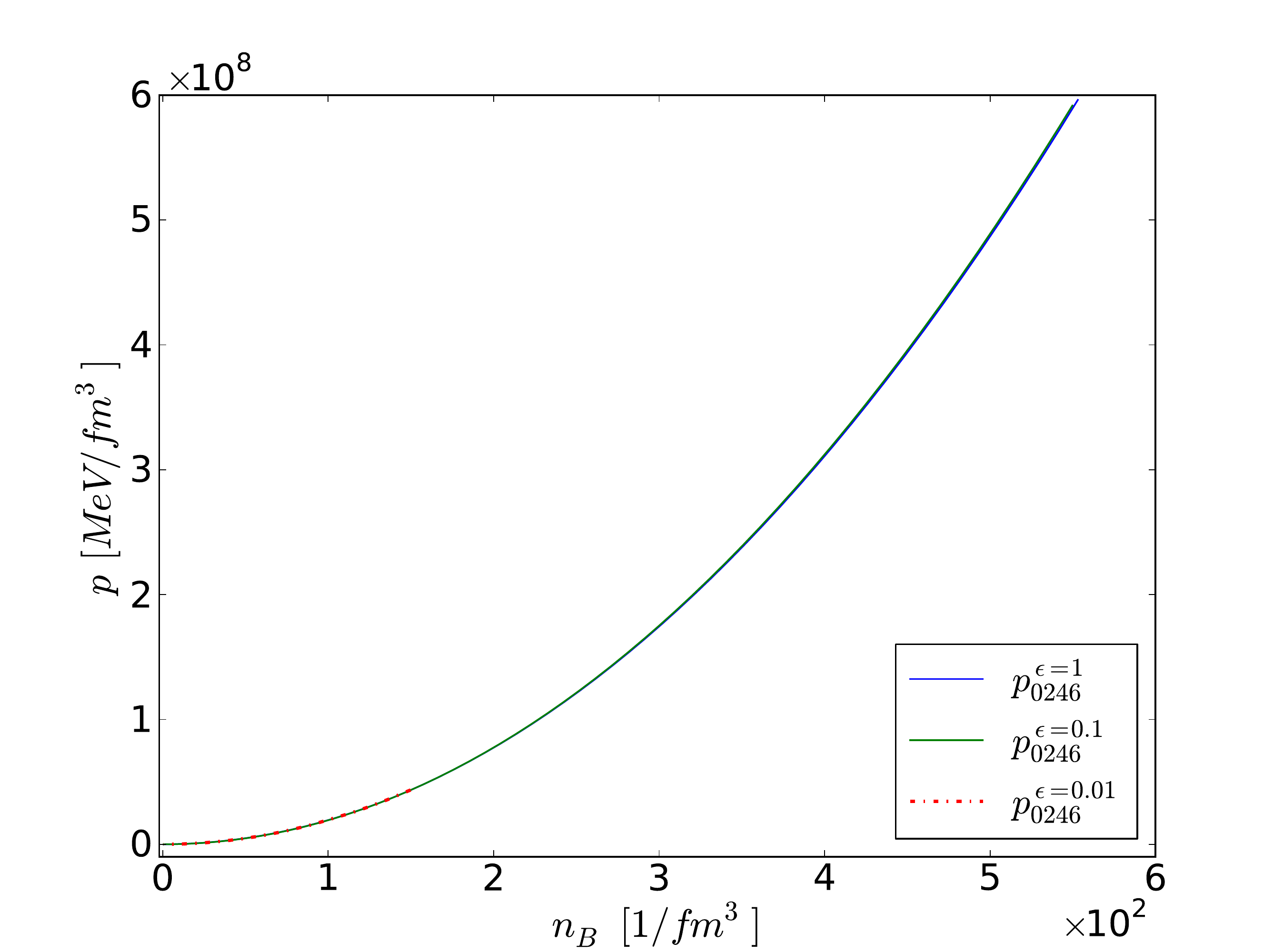}}
\subfigure[]{\includegraphics[totalheight=5.cm]{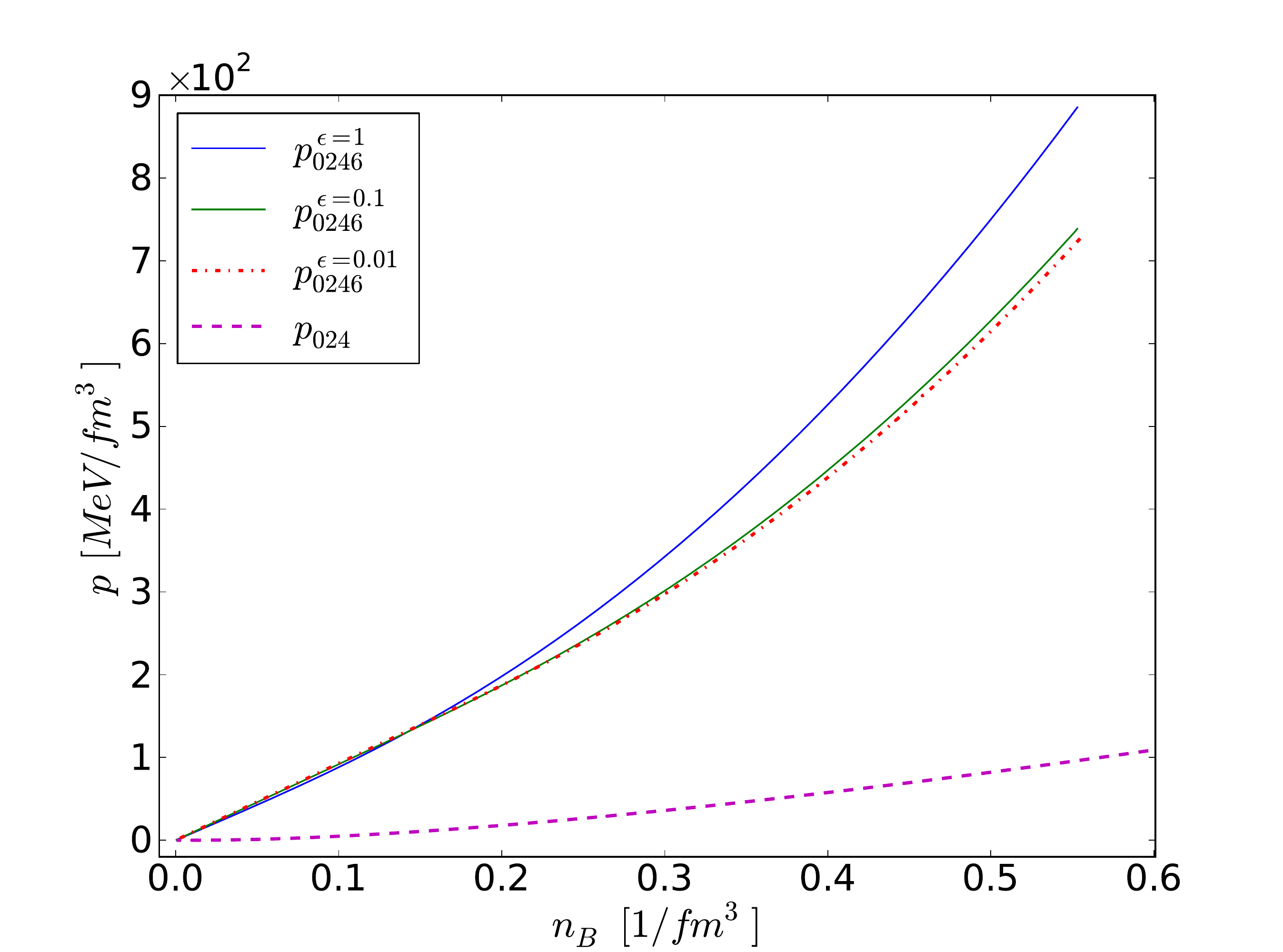}}
\subfigure[]{\includegraphics[totalheight=5.cm]{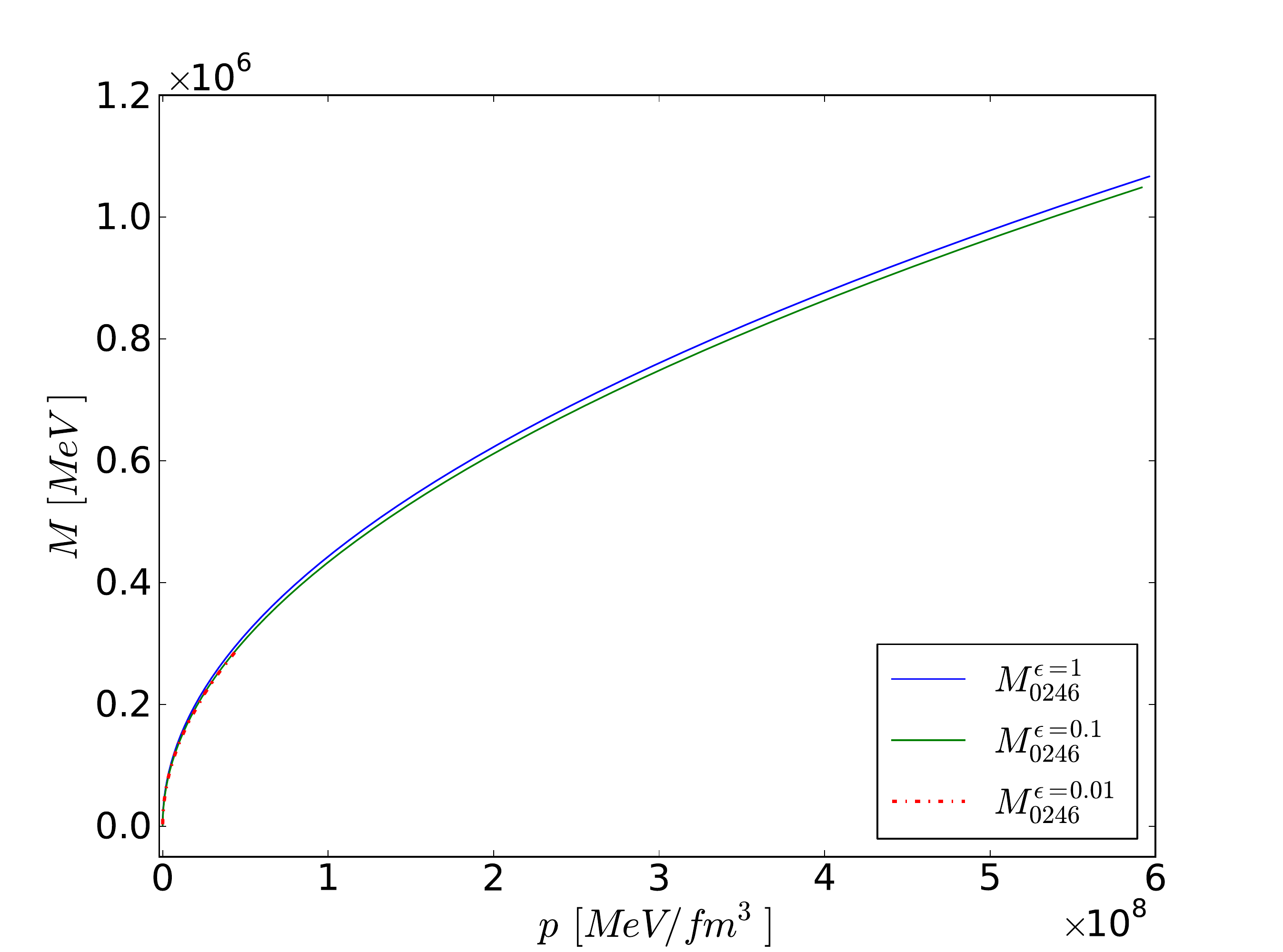}}
\subfigure[]{\includegraphics[totalheight=5.cm]{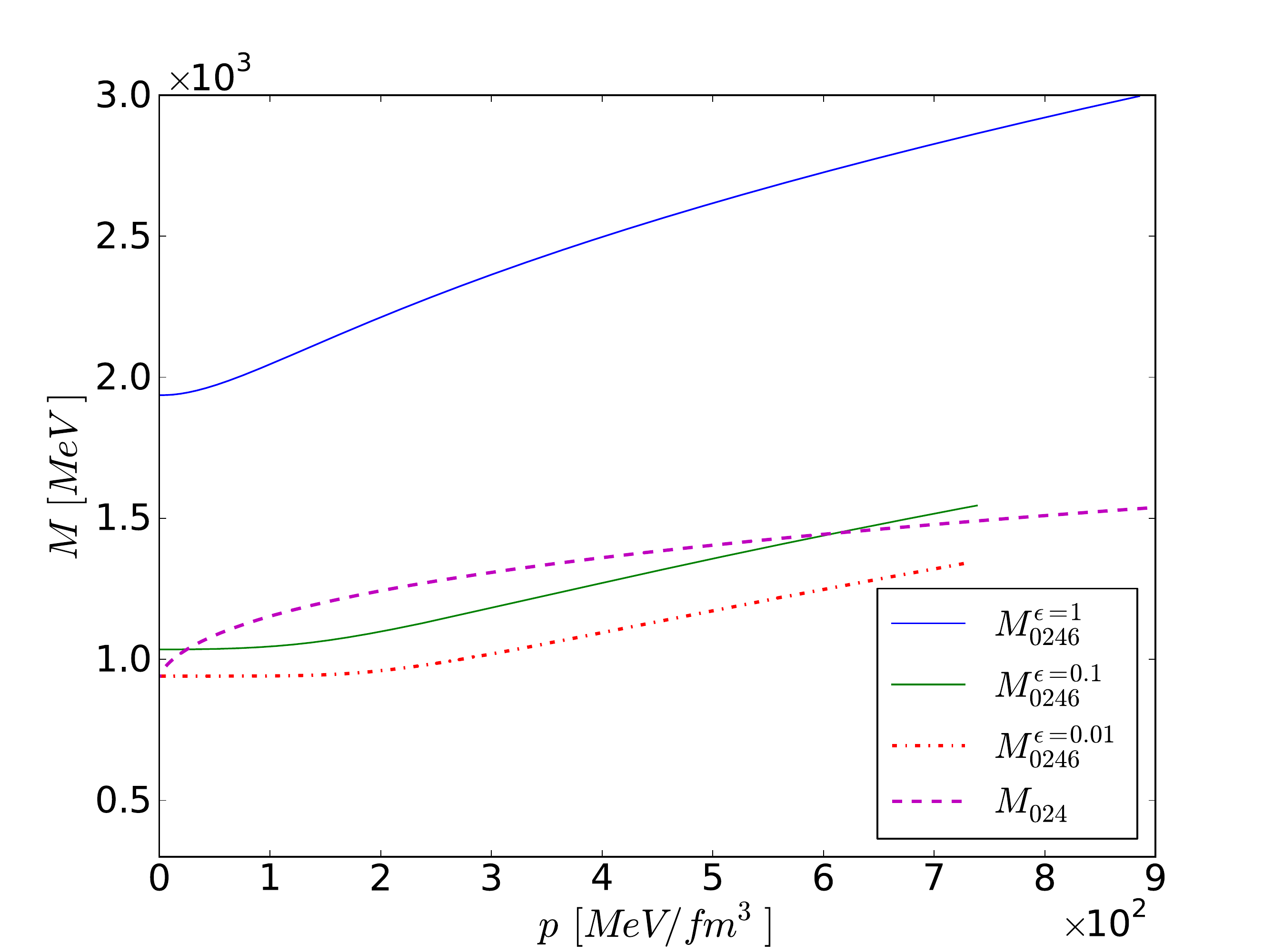}}
\subfigure[]{\includegraphics[totalheight=5.cm]{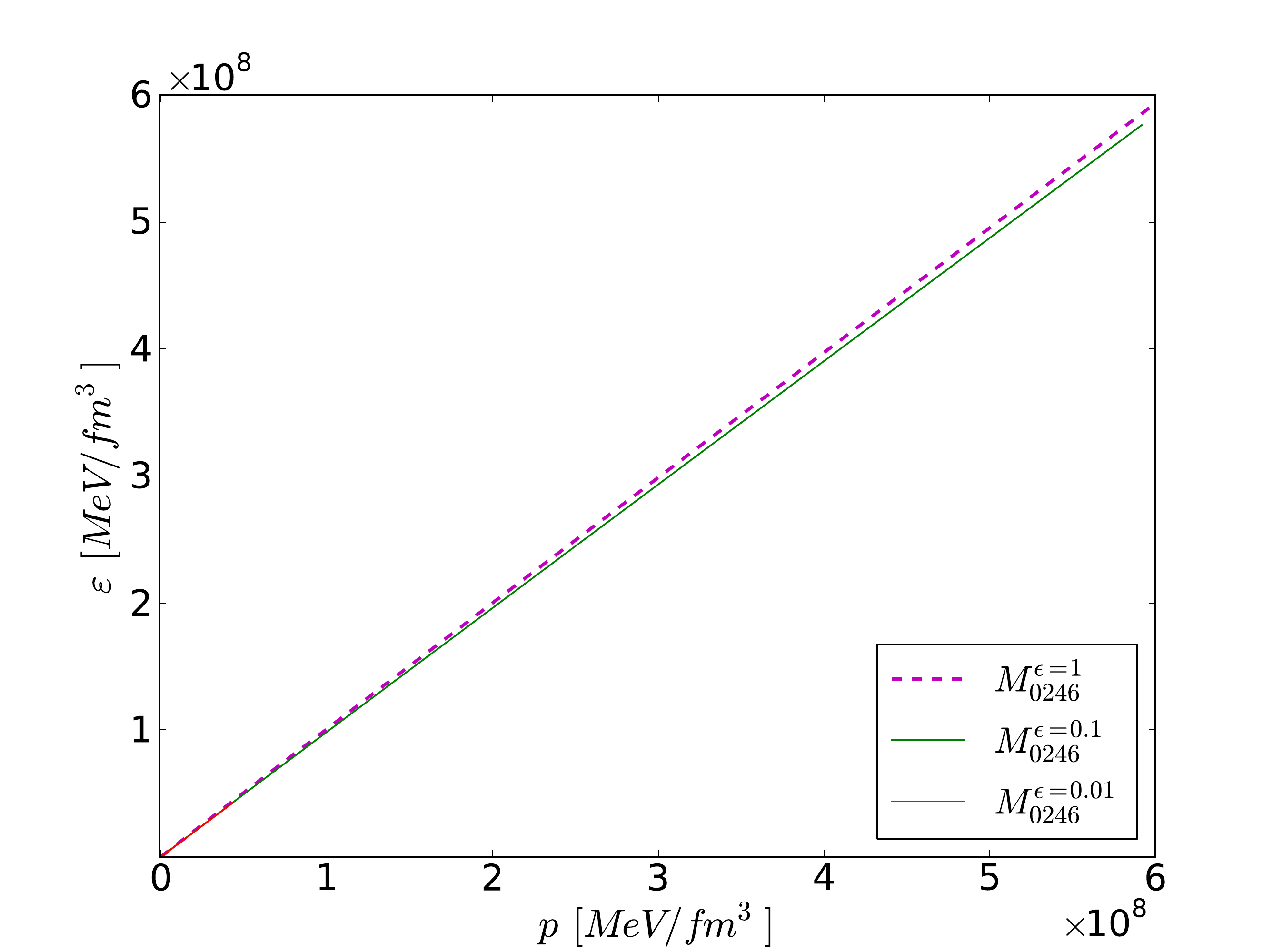}}
\subfigure[]{\includegraphics[totalheight=5.cm]{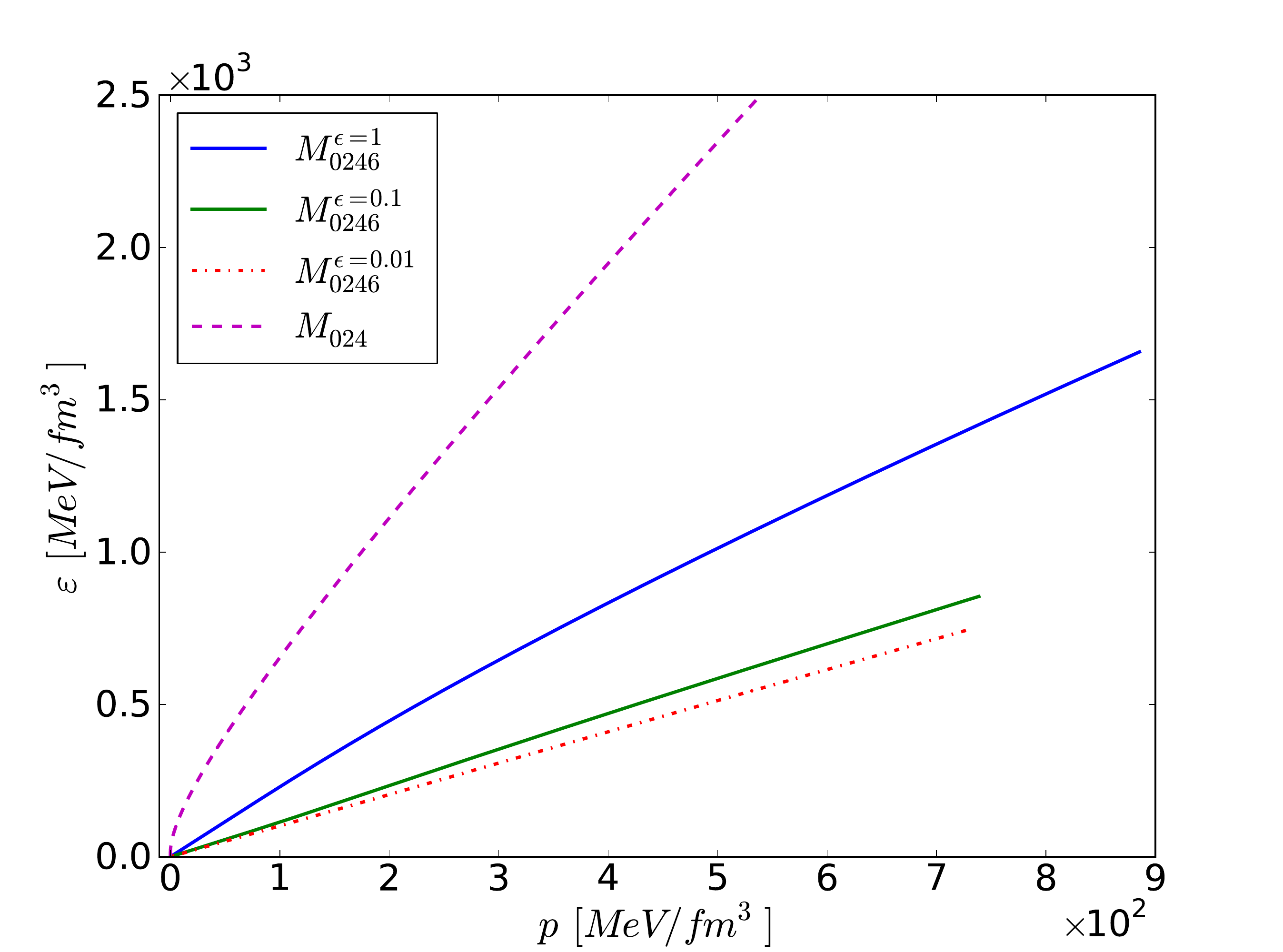}}
\caption{(Color online) Comparison plots for the B=1 skyrmion: (a)-(d)  mass and pressure as a function of the average baryon density; (e)-(f) mass as a function of pressure; (g)-(h) equation of state.}
\label{full_comp}
\end{figure}

\subsubsection{$\mbox{E}_{0246}$ model} \label{Sec_NumRes_nearBPS_E0246}
Finally, we consider the full near-BPS Skyrme model. As we are close to the BPS limit (for $\epsilon=0.1$ and $\epsilon=0.01$) it is reasonable to apply a different choice for the calibration. Namely, we fit the parameters of the BPS part of the model to the mass of the helium nucleon $m_{He}/4=931.75\; \mbox{MeV}$ and size of the nucleon $r_N=1.25 \; \mbox{fm}$. For the BPS Skyrme model we have
\be
E_{BPS}=\frac{2}{12\pi^2} 2\pi^2 \lambda \mu,
\ee
\be
R= \left( \frac{3\pi}{2}\right)^{1/3} \left( \frac{\lambda}{2\mu} \right)^{1/3}.
\ee
Then, $\lambda^2 = 2317 \; \mbox{Mev fm}^3$, $\mu^2=3372 \; \mbox{MeV fm}^{-3}$ in physical units or $\lambda^2=8.14$ and $\mu^2=2.19$ in the Skyrme units. In our numerical simulations we find the following values of the leading term in the energy-mass relation
\be
\tilde{\alpha}_{\epsilon=1}= 1929\; \mbox{MeV fm}^{3}, \;\;\; \tilde{\alpha}_{\epsilon=0.1}= 1908\; \mbox{MeV fm}^{3}, \;\;\; \tilde{\alpha}_{\epsilon=0.01}=1905 \; \mbox{MeV fm}^{3},
\ee
which can be compared with the theoretical value $\tilde{\alpha} = \frac{\lambda^2 \pi^2}{12} = 1905 \; \mbox{MeV fm}^3$. 

As expected, the sextic term gives the leading behavior for the mass-volume formula in the asymptotic regime. The subleading contribution, emerging from the quartic part of the action, has the form $ \alpha V^{-1/3}$, with the following values for the constant Fig.~\ref{E0246}
\be
\alpha_{\epsilon=1}= 924.7\; \mbox{MeV fm}, \;\;\; \alpha_{\epsilon=0.1}=92.7 \; \mbox{MeV fm}, \;\;\; \alpha_{\epsilon=0.01}= 9.8\; \mbox{MeV fm},
\ee  
which to a reasonable precision satisfies an obvious relation $\alpha_{\epsilon}=\epsilon \;  \alpha_{\epsilon=1}$. This can be treated as a test for our numerics. 

In Fig.~\ref{full_comp} we compare the near-BPS Skyrme model (for three values of $\epsilon$) with the usual (perturbative) Skyrme model. Of course, for a sufficiently large value of the average particle density the total energy (mass) as well as the pressure obey the relation which follows from the pure BPS model (sextic part), see Fig.~\ref{full_comp} (a), (c). For smaller values of $n_B$ (less squeezed configurations) i.e., closer to the equilibrium the relations are modified by a subleading behavior related to other terms of the action. In general, taking smaller $\epsilon$ results in a flatter $M-n_B$ curve for small $n_B$. This is perhaps related to the fact that the quartic term is more important for small densities and its strength is increased or reduced by a particular value of $\epsilon$. Furthermore, the $M-n_B$ and $p-n_B$ curves grow much faster for the full model ($M \sim n_B$ and $p \sim n_B^2$) than in the perturbative Skyrme model ($M\sim n_B^{1/3}$ and $p \sim n_B^{4/3}$). 
\section{Comparison to the Walecka model}
 Let us now understand the results from the point of view of the Walecka effective model \cite{walecka}, \cite{nuclmat}. The pressure and energy density are given as
 \be
 P=\frac{1}{2}\frac{g_\omega^2}{m_\omega^2} \bar{\rho}_B^2 - \frac{1}{2} \frac{g_\sigma^2}{m_\sigma^2} n_s^2+\frac{1}{4\pi^2} \left[ \left( \frac{2}{3} k_F^3-(m^*)^2k_F\right)E^*_F +(m^*)^4 \ln \frac{k_F+E^*_F}{m^*} \right],
 \ee
  \be
 \bar{\varepsilon}=\frac{1}{2}\frac{g_\omega^2}{m_\omega^2} \bar{\rho}_B^2 + \frac{1}{2} \frac{g_\sigma^2}{m_\sigma^2} n_s^2+\frac{1}{4\pi^2} \left[ \left( 2 k_F^3+(m^*)^2k_F\right)E^*_F -(m^*)^4 \ln \frac{k_F+E^*_F}{m^*} \right],
 \ee
 where the in-medium Fermi energy (chemical potential) and in-medium nucleon mass are
 \be
 E^*_F=\sqrt{k_F^2+(m^*)^2}, \;\;\;\; m^*=m - \frac{g_\sigma^2}{m_\sigma^2}n_s.
 \ee
The baryon and scalar densities at $T=0$ are
\be
\bar{\rho}_B=\frac{2k_F^3}{3\pi^2}, \;\;\;\; n_s=\frac{m^*}{\pi^2} \left[ k_FE^*_F-(m^*)^2\ln \frac{k_F+E_F^*}{m^*}\right].
\ee
It is straightforward to find the leading behaviour of the mean-field energy density as a function of the mean-field baryon density
\be
\bar{\varepsilon}=\frac{1}{2} \frac{g_\omega^2}{m_\omega^2} \bar{\rho}_B^2+   \frac{3}{4} \left( \frac{3\pi^2}{2} \right)^{1/3} \bar{\rho}_B^{4/3} +\frac{1}{2} \frac{m^2_Nm^2_\sigma}{g^2_\sigma},
\ee
where the leading part comes from the $\omega$ meson repulsion, the first subleading term is due to the free Fermion gas  while the second subleading term comes from the scalar meson (more specifically, we use that in the asymptotic regime the scalar condensate tends to a constant value),
\be
n_s=\frac{m_N m^2_\sigma}{g_\sigma^2}.
\ee
The results can be summarised as follows:
\begin{enumerate}
\item The asymptotic formulas for the equations of state in the full (i.e., containing the BPS part) Skyrme model and in the Walecka model coincide. Their functional dependence is exactly the same. The
leading part implying $\bar{\varepsilon}\simeq P$ comes from the baryon density current. In the Skyrme model it
enters via the sextic term while in the Walecka model it appears due to the $\omega$ meson repulsive
interaction dominating at high density. So, we may detect the $\omega$ meson in the Skyrme action - 
as an emergent object hidden in the sextic term. 

\item Furthermore, there is a coincidence between the first subleading terms, which behave as $\bar{\rho}_B^{4/3}$. In the Skyrme model it is generated by the usual Skyrme (quartic) part of the action while in the Walecka it is due to the ultra relativistic free fermion (nucleon) behaviour. 

\item Moreover, we got a constant term whose origin is found to be the potential (the Skyrme model) or scalar density condensation (the Walecka model). This gives further support to the idea that the potential corresponds to the $\sigma$ meson and provides the attractive long distance force.

\item Let us note that there are no further terms in the large density limit of the Walecka model which could correspond to the term generated by the sigma model part of the Skyrme model. Namely, a term behaving as $\bar{\rho}^a$ with $ 4/3 > a >0$.
\end{enumerate} 
 
The observed similarity between the equations of state can be used to get some quantitative insight into $\omega$ and $\sigma$ mesons within  the Skyrme model framework, where such low energy particles are not explicitly included. If we compare the coefficients in the expression above we find
\be
\pi^4\lambda^2=\frac{1}{2} \frac{g_\omega^2}{m^2_\omega}, 
\ee
\be
\alpha=\frac{3}{4} \left( \frac{3\pi^2}{2} \right)^{1/3},
\ee
\be
\tilde{\beta} = \frac{1}{2} \frac{m_N^2m^2_\sigma}{g^2_\sigma}.
\ee
The typical parameter values in the Walecka model are
\be
m_N=939 \;\mbox{MeV}, \;\;\; m_\omega=783 \;\mbox{MeV}, \;\;\; m_\sigma = (500-600) \;\mbox{MeV},
\ee
\be
\frac{g^2_\omega}{4\pi}= 14.717, \;\;\;\; \frac{g^2_\sigma}{4\pi}=9.537.
\ee
This gives (after taking into account the Plank constant $\hbar = 197.3 \;\mbox{MeV fm}$)
\be
\frac{1}{2\pi^4} \frac{g_\omega^2}{m^2_\omega}= 12\; \mbox{MeV fm}^3, \;\;\; \frac{3}{4} \left( \frac{3\pi^2}{2} \right)^{1/3} = 362 \; \mbox{MeV fm},\;\;\; \frac{1}{2} \frac{m_N^2m^2_\sigma}{g^2_\sigma} = (120-175)\; \mbox{MeV fm}^{-3}.
\ee
This can be compared with values derived for the generalized Skyrme model. For the BPS model (or near-BPS version with small $\epsilon$, for example equals to 0.01)  $\lambda^2$ and $\mu^2$ for the pure BPS Skyrme model in the case of the step-function potential (with the parameter fit to the nuclear saturation density and the binding energy per nucleon of infinite nuclear matter)
\be
\lambda^2 = 31\; \mbox{MeV fm}^3, \;\;\;  \beta= 71 \; \mbox{MeV fm}^{-3}.
\ee
As we know, in the case of the usual (perturbative) Skyrme model, with the calibration assumed here one gets
\be
\alpha = 924 \; \mbox{MeV fm}^3.
\ee 
Thus, for near-BPS Skyrme models this should be multiplied by the $\epsilon$ constant, which leads to a rather small number in comparison to the Walecka model. 
\section{Conclusions}
In the present work we have analyzed thermodynamical properties of skyrmionic matter at zero temperature for the most general Skyrme theory that possesses both Poincare invariance and a standard hamiltonian.  

The main result is that the sextic term is responsible for the leading behavior at high pressure and therefore is unavoidable for any realistic application of the Skyrme model for a description of nuclear matter at high densities. As we found, the dependence of all global characteristics (total energy, pressure, average energy density) on the geometric volume tends to analytical relations derived for the BPS Skyrme model as the volume decreases. The same happens with the mean-field equation of state which approaches a much stiffer form, $\bar{\varepsilon}=P$, instead of $\bar{\varepsilon}=3P$ for the usual Skyrme model without the sextic part. Similarly, the mean-field energy density grows much more rapidly with the average baryon density - the previously found $\bar{\varepsilon} \sim \bar{\rho}_B^{4/3}$ relation \cite{kut} is replaced by $\bar{\varepsilon} \sim \bar{\rho}_B^2$. This resolves a long standing discrepancy between the Skyrme model and other conventional models of nuclear matter. Indeed, the behavior generated by the sextic term coincides with the corresponding relation derived by Bethe and Johnson \cite{bethe} (non-relativistic) as well as Walecka \cite{walecka} (relativistic).
Moreover, the inclusion of this term might not only modify the phase diagram of cold nuclear matter or neutron stars, but can also influence the scattering of nuclei \cite{foster}. 
It is also an intriguing observation that an effective low energy model of QCD, which here is represented by the Skyrme model, has an solvable and integrable limit (in the sense of the generalised integrability) at asymptotically high density (pressure). 

The first subleading contribution is provided by the Skyrme term, i.e., by the quartic part of the action, and it can modify the thermodynamical properties at medium densities (pressure). Although we analytically derived the functional dependence for this contribution, we found only a lower bound for the true value of the proportionality constant $\alpha$. The true value was obtained numerically. The next subleading terms were found to be related to the attractive channels of the model, which emerge from the sigma model term and the potential. However, due to the strong nonlinearity of the theory, there is a significant mixing in these subleading contributions which results in the appearance of some new, effective terms in the energy-volume relation. In particular, in the full $E_{0246}$ model, instead of terms $V^{1/3}$ and $V$ - which are clearly visible in the $E_{24}$ and $E_{04}$ submodels, we find an effective $V^a$, $a \in (0,1)$ dependence. 

Let us also note that, from the point of view of nuclear matter properties, our results are relevant for the high and medium pressure regimes. Definitely, the obtained mean-field equations of states are not applicable at the saturation density, where we got zero energy density. In fact, it is a common feature of Skyrme type models (with the exception of the BPS Skyrme model and the $\mbox{E}_{04}$ submodel) that $B=1$ solitons are infinitely extended at equilibrium ($P=0)$. Then the equilibrium volume is infinite, which leads to a vanishing mean-field energy density. This is not the expected behaviour for infinite $B$. In the original Skyrme model, e.g, at infinite $B$ a crystal-like configuration with a finite volume per baryon number is conjectured to be the true minimizer even for small $P$. 
This does not mean that one cannot use finite charge skyrmions to understand thermodynamical properties of nuclear matter or neutron stars. It simply means that one has to be very careful with the mean-field approximation. The full (non-mean field) densities are still fine. We want to underline that in our computations we solve the full field equations which provides us with full (spatially dependent) densities. Then they are averaged leading to the mean-field equation of state. This is an alternative approach to mean-field computations previously done for Skyrme-related models \cite{eos-sk}.

If compared with another widely used effective model, i.e., the Walecka model, we found a correspondence between the $\omega$ meson interaction and the sextic term in the Skyrme model. Both generate the same mean-field equation of state, $\bar{\varepsilon}=P$. This again supports the crucial role played by the sextic term (and the BPS Skyrme model part) in the Skyrme framework. There is also a clear correspondence between the main subleading terms related to the ultra relativistic free fermion gas (Walecka) and the quartic Skyrme term (Skyrme model). The next subleading term in the mean-field equation of state in the Walecka model emerges due to the attractive  scalar meson interaction. A similar constant term was found in the Skyrme model as well and it is generated by the potential. This provides some support to the natural expectation that the potential in the Skyrme model corresponds to the $\sigma$ meson in the Walecka model. However, the situation is much more involved as in the Skyrme model the $\bar{\rho}_B^0$ term is accompanied by a $\bar{\rho}_B^{2/3}$ term, which is generated by another source of the attractive force, namely the sigma model part. Furthermore, there is no such term in the Walecka model. In addition, the observed effective mixing between these attractive channels does not allow for a clear understanding of a possible emergence of the hidden scalar meson in the Skyrme model, in a similar fashion as it has been recently understood in the case of the emergent  $\omega$ meson \cite{omegaBPS,foster2}. In any case, it is an interesting feature of the Skyrme model that it does include some effects related to low lying mesons (here the $\omega$ and $\sigma$) although they are not introduced at the level of fields.

\vspace*{0.2cm} 

There are several natural directions in which the present investigation can and should be further developed. 
First of all, one should find the mean-field equation of state in the limit of infinite nuclear matter. This can be achieved by considering the general Skyrme model, with the sextic term included, on a torus. The main advantage of this set-up follows from the fact that the minimum energy solution occurs for a particular size (volume) of the torus. Therefore, the equilibrium solutions (at zero pressure) do not correspond to infinite geometrical volume which, as we already commented, is a drawback of finite charge solutions. As a consequence, equilibrium solutions possess finite mean-field energy and particle densities. Besides the derivation of the equation of state and understanding how it differs from the charge one results (which should be negligible in the high pressure limit), one can expect the appearance of some new phases (as, for example, a half-skyrmion phase). It would be desirable to fully discover the Skyrme phase diagram with the coupling constants as free parameters. 

Second, one can ask the more mathematical question what is the energy minimiser in a given topological sector on a given compact manifold with a certain volume (see for example \cite{finite}). Obviously, the form of the solution does not depend only on the volume of the manifold but also on its geometry (shape), although, as has been shown here, the asymptotic (small volume) behaviour is geometry independent. 

Third, it would also be desirable to extend our results to non-zero temperature, $T>0$. The simplest way to accomplish this problem is to apply the caloron approach, i.e., to use euclidean periodic solutions in 4 dimension \cite{temp}.  Alternatively, at least in the BPS limit, one can try to compute the partition function which is essentially given by an integration over the moduli space \cite{manton-t}. Unfortunately, for the BPS Skyrme model this group has infinite dimension which makes the problem quite involved.  

\section*{Acknowledgement}
The authors acknowledge financial support from the Ministry of Education, Culture, and Sports, Spain (Grant No. FPA2008-01177), the Xunta de Galicia (Grant No. INCITE09.296.035PR and Conselleria de Educacion), the Spanish Consolider-Ingenio 2010 Programme CPAN (CSD2007-00042), and FEDER.  AW thanks S. Krusch for discussion and D. Harland for inspiring remarks on the average pressure issue. Further, the authors thank J. Sanchez-Guillen for helpful comments.

\end{document}